\documentclass[12pt]{article}
\usepackage{ssi}
\usepackage{times}
\usepackage{epsfig}

\newcommand{\postscript}[2]{\setlength{\epsfxsize}{#2\hsize}
   \centerline{\epsfbox{#1}}}
\newcommand{\sign}{\:\!\text{sign}\:\!}

\newcommand{\pb}{ \text{pb}}

\newcommand{\cm}{\text{cm}}
\newcommand{\km}{\text{km}}
\newcommand{\kg}{\text{kg}}
\newcommand{\s}{\text{s}}
\newcommand{\yr}{\text{yr}}
\newcommand{\sr}{\text{sr}}
\newcommand{\ev}{\text{eV}}
\newcommand{\kev}{\text{keV}}
\newcommand{\mev}{\text{MeV}}
\newcommand{\gev}{\text{GeV}}
\newcommand{\tev}{\text{TeV}}
\newcommand{\K}{\text{K}}

\newcommand{\mgaugino}{M_{1/2}}
\newcommand{\mweak}{M_{\text{weak}}}

\newcommand{\mgut}{M_{\text{GUT}}}
\newcommand{\mstar}{M_*}

\newcommand{\mmess}{M_m}
\newcommand{\tb}{\tan\beta}

\newcommand{\photino}{\tilde{\gamma}}
\newcommand{\gravitino}{\tilde{G}}
\newcommand{\stau}{\tilde{\tau}}
\newcommand{\mchi}{m_{\chi}}
\newcommand{\ethr}{E_{\text{thr}}}
\newcommand{\NLSP}{{\text{NLSP}}}

\newcommand{\OmegaDM}{\Omega_{\text{DM}}}
\newcommand{\Omegachi}{\Omega_{\chi}}
\newcommand{\nequ}{n_{\text{eq}}}
\newcommand{\Yeq}{Y_{\text{eq}}}
\newcommand{\zetaEM}{\zeta_{\text{EM}}}
\newcommand{\zetahad}{\zeta_{\text{had}}}
\newcommand{\epsEM}{\varepsilon_{\text{EM}}}
\newcommand{\epshad}{\varepsilon_{\text{had}}}

\newcommand{\etal}{{\em et al.}}

\newcommand{\eqref}[1]{Eq.~(\ref{#1})}

\newcommand{\secref}[1]{Sec.~\ref{sec:#1}}
\newcommand{\secsref}[2]{Secs.~\ref{sec:#1} and \ref{sec:#2}}

\newcommand{\figref}[1]{Fig.~\ref{fig:#1}}
\newcommand{\figsref}[2]{Figs.~\ref{fig:#1} and \ref{fig:#2}}

\newcommand{\bold}[1]{\mbox{\boldmath $#1$}}
\newcommand{\text}[1]{{\rm #1}}
\newcommand{\mtext}[1]{\mbox{{\rm #1}}}
\newcommand{\agt}{ \mathop{}_{\textstyle \sim}^{\textstyle >} }
\newcommand{\alt}{ \mathop{}_{\textstyle \sim}^{\textstyle <} }

\def\APJ#1#2#3{Ap. J. {\bf #1}, #2 (#3)}

\begin{document}

\sloppy

\title{SUPERSYMMETRY AND COSMOLOGY}

\author{Jonathan L.~Feng\thanks{
\copyright\ 2004 by Jonathan L.~Feng. }\\ [0.4cm]
Department of Physics and Astronomy \\
University of California, Irvine, CA 92697 \\[0.4cm]
}

\maketitle
\begin{abstract}%
\baselineskip 16pt 

Cosmology now provides unambiguous, quantitative evidence for new
particle physics.  I discuss the implications of cosmology for
supersymmetry and vice versa. Topics include: motivations for
supersymmetry; supersymmetry breaking; dark energy; freeze out and
WIMPs; neutralino dark matter; cosmologically preferred regions of
minimal supergravity; direct and indirect detection of neutralinos;
the DAMA and HEAT signals; inflation and reheating; gravitino dark
matter; Big Bang nucleosynthesis; and the cosmic microwave background.
I conclude with speculations about the prospects for a microscopic
description of the dark universe, stressing the necessity of diverse
experiments on both sides of the particle physics/cosmology interface.
\end{abstract}


\pagestyle{plain}

\tableofcontents

\section{Introduction}
\label{sec:introduction}

Not long ago, particle physicists could often be heard bemoaning the
lack of unambiguous, quantitative evidence for physics beyond their
standard model.  Those days are gone.  Although the standard model of
particle physics remains one of the great triumphs of modern science,
it now appears that it fails at even the most basic level ---
providing a reasonably complete catalog of the building blocks of our
universe.

Recent cosmological measurements have pinned down the amount of
baryon, matter, and dark energy in the
universe~\cite{Spergel:2003cb,Tegmark:2003ud}.  In units of the
critical density, these energy densities are
\begin{eqnarray}
\Omega_B &=& 0.044 \pm 0.004  \\
\Omega_{\text{matter}} &=& 0.27 \pm 0.04 \\
\Omega_{\Lambda} &=& 0.73 \pm 0.04 \ ,
\end{eqnarray}
implying a non-baryonic dark matter component with
\begin{equation}
0.094 < \OmegaDM h^2 < 0.129 ~~\text{(95\%\ CL)} \ ,
\end{equation}
where $h \simeq 0.71$ is the normalized Hubble expansion rate.  Both
the central values and uncertainties were nearly unthinkable even just
a few years ago.  These measurements are clear and surprisingly
precise evidence that the known particles make up only a small
fraction of the total energy density of the universe.  Cosmology now
provides overwhelming evidence for new particle physics.

At the same time, the microscopic properties of dark matter and dark
energy are remarkably unconstrained by cosmological and astrophysical
observations.  Theoretical insights from particle physics are
therefore required, both to suggest candidates for dark matter and
dark energy and to identify experiments and observations that may
confirm or exclude these speculations.

Weak-scale supersymmetry is at present the most well-motivated
framework for new particle physics.  Its particle physics motivations
are numerous and are reviewed in \secref{susyessentials}.  More than
that, it naturally provides dark matter candidates with approximately
the right relic density.  This fact provides a strong, fundamental,
and completely independent motivation for supersymmetric theories.
For these reasons, the implications of supersymmetry for cosmology,
and vice versa, merit serious consideration.

Many topics lie at the interface of particle physics and cosmology,
and supersymmetry has something to say about nearly every one of them.
Regrettably, spacetime constraints preclude detailed discussion of
many of these topics.  Although the discussion below will touch on a
variety of subjects, it will focus on dark matter, where the
connections between supersymmetry and cosmology are concrete and rich,
the above-mentioned quantitative evidence is especially tantalizing,
and the role of experiments is clear and promising.

Weak-scale supersymmetry is briefly reviewed in
\secref{susyessentials} with a focus on aspects most relevant to
astrophysics and cosmology.  In \secsref{neutralino}{gravitino} the
possible roles of neutralinos and gravitinos in the early universe are
described.  As will be seen, their cosmological and astrophysical
implications are very different; together they illustrate the wealth
of possibilities in supersymmetric cosmology. I conclude in
\secref{synergy} with speculations about the future prospects for a
microscopic understanding of the dark universe.

\section{Supersymmetry Essentials}
\label{sec:susyessentials}

\subsection{A New Spacetime Symmetry}

Supersymmetry is an extension of the known spacetime
symmetries~\cite{susyintros}.  The spacetime symmetries of rotations,
boosts, and translations are generated by angular momentum operators
$L_i$, boost operators $K_i$, and momentum operators $P_{\mu}$,
respectively.  The $L$ and $K$ generators form Lorentz symmetry, and
all 10 generators together form Poincare symmetry.  Supersymmetry is
the symmetry that results when these 10 generators are further
supplemented by fermionic operators $Q_{\alpha}$.  It emerges
naturally in string theory and, in a sense that may be made
precise~\cite{Haag:1974qh}, is the maximal possible extension of
Poincare symmetry.

If a symmetry exists in nature, acting on a physical state with any
generator of the symmetry gives another physical state.  For example,
acting on an electron with a momentum operator produces another
physical state, namely, an electron translated in space or time.
Spacetime symmetries leave the quantum numbers of the state invariant
--- in this example, the initial and final states have the same mass,
electric charge, etc.

In an exactly supersymmetric world, then, acting on any physical state
with the supersymmetry generator $Q_{\alpha}$ produces another
physical state. As with the other spacetime generators, $Q_{\alpha}$
does not change the mass, electric charge, and other quantum numbers
of the physical state.  In contrast to the Poincare generators,
however, a supersymmetric transformation changes bosons to fermions
and vice versa.  The basic prediction of supersymmetry is, then, that
for every known particle there is another particle, its superpartner,
with spin differing by $\frac{1}{2}$.

One may show that no particle of the standard model is the
superpartner of another.  Supersymmetry therefore predicts a plethora
of superpartners, none of which has been discovered. Mass degenerate
superpartners cannot exist --- they would have been discovered long
ago --- and so supersymmetry cannot be an exact symmetry.  The only
viable supersymmetric theories are therefore those with non-degenerate
superpartners.  This may be achieved by introducing
supersymmetry-breaking contributions to superpartner masses to lift
them beyond current search limits.  At first sight, this would appear
to be a drastic step that considerably detracts from the appeal of
supersymmetry.  It turns out, however, that the main virtues of
supersymmetry are preserved even if such mass terms are introduced.
In addition, the possibility of supersymmetric dark matter emerges
naturally and beautifully in theories with broken supersymmetry.

\subsection{Supersymmetry and the Weak Scale}

\begin{figure}[tb]
\postscript{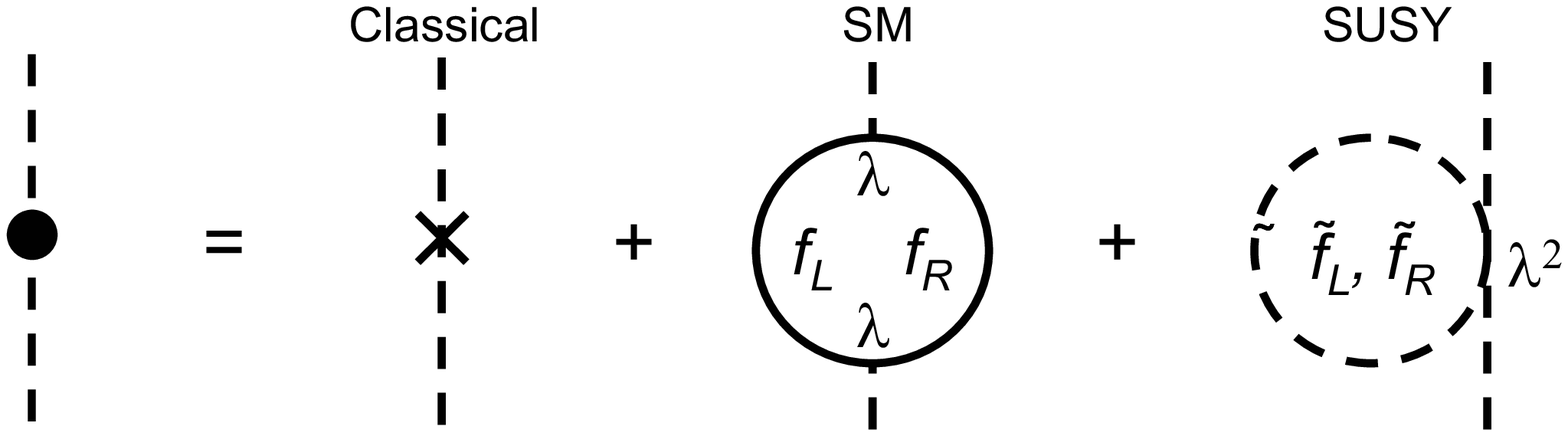}{0.85}
\caption{Contributions to the Higgs boson mass in the standard model
  and in supersymmetry.
\label{fig:higgsmass} }
\end{figure}

Once supersymmetry is broken, the mass scale for superpartners is
unconstrained.  There is, however, a strong motivation for this scale
to be the weak scale: the gauge hierarchy problem.  In the standard
model of particle physics, the classical mass of the Higgs boson
$(m_h^2)_0$ receives quantum corrections.  (See \figref{higgsmass}.)
Including quantum corrections from standard model fermions $f_L$ and
$f_R$, one finds that the physical Higgs boson mass is
\begin{equation}
m_h^2 = (m_h^2)_0 - \frac{1}{16 \pi^2} \lambda^2 \Lambda^2 + \ldots \ ,
\end{equation}
where the last term is the leading quantum correction, with $\lambda$
the Higgs-fermion coupling.  $\Lambda$ is the ultraviolet cutoff of
the loop integral, presumably some high scale well above the weak
scale.  If $\Lambda$ is of the order of the Planck scale $\sim
10^{19}~\gev$, the classical Higgs mass and its quantum correction
must cancel to an unbelievable 1 part in $10^{34}$ to produce the
required weak-scale $m_h$.  This unnatural fine-tuning is the gauge
hierarchy problem.

In the supersymmetric standard model, however, for every quantum
correction with standard model fermions $f_L$ and $f_R$ in the loop,
there are corresponding quantum corrections with superpartners
$\tilde{f}_L$ and $\tilde{f}_R$.  The physical Higgs mass then becomes
\begin{eqnarray}
m_h^2 &=& (m_h^2)_0 - \frac{1}{16 \pi^2} \lambda^2 \Lambda^2
+ \frac{1}{16 \pi^2} \lambda^2 \Lambda^2 + \ldots \nonumber \\
&\approx& (m_h^2)_0 + \frac{1}{16 \pi^2} (m_{\tilde{f}}^2 - m_f^2) \ln
(\Lambda/m_{\tilde{f}}) \ , 
\label{higgsmass}
\end{eqnarray}
where the terms quadratic in $\Lambda$ cancel, leaving a term
logarithmic in $\Lambda$ as the leading contribution.  In this case,
the quantum corrections are reasonable even for very large $\Lambda$,
and no fine-tuning is required.

In the case of exact supersymmetry, where $m_{\tilde{f}} = m_f$, even
the logarithmically divergent term vanishes.  In fact, quantum
corrections to masses vanish to all orders in perturbation theory, an
example of powerful non-renormalization theorems in supersymmetry.
{}From \eqref{higgsmass}, however, we see that exact mass degeneracy
is not required to solve the gauge hierarchy problem.  What {\em is}
required is that the dimensionless couplings $\lambda$ of standard
model particles and their superpartners are identical, and that the
superpartner masses be not too far above the weak scale (or else even
the logarithmically divergent term would be large compared to the weak
scale, requiring another fine-tuned cancellation).  This can be
achieved simply by adding supersymmetry-breaking weak-scale masses for
superpartners.  In fact, other terms, such as some cubic scalar
couplings, may also be added without re-introducing the fine-tuning.
All such terms are called ``soft,'' and the theory with weak-scale
soft supersymmetry-breaking terms is ``weak-scale supersymmetry.''

\subsection{The Neutral Supersymmetric Spectrum}
\label{sec:neutralspectrum}

\begin{figure}[tb]
\postscript{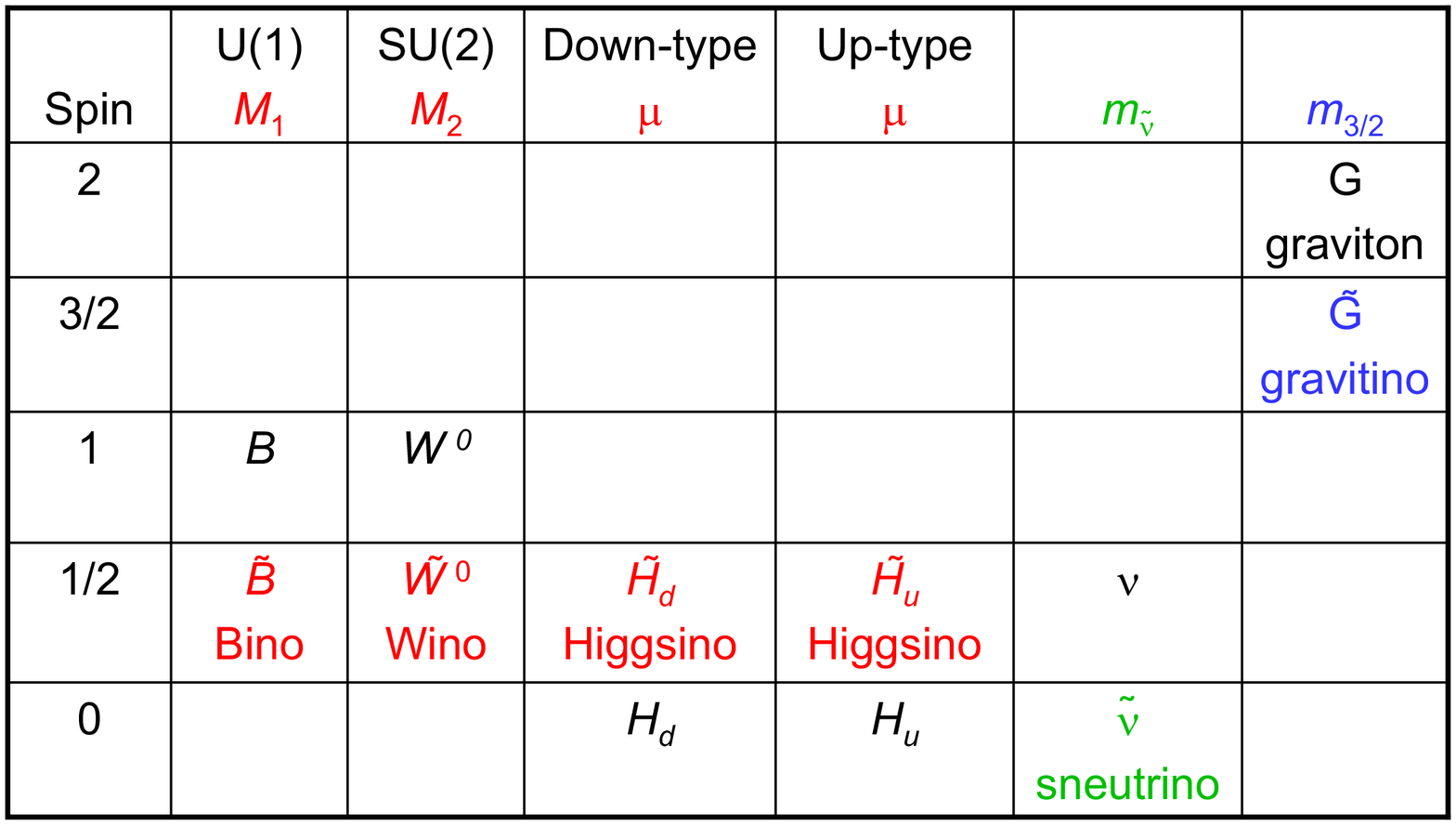}{0.85}
\caption{Neutral particles in the supersymmetric spectrum.  $M_1$,
  $M_2$, $\mu$, $m_{\tilde{\nu}}$, and $m_{3/2}$ are unknown
  weak-scale mass parameters.  The Bino, Wino, and down- and up-type
  Higgsinos mix to form neutralinos.
\label{fig:neutralspectrum} }
\end{figure}

Supersymmetric particles that are electrically neutral, and so
promising dark matter candidates, are shown with their standard model
partners in \figref{neutralspectrum}.  In supersymmetric models, two
Higgs doublets are required to give mass to all fermions.  The two
neutral Higgs bosons are $H_d$ and $H_u$, which give mass to the
down-type and up-type fermions, respectively, and each of these has a
superpartner.  Aside from this subtlety, the superpartner spectrum is
exactly as one would expect.  It consists of spin 0 sneutrinos, one
for each neutrino, the spin $\frac{3}{2}$ gravitino, and the spin
$\frac{1}{2}$ Bino, neutral Wino, and down- and up-type Higgsinos.
These states have masses determined (in part) by the corresponding
mass parameters listed in the top row of \figref{neutralspectrum}.
These parameters are unknown, but are presumably of the order of the
weak scale, given the motivations described above.

The gravitino is a mass eigenstate with mass $m_{3/2}$.  The
sneutrinos are also mass eigenstates, assuming flavor and $R$-parity
conservation. (See \secref{Rparity}.)  The spin $\frac{1}{2}$ states
are differentiated only by their electroweak quantum numbers.  After
electroweak symmetry breaking, these gauge eigenstates therefore mix
to form mass eigenstates.  In the basis $(-i\tilde{B}, -i\tilde{W}^3,
\tilde{H}_d, \tilde{H}_u)$ the mixing matrix is
\begin{equation}
\label{neumass}
\bold{M}_{\chi} =
\left( \begin{array}{cccc}
M_1             &0            &-M_Z \cos\beta\, s_W & M_Z \sin\beta\, s_W \\
0               &M_2          & M_Z \cos\beta\, c_W &-M_Z \sin\beta\, c_W \\
-M_Z \cos\beta\, s_W  & M_Z \cos\beta\, c_W &0            &-\mu           \\
 M_Z \sin\beta\, s_W  &-M_Z \sin\beta\, c_W &-\mu         &0     \end{array}
\right) ,
\end{equation}
where $c_W \equiv \cos \theta_W$, $s_W \equiv \sin \theta_W$, and
$\beta$ is another unknown parameter defined by $\tan\beta \equiv
\langle H_u \rangle / \langle H_d \rangle$, the ratio of the up-type
to down-type Higgs scalar vacuum expectation values (vevs).  The mass
eigenstates are called neutralinos and denoted $\{ \chi \equiv \chi_1,
\chi_2, \chi_3, \chi_4 \}$, in order of increasing mass.  If $M_1 \ll
M_2, |\mu|$, the lightest neutralino $\chi$ has a mass of
approximately $M_1$ and is nearly a pure Bino.  However, for $M_1 \sim
M_2 \sim |\mu|$, $\chi$ is a mixture with significant components of
each gauge eigenstate.

Finally, note that neutralinos are Majorana fermions; they are their
own anti-particles.  This fact has important consequences for
neutralino dark matter, as will be discussed below.

\subsection{$R$-Parity}
\label{sec:Rparity}

\begin{figure}[tb]
\postscript{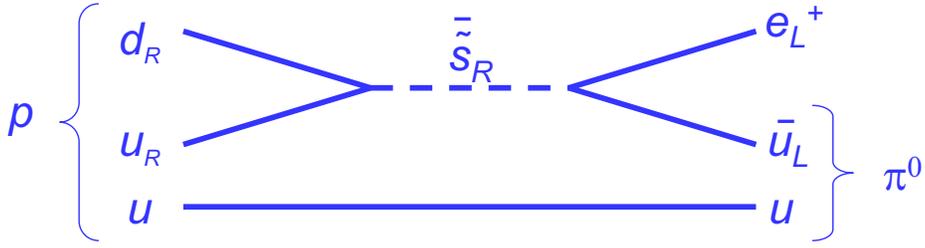}{0.85}
\caption{Proton decay mediated by squark.
\label{fig:protondecay} }
\end{figure}

Weak-scale superpartners solve the gauge hierarchy problem through
their virtual effects.  However, without additional structure, they
also mediate baryon and lepton number violation at unacceptable
levels.  For example, proton decay $p \to \pi^0 e^+$ may be mediated
by a squark as shown in \figref{protondecay}.

An elegant way to forbid this decay is to impose the conservation of
$R$-parity $R_p \equiv (-1)^{3(B-L)+2S}$, where $B$, $L$, and $S$ are
baryon number, lepton number, and spin, respectively.  All standard
model particles have $R_p = 1$, and all superpartners have $R_p=-1$.
$R$-parity conservation implies $\Pi R_p = 1$ at each vertex, and so
both vertices in \figref{protondecay} are forbidden.  Proton decay may
be eliminated without $R$-parity conservation, for example, by
forbidding $B$ or $L$ violation, but not both.  However, in these
cases, the non-vanishing $R$-parity violating couplings are typically
subject to stringent constraints from other processes, requiring some
alternative explanation.

An immediate consequence of $R$-parity conservation is that the
lightest supersymmetric particle (LSP) cannot decay to standard model
particles and is therefore stable.  Particle physics constraints
therefore naturally suggest a symmetry that provides a new stable
particle that may contribute significantly to the present energy
density of the universe.

\subsection{Supersymmetry Breaking and Dark Energy}
\label{sec:susybreaking}

Given $R$-parity conservation, the identity of the LSP has great
cosmological importance.  The gauge hierarchy problem is no help in
identifying the LSP, as it may be solved with any superpartner masses,
provided they are all of the order of the weak scale.  What is
required is an understanding of supersymmetry breaking, which governs
the soft supersymmetry-breaking terms and the superpartner spectrum.

The topic of supersymmetry breaking is technical and large.  However,
the most popular models have ``hidden sector'' supersymmetry breaking,
and their essential features may be understood by analogy to
electroweak symmetry breaking in the standard model.

\begin{figure}[tb]
\postscript{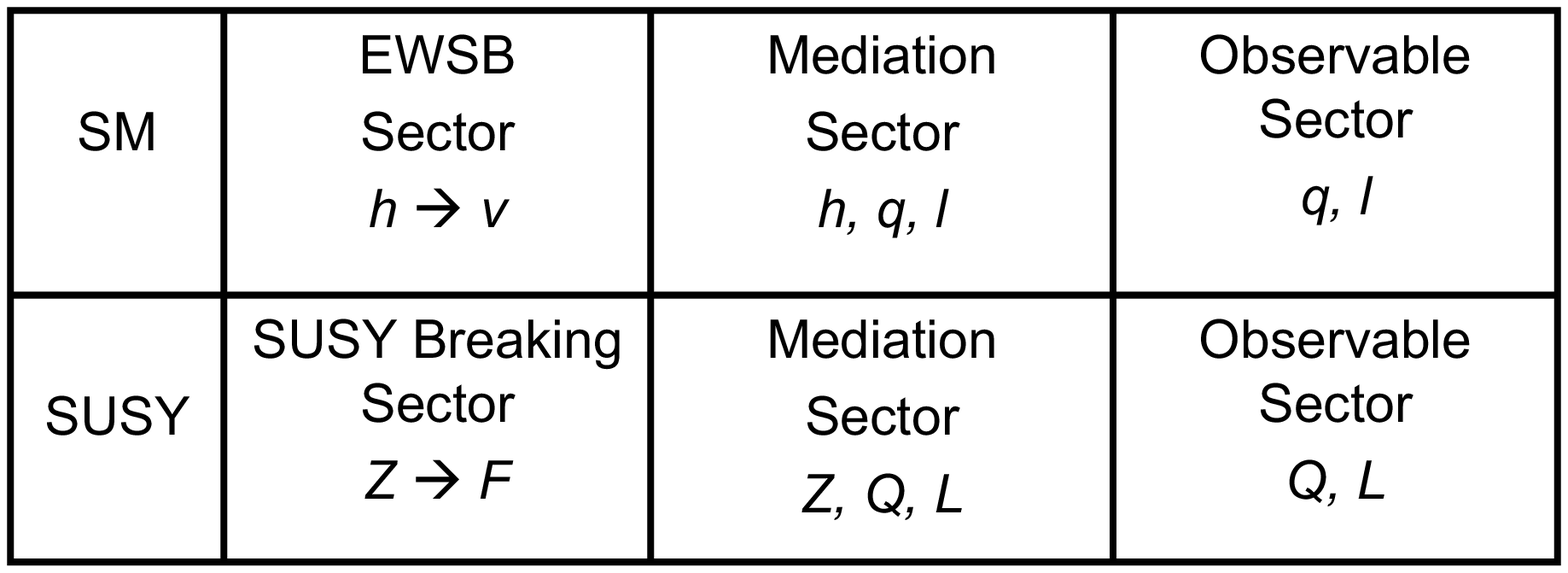}{0.85}
\caption{Sectors of interactions for electroweak symmetry breaking in
  the standard model and supersymmetry breaking in hidden sector
  supersymmetry breaking models.
\label{fig:hidden} }
\end{figure}

The interactions of the standard model may be divided into three
sectors. (See \figref{hidden}.)  The electroweak symmetry breaking
(EWSB) sector contains interactions involving only the Higgs boson
(the Higgs potential); the observable sector contains interactions
involving only what we might call the ``observable fields,'' such as
quarks $q$ and leptons $l$; and the mediation sector contains all
remaining interactions, which couple the Higgs and observable fields
(the Yukawa interactions).  Electroweak symmetry is broken in the EWSB
sector when the Higgs boson obtains a non-zero vev: $h \to v$.  This
is transmitted to the observable sector by the mediating interactions.
The EWSB sector determines the overall scale of EWSB, but the
interactions of the mediating sector determine the detailed spectrum
of the observed particles, as well as much of their phenomenology.

Models with hidden sector supersymmetry breaking have a similar
structure.  They have a supersymmetry breaking sector, which contains
interactions involving only fields $Z$ that are not part of the
standard model; an observable sector, which contains all interactions
involving only standard model fields and their superpartners; and a
mediation sector, which contains all remaining interactions coupling
fields $Z$ to the standard model. Supersymmetry is broken in the
supersymmetry breaking sector when one or more of the $Z$ fields
obtains a non-zero vev: $Z \to F$.  This is then transmitted to the
observable fields through the mediating interactions.  In contrast to
the case of EWSB, the supersymmetry-breaking vev $F$ has mass
dimension 2.  (It is the vev of the auxiliary field of the $Z$
supermultiplet.)

In simple cases where only one non-zero $F$ vev develops, the
gravitino mass is
\begin{equation}
m_{3/2} = \frac{F}{\sqrt{3} \mstar} \ ,
\end{equation}
where $\mstar \equiv (8\pi G_N)^{-1/2} \simeq 2.4 \times 10^{18}~\gev$
is the reduced Planck mass.  The standard model superpartner masses
are determined through the mediating interactions by terms such as
\begin{equation}
c_{ij} \frac{Z^\dagger Z}{\mmess^2} \tilde{f}^*_i \tilde{f}_j \quad 
\text{and} \quad c_a \frac{Z}{\mmess} \lambda_a \lambda_a \ ,
\end{equation}
where $c_{ij}$ and $c_a$ are constants, $\tilde{f}_i$ and $\lambda_a$
are superpartners of standard model fermions and gauge bosons,
respectively, and $\mmess$ is the mass scale of the mediating
interactions.  When $Z \to F$, these terms become mass terms for
sfermions and gauginos.  Assuming order one constants,
\begin{equation}
m_{\tilde{f}} , m_{\lambda} \sim \frac{F}{\mmess} \ .
\end{equation}

In supergravity models, the mediating interactions are gravitational,
and so $\mmess \sim \mstar$.   We then have 
\begin{equation}
m_{3/2}, m_{\tilde{f}} , m_{\lambda} \sim \frac{F}{\mstar} \ ,
\end{equation}
and $\sqrt{F} \sim \sqrt{\mweak \mstar} \sim 10^{10}~\gev$.  In such
models with ``high-scale'' supersymmetry breaking, the gravitino or
any standard model superpartner could in principle be the LSP.  In
contrast, in ``low-scale'' supersymmetry breaking models with $\mmess
\ll \mstar$, such as gauge-mediated supersymmetry breaking models,
\begin{equation}
m_{3/2} = \frac{F}{\sqrt{3} \mstar} \ll m_{\tilde{f}} ,
m_{\lambda} \sim \frac{F}{\mmess} \ ,
\end{equation}
$\sqrt{F} \sim \sqrt{\mweak \mmess} \ll 10^{10}~\gev$, and the
gravitino is necessarily the LSP.

As with electroweak symmetry breaking, the dynamics of supersymmetry
breaking contributes to the energy density of the vacuum, that is,
to dark energy.  In non-supersymmetric theories, the vacuum energy
density is presumably naturally $\Lambda \sim \mstar^4$ instead of
its measured value $\sim \text{meV}^4$, a discrepancy of $10^{120}$.
This is the cosmological constant problem.  In supersymmetric
theories, the vacuum energy density is naturally $F^2$.  For
high-scale supersymmetry breaking, one finds $\Lambda \sim \mweak^2
\mstar^2$, reducing the discrepancy to $10^{90}$.  Lowering the
supersymmetry breaking scale as much as possible to $F \sim \mweak^2$
gives $\Lambda \sim \mweak^4$, still a factor of $10^{60}$ too big.
Supersymmetry therefore eliminates from 1/4 to 1/2 of the fine-tuning
in the cosmological constant, a truly underwhelming achievement.  One
must look deeper for insights about dark energy and a solution to the
cosmological constant problem.

\subsection{Minimal Supergravity}

To obtain detailed information regarding the superpartner spectrum,
one must turn to specific models.  These are motivated by the
expectation that the weak-scale supersymmetric theory is derived from
a more fundamental framework, such as a grand unified theory or string
theory, at smaller length scales.  This more fundamental theory should
be highly structured for at least two reasons.  First, unstructured
theories lead to violations of low energy constraints, such as bounds
on flavor-changing neutral currents and CP-violation in the kaon
system and in electric dipole moments. Second, the gauge coupling
constants unify at high energies in supersymmetric
theories~\cite{Dimopoulos:1981yj}, and a more fundamental theory
should explain this.

{}From this viewpoint, the many parameters of weak-scale supersymmetry
should be derived from a few parameters defined at smaller length
scales through renormalization group evolution.  Minimal
supergravity~\cite{Chamseddine:jx,Barbieri:1982eh,Ohta:1982wn,Hall:iz,%
Alvarez-Gaume:1983gj}, the canonical model for studies of
supersymmetry phenomenology and cosmology, is defined by 5 parameters:
\begin{equation}
m_0, \mgaugino, A_0, \tan\beta, \sign(\mu) \ ,
\end{equation}
where the most important parameters are the universal scalar mass
$m_0$ and the universal gaugino mass $\mgaugino$, both defined at the
grand unified scale $\mgut \simeq 2 \times 10^{16}~\gev$.  In fact,
there is a sixth free parameter, the gravitino mass
\begin{equation}
m_{3/2} \ .
\end{equation}
As noted in \secref{susybreaking}, the gravitino may naturally be the
LSP.  It may play an important cosmological role, as we will see in
\secref{gravitino}.  For now, however, we follow most of the
literature and assume the gravitino is heavy and so irrelevant for
most discussions.

\begin{figure}[tb]
\postscript{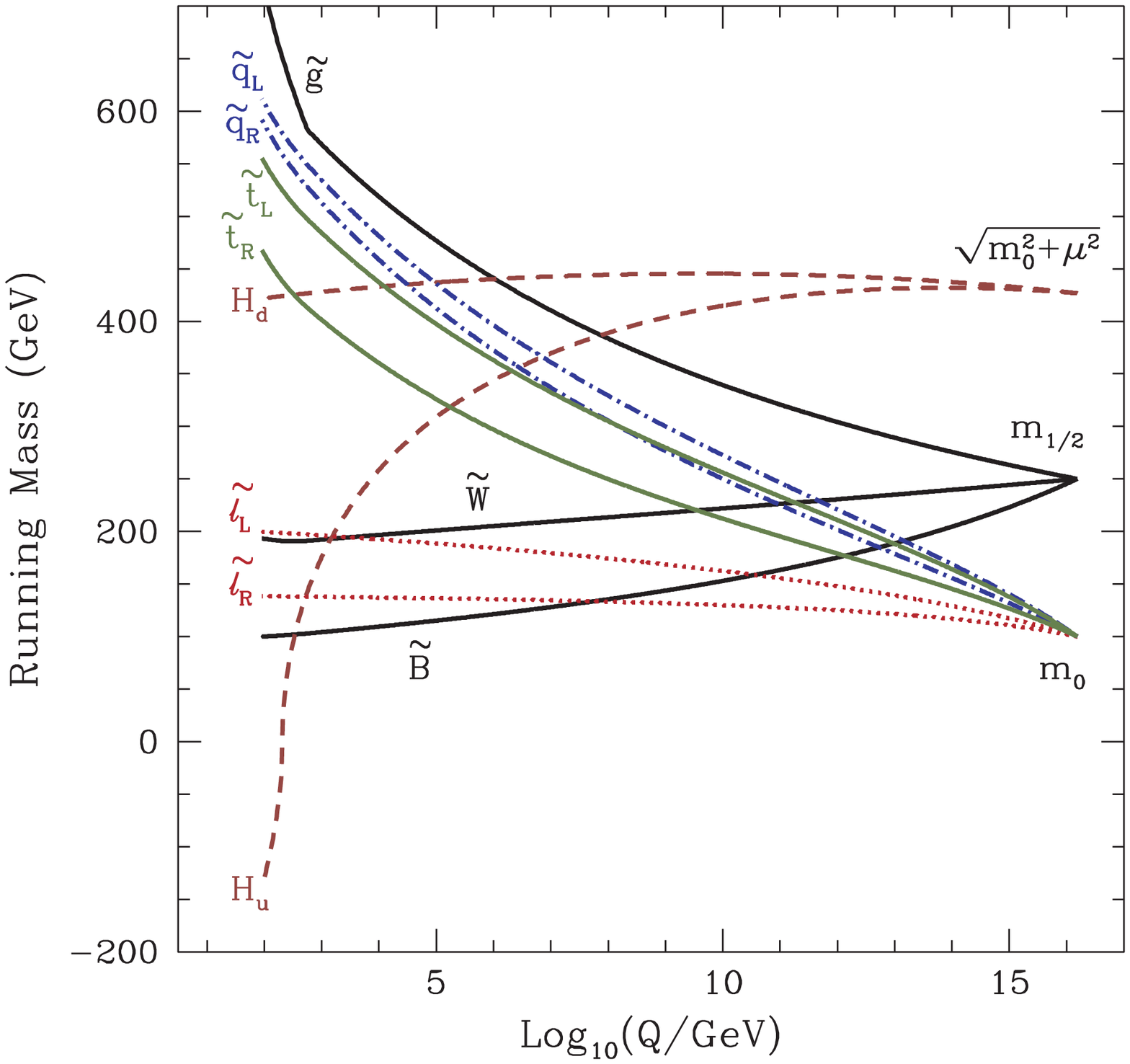}{0.65}
\caption{Renormalization group evolution of supersymmetric mass
  parameters.  From Ref.~\protect\citenum{Olive:2003iq}.
\label{fig:RGE} }
\end{figure}

The renormalization group evolution of supersymmetry parameters is
shown in \figref{RGE} for a particular point in minimal supergravity
parameter space.  This figure illustrates several key features that
hold more generally.  First, as superpartner masses evolve from
$\mgut$ to $\mweak$, gauge couplings increase these parameters, while
Yukawa couplings decrease them.  At the weak scale, colored particles
are therefore expected to be heavy, and unlikely to be the LSP.  The
Bino is typically the lightest gaugino, and the right-handed sleptons
(more specifically, the right-handed stau $\tilde{\tau}_R$) are
typically the lightest scalars.

Second, the mass parameter $m_{H_u}^2$ is typically driven negative by
the large top Yukawa coupling.  This is a requirement for electroweak
symmetry breaking: at tree-level, minimization of the electroweak
potential at the weak scale requires
\begin{eqnarray}
|\mu|^2 &=& \frac{m_{H_d}^2 - m_{H_u}^2 \tan^2\beta }
{\tan^2\beta -1} - \frac{1}{2} m_Z^2  \nonumber \\
&\approx& - m_{H_u}^2 - \frac{1}{2} m_Z^2 \ ,
\label{finetune}
\end{eqnarray}
where the last line follows for all but the lowest values of
$\tan\beta$, which are phenomenologically disfavored anyway.  Clearly,
this equation can only be satisfied if $m_{H_u}^2 < 0$.  This property
of evolving to negative values is unique to $m_{H_u}^2$; all other
mass parameters that are significantly diminished by the top Yukawa
coupling also experience a compensating enhancement from the strong
gauge coupling.  This behavior naturally explains why SU(2) is broken
while the other gauge symmetries are not.  It is a beautiful feature
of supersymmetry derived from a simple high energy framework and lends
credibility to the extrapolation of parameters all the way up to a
large mass scale like $\mgut$.

\begin{figure}[tb]
\postscript{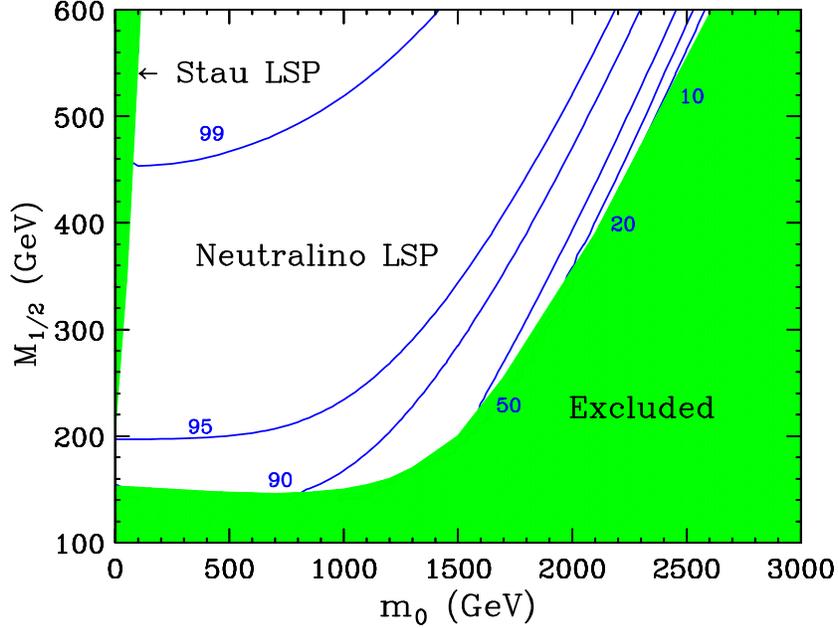}{0.75}
\caption{Regions of the $(m_0, \mgaugino)$ parameter space in minimal
supergravity with $A_0 = 0$, $\tan\beta=10$, and $\mu >0$.  The lower
shaded region is excluded by the LEP chargino mass limit.  The stau is
the LSP in the narrow upper shaded region.  In the rest of parameter
space, the LSP is the lightest neutralino, and contours of its
gaugino-ness $R_{\chi}$ (in percent) are shown. From
Ref.~\protect\citenum{Feng:2000gh}.
\label{fig:mSUGRA_LSP} }
\end{figure}

Given a particular high energy framework, one may then scan parameter
space to determine what possibilities exist for the LSP.  The results
for a slice through minimal supergravity parameter space are shown in
\figref{mSUGRA_LSP}.  They are not surprising.  The LSP is either the
the lightest neutralino $\chi$ or the right-handed stau
$\tilde{\tau}_R$.  In the $\chi$ LSP case, contours of gaugino-ness
\begin{equation}
R_{\chi} \equiv |a_{\tilde{B}}|^2 + |a_{\tilde{W}}|^2 \ ,
\end{equation}
where 
\begin{equation}
\chi = a_{\tilde{B}} (- i \tilde{B}) + a_{\tilde{W}} (- i \tilde{W})
+ a_{\tilde{H}_d} \tilde{H}_d + a_{\tilde{H}_u} \tilde{H}_u \ ,
\end{equation}
are also shown.  The neutralino is nearly pure Bino in much of
parameter space, but may have a significant Higgsino mixture for $m_0
\agt 1~\tev$, where \eqref{finetune} implies $|\mu| \sim M_1$.

There are, of course, many other models besides minimal supergravity.
Phenomena that do not occur in minimal supergravity may very well
occur or even be generic in other supersymmetric frameworks.  On the
other hand, if one looks hard enough, minimal supergravity contains a
wide variety of dark matter possibilities, and it will serve as a
useful framework for illustrating many points below.

\subsection{Summary}

\begin{itemize}
\item Supersymmetry is a new spacetime symmetry that predicts the
  existence of a new boson for every known fermion, and a new fermion
  for every known boson.
\item The gauge hierarchy problem may be solved by supersymmetry, but
  requires that all superpartners have masses at the weak scale.
\item The introduction of superpartners at the weak scale mediates
  proton decay at unacceptably large rates unless some symmetry is
  imposed.  An elegant solution, $R$-parity conservation, implies that
  the LSP is stable.  Electrically neutral superpartners, such as the
  neutralino and gravitino, are therefore promising dark matter
  candidates.
\item The superpartner masses depend on how supersymmetry is broken.
  In models with high-scale supersymmetry breaking, such as
  supergravity, the gravitino may or may not be the LSP; in models
  with low-scale supersymmetry breaking, the gravitino is the LSP.
\item Among standard model superpartners, the lightest neutralino
  naturally emerges as the dark matter candidate from the simple high
  energy framework of minimal supergravity.
\item Supersymmetry reduces fine tuning in the cosmological constant
  from 1 part in $10^{120}$ to 1 part in $10^{60}$ to $10^{90}$, and
  so does not provide much insight into the problem of dark energy.
\end{itemize}

\section{Neutralino Cosmology}
\label{sec:neutralino}

Given the motivations described in \secref{susyessentials} for stable
neutralino LSPs, it is natural to consider the possibility that
neutralinos are the dark
matter~\cite{Goldberg:1983nd,Ellis:1983ew,DMintros}.  In this section,
we review the general formalism for calculating thermal relic
densities and its implications for neutralinos and supersymmetry.  We
then describe a few of the more promising methods for detecting
neutralino dark matter.

\subsection{Freeze Out and WIMPs}
\label{sec:freezeout}

Dark matter may be produced in a simple and predictive manner as a
thermal relic of the Big Bang. The very early universe is a very
simple place --- all particles are in thermal equilibrium.  As the
universe cools and expands, however, interaction rates become too low
to maintain this equilibrium, and so particles ``freeze out.''
Unstable particles that freeze out disappear from the universe.
However, the number of stable particles asymptotically approaches a
non-vanishing constant, and this, their thermal relic density,
survives to the present day.

This process is described quantitatively by the Boltzmann equation
\begin{equation}
\frac{dn}{dt} = -3 H n - \langle \sigma_A v \rangle 
\left( n^2 - \nequ^2 \right) \ ,
\label{Boltzmann}
\end{equation}
where $n$ is the number density of the dark matter particle $\chi$,
$H$ is the Hubble parameter, $\langle \sigma_A v \rangle$ is the
thermally averaged annihilation cross section, and $\nequ$ is the
$\chi$ number density in thermal equilibrium.  On the right-hand side
of \eqref{Boltzmann}, the first term accounts for dilution from
expansion.  The $n^2$ term arises from processes $\chi \chi \to f
\bar{f}$ that destroy $\chi$ particles, and the $\nequ^2$ term arises
from the reverse process $f \bar{f} \to \chi \chi$, which creates
$\chi$ particles.

\begin{figure}[tb]
\postscript{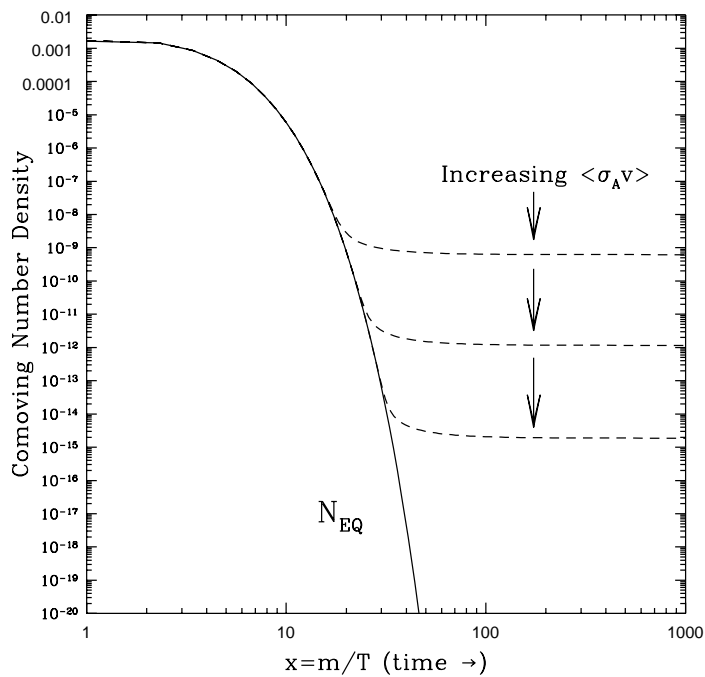}{0.65}
\caption{The co-moving number density $Y$ of a dark matter particle as
a function of temperature and time.  From
Ref.~\protect\citenum{Jungman:1995df}.
\label{fig:relicabund} }
\end{figure}

It is convenient to change variables from time to temperature,
\begin{equation}
t \to x \equiv \frac{m}{T} \ ,
\end{equation}
where $m$ is the $\chi$ mass, and to replace the number density by the
co-moving number density
\begin{equation}
n \to Y \equiv \frac{n}{s} \ ,
\label{comoving}
\end{equation}
where $s$ is the entropy density.  The expansion of the universe has
no effect on $Y$, because $s$ scales inversely with the volume of the
universe when entropy is conserved.  In terms of these new variables,
the Boltzmann equation is
\begin{equation}
\frac{x}{\Yeq} \frac{dY}{dx} = 
- \frac{\nequ \langle \sigma_A v \rangle } {H} 
\left( \frac{Y^2}{\Yeq^2} - 1 \right) \ .
\end{equation}
In this form, it is clear that before freeze out, when the
annihilation rate is large compared with the expansion rate, $Y$
tracks its equilibrium value $\Yeq$.  After freeze out, $Y$ approaches
a constant.  This constant is determined by the annihilation cross
section $\langle \sigma_A v \rangle$.  The larger this cross section,
the longer $Y$ follows its exponentially decreasing equilibrium value,
and the lower the thermal relic density. This behavior is shown in
\figref{relicabund}.

Let us now consider WIMPs --- weakly interacting massive particles
with mass and annihilation cross section set by the weak scale:
$m^2 \sim \langle \sigma_A v \rangle^{-1} \sim \mweak^2$.  Freeze out
takes place when
\begin{equation}
\nequ \langle \sigma_A v \rangle \sim H \ .
\end{equation}
Neglecting numerical factors, $\nequ \sim (mT)^{3/2} e^{-m/T}$ for a
non-relativistic particle, and $H \sim T^2 / \mstar$.  From these
relations, we find that WIMPs freeze out when
\begin{equation}
\frac{m}{T} \sim \ln \left[ \langle \sigma_A v \rangle m \mstar
\left( \frac{m}{T} \right) ^{1/2} \right] \sim 30 \ .
\end{equation}
Since $\frac{1}{2} m v^2 = \frac{3}{2} T$, WIMPs freeze out with
velocity $v \sim 0.3$.

One might think that, since the number density of a particle falls
exponentially once the temperature drops below its mass, freeze out
should occur at $T \sim m$.  This is not the case.  Because gravity is
weak and $\mstar$ is large, the expansion rate is extremely slow, and
freeze out occurs much later than one might naively expect.  For a $m
\sim 300~\gev$ particle, freeze out occurs not at $T \sim 300~\gev$
and time $t \sim 10^{-12}~\s$, but rather at temperature $T \sim
10~\gev$ and time $t \sim 10^{-8}~\s$.

With a little more work~\cite{Kolb:vq}, one can find not just the
freeze out time, but also the freeze out density
\begin{equation}
\Omega_{\chi} = m s Y( x = \infty) \sim \frac{10^{-10} ~\gev^{-2}} 
{\langle \sigma_A v \rangle} \ .
\end{equation}
A typical weak cross section is
\begin{equation}
\langle \sigma_A v \rangle \sim \frac{\alpha^2}{\mweak^2} 
\sim 10^{-9}~\gev^{-2} \ ,
\end{equation}
corresponding to a thermal relic density of $\Omega h^2 \sim 0.1$.
WIMPs therefore naturally have thermal relic densities of the observed
magnitude.  The analysis above has ignored many numerical factors, and
the thermal relic density may vary by as much as a few orders of
magnitude.  Nevertheless, in conjunction with the other strong
motivations for new physics at the weak scale, this coincidence is an
important hint that the problems of electroweak symmetry breaking and
dark matter may be intimately related.

\subsection{Thermal Relic Density}
\label{sec:thermal}

We now want to apply the general formalism above to the specific case
of neutralinos.  This is complicated by the fact that neutralinos may
annihilate to many final states: $f\bar{f}$, $W^+ W^-$, $ZZ$, $Zh$,
$hh$, and states including the heavy Higgs bosons $H$, $A$, and
$H^{\pm}$. Many processes contribute to each of these final states,
and nearly every supersymmetry parameter makes an appearance in at
least one process.  The full set of annihilation diagrams is discussed
in Ref.~\citenum{Drees:1992am}.  Codes to calculate the relic density
are publicly available~\cite{Gondolo:2000ee}.

Given this complicated picture, it is not surprising that there are a
variety of ways to achieve the desired relic density for neutralino
dark matter.  What is surprising, however, is that many of these
different ways may be found in minimal supergravity, provided one
looks hard enough.  We will therefore consider various regions of
minimal supergravity parameter space where qualitatively distinct
mechanisms lead to neutralino dark matter with the desired thermal
relic density.

\subsubsection{Bulk Region}

As evident from \figref{mSUGRA_LSP}, the LSP is a Bino-like
neutralino in much of minimal supergravity parameter space.  It is
useful, therefore, to begin by considering the pure Bino limit.  In
this case, all processes with final state gauge bosons vanish.  This
follows from supersymmetry and the absence of 3-gauge boson vertices
involving the hypercharge gauge boson.

\begin{figure}[tb]
\postscript{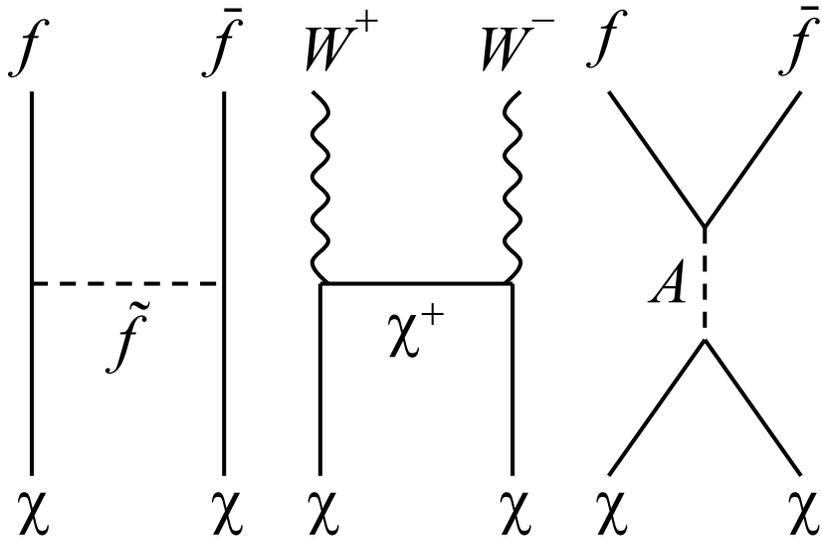}{0.75}
\caption{Three representative neutralino annihilation diagrams.
\label{fig:annih} }
\end{figure}

The process $\chi \chi \to f \bar{f}$ through a $t$-channel sfermion
does not vanish in the Bino limit.  This process is the first shown in
\figref{annih}.  This reaction has an interesting structure.  Recall
that neutralinos are Majorana fermions.  If the initial state
neutralinos are in an $S$-wave state, the Pauli exclusion principle
implies that the initial state is CP-odd, with total spin $S=0$ and
total angular momentum $J=0$.  If the neutralinos are gauginos, the
vertices preserve chirality, and so the final state $f\bar{f}$ has
spin $S=1$.  This is compatible with $J=0$ only with a mass insertion
on the fermion line.  This process is therefore either
$P$-wave-suppressed (by a factor $v^2 \sim 0.1$) or chirality
suppressed (by a factor $m_f/M_W$).  In fact, this conclusion holds
also for mixed gaugino-Higgsino neutralinos and for all other
processes contributing to the $f\bar{f}$ final
state~\cite{Drees:1992am}.  (It also has important implications for
indirect detection. See \secref{indirect}.)

\begin{figure}[tb]
\postscript{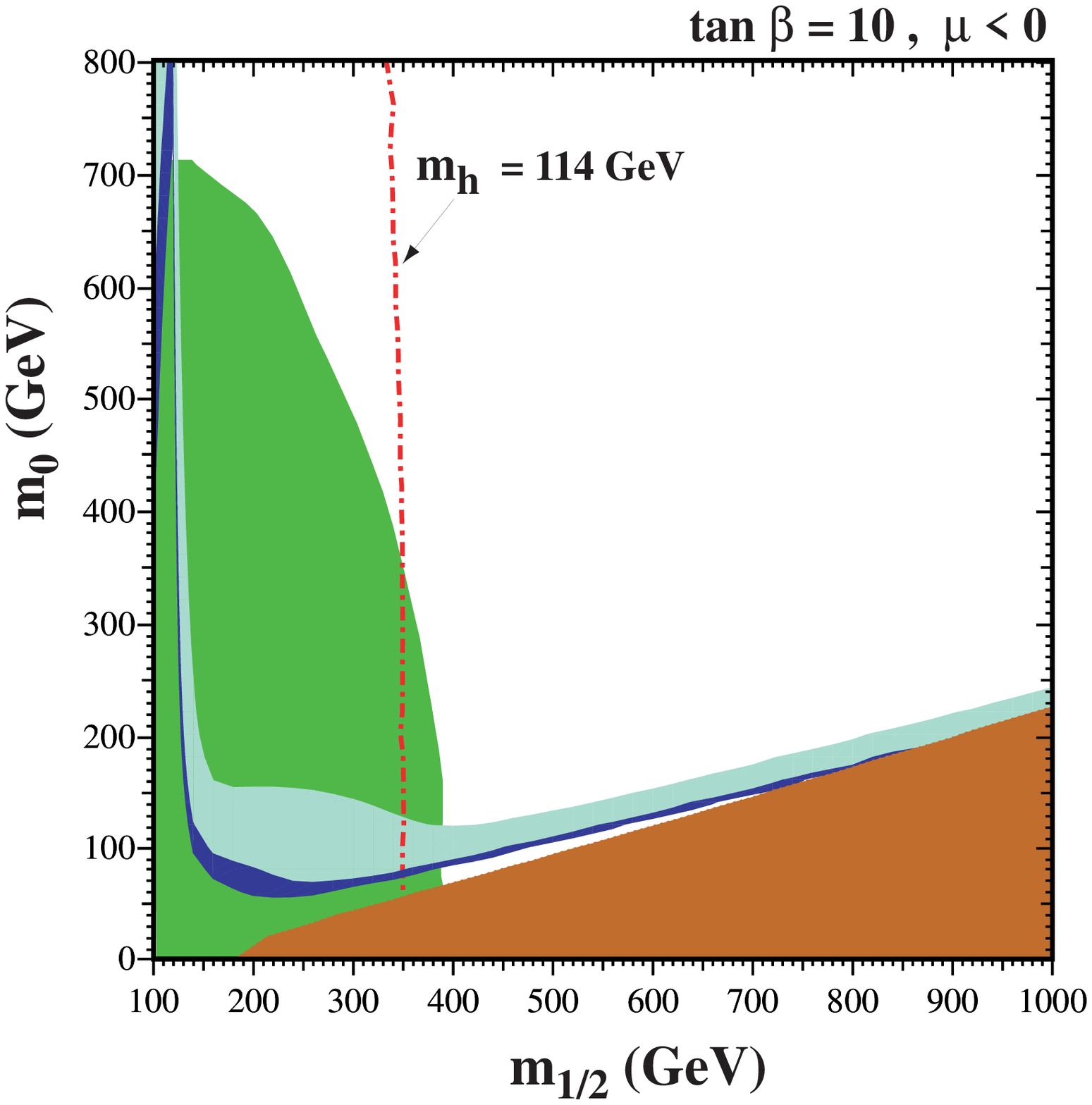}{0.65}
\caption{The bulk and co-annihilation regions of minimal supergravity
with $A_0 = 0$, $\tan \beta = 10$ and $\mu < 0$.  In the light blue
region, the thermal relic density satisfies the pre-WMAP constraint
$0.1 < \OmegaDM h^2 < 0.3$. In the dark blue region, the neutralino
density is in the post-WMAP range $0.094 < \OmegaDM h^2 < 0.129$. The
bulk region is the dark blue region with $(m_0, \mgaugino) \sim
(100~\gev, 200~\gev)$.  The stau LSP region is given in dark red, and
the co-annihilation region is the dark blue region along the stau LSP
border.  Current bounds on $b \to s \gamma$ exclude the green shaded
region, and the Higgs mass is too low to the left of the $m_h =
114~\gev$ contour.  From Ref.~\protect\citenum{Ellis:2003cw}.
\label{fig:coannih} }
\end{figure}

The region of minimal supergravity parameter space with a Bino-like
neutralino where $\chi \chi \to f \bar{f}$ yields the right relic
density is the $(m_0, \mgaugino) \sim (100~\gev, 200~\gev)$ region
shown in \figref{coannih}.  It is called the ``bulk region,'' as, in
the past, there was a wide range of parameters with $m_0, \mgaugino
\alt 300~\gev$ that predicted dark matter within the observed
range. The dark matter energy density has by now become so tightly
constrained, however, that the ``bulk region'' has now been reduced to
a thin ribbon of acceptable parameter space.

Moving from the bulk region by increasing $m_0$ and keeping all other
parameters fixed, one finds too much dark matter.  This behavior is
evident in \figref{coannih} and not difficult to understand: in the
bulk region, a large sfermion mass suppresses $\langle \sigma_A v
\rangle$, which implies a large $\OmegaDM$.  In fact, sfermion masses
not far above current bounds are required to offset the $P$-wave
suppression of the annihilation cross section.  This is an interesting
fact --- cosmology seemingly provides an upper bound on superpartner
masses!  If this were true, one could replace subjective naturalness
arguments by the fact that the universe cannot be overclosed to
provide upper bounds on superpartner masses.

Unfortunately, this line of reasoning is not airtight even in the
constrained framework of minimal supergravity.  The discussion above
assumes that $\chi \chi \to f \bar{f}$ is the only annihilation
channel.  In fact, however, for non-Bino-like neutralinos, there are
many other contributions.  Exactly this possibility is realized in the
focus point region, which we describe next.

In passing, it is important to note that the bulk region, although the
most straightforward and natural in many respects, is also severely
constrained by other data.  The existence of a light superpartner
spectrum in the bulk region implies a light Higgs boson mass, and
typically significant deviations in low energy observables such as $b
\to s \gamma$ and $(g-2)_{\mu}$.  Current bounds on the Higgs boson
mass, as well as concordance between experiments and standard model
predictions for $b \to s \gamma$ and (possibly) $(g-2)_{\mu}$,
therefore disfavor this region, as can be seen in \figref{coannih}.
For this reason, it is well worth considering other possibilities,
including the three we now describe.

\subsubsection{Focus Point Region}

As can be seen in \figref{mSUGRA_LSP}, a Bino-like LSP is not a
definitive prediction of minimal supergravity. For large scalar mass
parameter $m_0$, the Higgsino mass parameter $|\mu|$ drops to
accommodate electroweak symmetry breaking, as required by
\eqref{finetune}.  The LSP then becomes a gaugino-Higgsino mixture.
The region where this happens is called the focus point region, a name
derived from peculiar properties of the renormalization group
equations which suggest that large scalar masses do not necessarily
imply fine-tuning~\cite{Feng:1999mn,Feng:1999zg,Feng:2000bp}.

\begin{figure}[tb]
\postscript{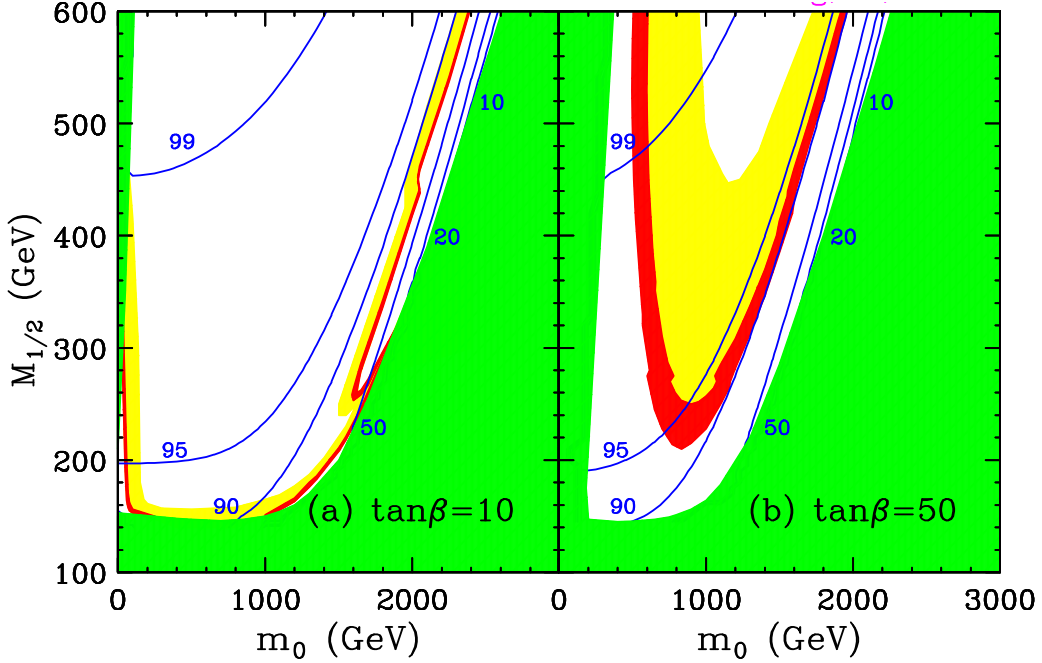}{0.95}
\caption{Focus point region of minimal supergravity for $A_0 = 0$,
$\mu>0$, and $\tan\beta$ as indicated.  The excluded regions and
contours are as in \figref{mSUGRA_LSP}.  In the light yellow region,
the thermal relic density satisfies the pre-WMAP constraint $0.1 <
\OmegaDM h^2 < 0.3$. In the medium red region, the neutralino density
is in the post-WMAP range $0.094 < \OmegaDM h^2 < 0.129$.  The focus
point region is the cosmologically favored region with $m_0 \agt
1~\tev$.  Updated from Ref.~\protect\citenum{Feng:2000gh}.
\label{fig:mSUGRA} }
\end{figure}

In the focus point region, the first diagram of \figref{annih} is
suppressed by very heavy sfermions.  However, the existence of
Higgsino components in the LSP implies that diagrams like the 2nd of
\figref{annih}, $\chi \chi \to W^+W^-$ through a $t$-channel chargino,
are no longer suppressed. This provides a second method by which
neutralinos may annihilate efficiently enough to produce the desired
thermal relic density.  The cosmologically preferred regions with the
right relic densities are shown in \figref{mSUGRA}.  The right amount
of dark matter can be achieved with arbitrarily heavy sfermions, and
so there is no useful cosmological upper bound on superpartner masses,
even in the framework of minimal supergravity.

\subsubsection{$A$ Funnel Region}

A third possibility realized in minimal supergravity is that the dark
matter annihilates to fermion pairs through an $s$-channel pole.  The
potentially dominant process is through the $A$ Higgs boson (the last
diagram of \figref{annih}), as the $A$ is CP-odd, and so may couple to
an initial $S$-wave state.  This process is efficient when $2 m_{\chi}
\approx m_A$.  In fact, the $A$ resonance may be broad, extending the
region of parameter space over which this process is important.

\begin{figure}[tb]
\postscript{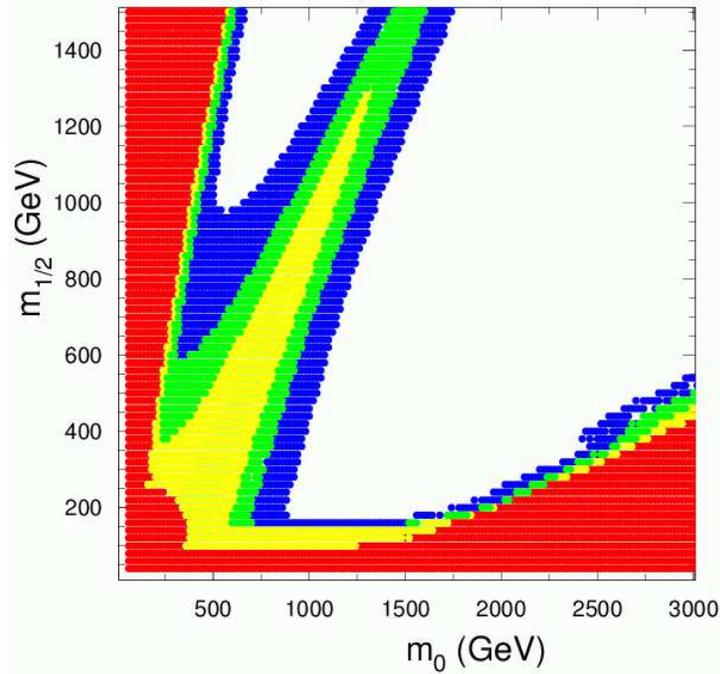}{0.65}
\caption{The $A$ funnel region of minimal supergravity with $A_0 =0$,
$\tan\beta = 45$, and $\mu < 0$.  The red region is excluded.  The
other shaded regions have $\OmegaDM h^2 < 0.1$ (yellow), $0.1 <
\OmegaDM h^2 < 0.3$ (green), and $0.3 < \OmegaDM h^2 < 1$ (blue).
{}From Ref.~\protect\citenum{Baer:2002ps}.
\label{fig:Afunnel} }
\end{figure}

The $A$ resonance region occurs in minimal supergravity for $\tan
\beta \agt 40$~\cite{Baer:2002fv,Baer:2002ps} and is shown in
\figref{Afunnel}.  Note that the resonance is so efficient that the
relic density may be reduced too much.  The desired relic density is
therefore obtained when the process is near resonance, but not exactly
on it.

\subsubsection{Co-annihilation Region}

Finally, the desired neutralino relic density may be obtained even if
$\chi\chi$ annihilation is inefficient if there are other particles
present in significant numbers when the LSP freezes out.  The
neutralino density may then be brought down through co-annihilation
with the other species~\cite{Binetruy:1983jf,Griest:1990kh}.  Naively,
the presence of other particles requires that they be mass degenerate
with the neutralino to within the temperature at freeze out, $T
\approx m_{\chi}/30$.  In fact, co-annihilation may be so enhanced
relative to the $P$-wave-suppressed $\chi \chi$ annihilation cross
section that co-annihilation may be important even with mass
splittings much larger than $T$.

The co-annihilation possibility is realized in minimal supergravity
along the $\tilde{\tau}$ LSP -- $\chi$ LSP border.  (See
\figref{coannih}.)  Processes such as $\chi \tilde{\tau} \to \tau^*
\to \tau \gamma$ are not $P$-wave suppressed, and so enhance the $\chi
\chi$ annihilation rate substantially.  There is therefore a narrow
finger extending up to masses $m_{\chi} \sim 600~\gev$ with acceptable
neutralino thermal relic densities.

\subsection{Direct Detection}

If dark matter is composed of neutralinos, it may be detected
directly, that is, by looking for signals associated with its
scattering in ordinary matter.  Dark matter velocity and spatial
distributions are rather poorly known and are an important source of
uncertainty~\cite{Sikivie:1996nn,Brhlik:1999tt,Belli:1999nz,%
Gelmini:2000dm,Calcaneo-Roldan:2000yt}.  A common assumption is that
dark matter has a local energy density of $\rho_{\chi} =
0.3~\gev/\cm^3$ with a velocity distribution characterized by a
velocity $v \approx 220~\km/\s$.  Normalizing to these values, the
neutralino flux is
\begin{equation}
\Phi_{\chi} = 6.6 \times 10^{4}~\cm^{-2}~\s^{-1}
\frac{\rho_{\chi}}{0.3~\gev/\cm^3} \frac{100~\gev}{m_{\chi}}
\frac{v}{220~\km/\s} \ .  
\end{equation}
Such values therefore predict a substantial flux of halo neutralinos
in detectors here on Earth.

The maximal recoil energy from a WIMP scattering off a nucleus $N$ is
\begin{equation}
E_{\text{recoil}}^{\text{max}} = \frac{2 m_{\chi}^2 m_N}
{\left( m_{\chi} + m_N \right)^2} \, v^2 \sim 100~\kev \ .
\end{equation}
With such low energies, elastic scattering is the most promising
signal at present, although the possibility of detecting inelastic
scattering has also been discussed.  As we will see below, event rates
predicted by supersymmetry are at most a few per kilogram per day.
Neutralino dark matter therefore poses a serious experimental
challenge, requiring detectors sensitive to extremely rare events with
low recoil energies.

\begin{figure}[tb]
\begin{minipage}[t]{0.4\textwidth}
\begin{center}
\postscript{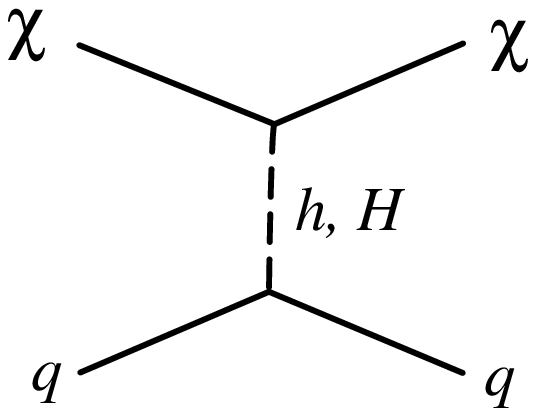}{0.9}
\end{center}
\end{minipage}
\hfil
\begin{minipage}[t]{0.5\textwidth}
\begin{center}
\postscript{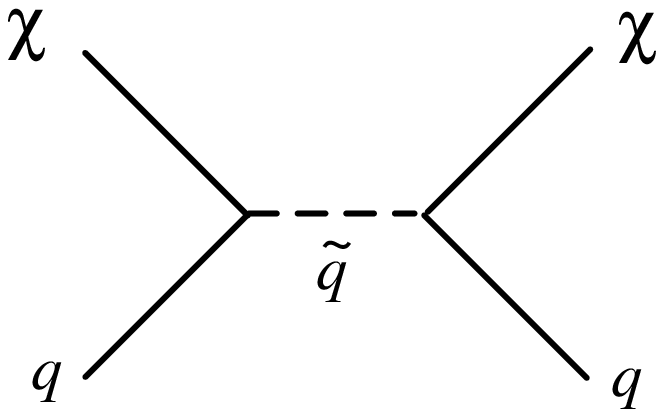}{0.9}
\end{center}
\end{minipage}
\caption{Feynman diagrams contributing to $\chi q \to \chi q$
  scattering.
\label{fig:scattering} }
\end{figure}

Neutralino-nucleus interactions take place at the parton level through
neutralino-quark interactions, such as those in \figref{scattering}.
Because neutralinos now have velocities $v \sim 10^{-3}$, we may take
the non-relativistic limit for these scattering amplitudes. In this
limit, only two types of neutralino-quark couplings are
non-vanishing~\cite{Goodman:1984dc}.  The interaction Lagrangian may
be parameterized as
\begin{equation}
{\cal L} = \sum_{q=u,d,s,c,b,t} \left( \alpha_q^{\text{SD}} \bar{\chi} 
\gamma^{\mu} \gamma^5 \chi
\bar{q} \gamma_{\mu} \gamma^5 q 
+ \alpha_q^{\text{SI}} \bar{\chi} \chi \bar{q} q \right) \ .
\end{equation}
The first term is the spin-dependent coupling, as it reduces to
$\bold{S}_{\chi} \cdot \bold{S}_{N}$ in the non-relativistic limit.
The second is the spin-independent coupling. All of the supersymmetry
model dependence is contained in the parameters $\alpha_q^{\text{SD}}$
and $\alpha_q^{\text{SI}}$.  The $t$-channel Higgs exchange diagram of
\figref{scattering} contributes solely to $\alpha_q^{\text{SI}}$,
while the $s$-channel squark diagram contributes to both
$\alpha_q^{\text{SD}}$ and $\alpha_q^{\text{SI}}$.

For neutralinos scattering off protons, the spin-dependent coupling is
dominant.  However, the spin-independent coupling is coherent and so
greatly enhanced for heavy nuclei, a fact successfully exploited by
current experiments.  As a result, spin-independent direct detection
is currently the most promising approach for neutralino dark matter,
and we focus on this below.

Given the parameters $\alpha_q^{\text{SI}}$, the spin-independent cross
section for $\chi N$ scattering is 
\begin{equation}
\sigma_{\text{SI}} = \frac{4}{\pi} \mu_N^2 
\sum_q \alpha_q^{\text{SI}\, 2} 
\left[ Z \frac{m_p}{m_q} f_{T_q}^p 
+ (A-Z) \frac{m_n}{m_q} f_{T_q}^n \right]^2 \ ,
\end{equation}
where 
\begin{equation}
\mu_N = \frac{m_{\chi} m_N}{m_{\chi} + m_N} 
\end{equation}
is the reduced mass of the $\chi$-$N$ system, $Z$ and $A$ are the
atomic number and weight of the nucleus, respectively, and
\begin{equation}
f_{T_q}^{p,n} = \frac{\langle p,n | m_q \bar{q}q | p,n
 \rangle}{m_{p,n}}
\end{equation}
are constants quantifying what fraction of the nucleon's mass is
carried by quark $q$.  For the light quarks~\cite{Ellis:2000ds},
\begin{eqnarray}
f_{T_u}^p = 0.020 \pm 0.004 && f_{T_u}^n = 0.014 \pm 0.003 \nonumber \\
f_{T_d}^p = 0.026 \pm 0.005 && f_{T_d}^n = 0.036 \pm 0.008 \nonumber \\
f_{T_s}^p = 0.118 \pm 0.062 && f_{T_s}^n = 0.118 \pm 0.062 \ . 
\end{eqnarray}
The contribution from neutralino-gluon couplings mediated by heavy
quark loops may be included by taking $f_{T_{c,b,t}}^{p,n} =
\frac{2}{27} f_{T_G}^{p,n} = \frac{2}{27} (1 - f_{T_u}^{p,n} -
f_{T_d}^{p,n} - f_{T_s}^{p,n})$~\cite{Shifman:zn}.

The number of dark matter scattering events is
\begin{eqnarray}
\lefteqn{{\cal N} = N_N T \frac{\rho_{\chi}}{m_{\chi}} \sigma_N v} \\
&=& 3.4 \times 10^{-6} \frac{M_D}{\kg} \frac{T}{\text{day}}
\frac{\rho_{\chi}}{0.3~\gev/\cm^3} \frac{100~\gev}{m_{\chi}}
\frac{v}{220~\km/\s} \frac{\mu_N^2 A}{m_p^2}
\frac{\sigma_p}{10^{-6}~\pb} \ ,
\end{eqnarray}
where $N_N$ is the number of target nuclei, $T$ is the experiment's
running time, $M_D$ is the mass of the detector, and the proton
scattering cross section $\sigma_p$ has been normalized to a
near-maximal supersymmetric value.  This is a discouragingly low event
rate.  However, for a detector with a fixed mass, this rate is
proportional to $\mu_N^2 A$.  For heavy nuclei with $A \sim
m_{\chi}/m_p$, the event rate is enhanced by a factor of $\sim A^3$,
providing the strong enhancement noted above.

\begin{figure}[tb]
\postscript{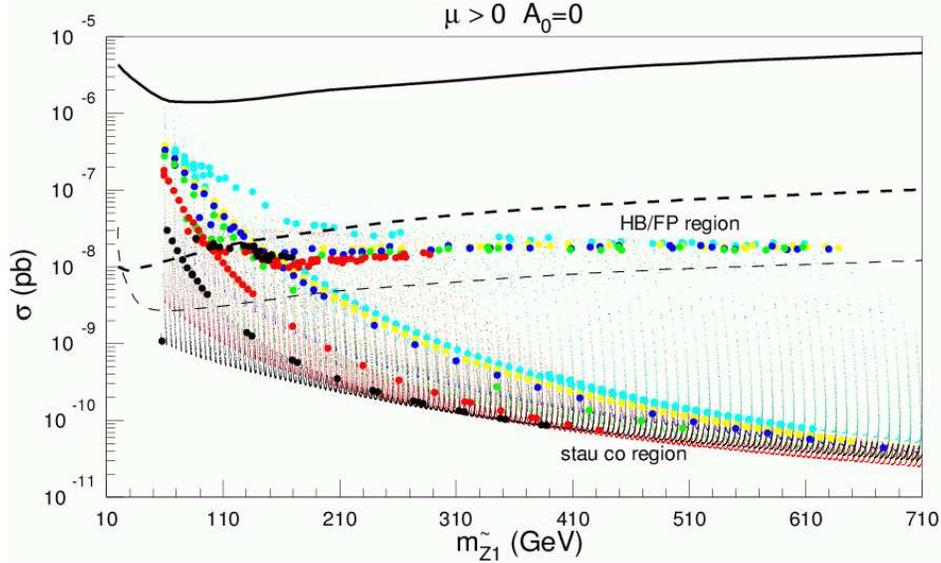}{0.85}
\caption{Spin-independent neutralino-proton cross sections for minimal
supergravity models with $A_0 = 0$, $\mu > 0$.  The colors correspond
to various values of $\tan\beta$ in the range $10 \le \tan \beta \le
55$.  Points with the preferred thermal relic density $0.094 <
\OmegaDM h^2 < 0.129$ are highlighted with enlarged circles, and those
in the focus point and co-annihilation regions are indicated.
Estimated reaches of current (CDMS, EDELWEISS, ZEPLIN1, DAMA), near
future (CDMS2, EDELWEISS2, ZEPLIN2, CRESST2), and future detectors
(GENIUS, ZEPLIN4,CRYOARRAY) are given by the solid, dark dashed, and
light dashed contours, respectively.  From
Ref.~\protect\citenum{Baer:2003jb}.
\label{fig:direct} }
\end{figure}

Comparisons between theory and experiment are typically made by
converting all results to proton scattering cross sections.  In
\figref{direct}, minimal supergravity predictions for spin-independent
cross sections are given.  These vary by several orders of magnitude.
In the stau co-annihilation region, these cross sections can be small,
as the neutralino is Bino-like, suppressing the Higgs diagram, and
squarks can be quite heavy, suppressing the squark diagram.  However,
in the focus point region, the neutralino is a gaugino-Higgsino
mixture, and the Higgs diagram is large.  Current and projected
experimental sensitivities are also shown in \figref{direct}.  Current
experiments are just now probing the interesting parameter region for
supersymmetry, but future searches will provide stringent tests of
some of the more promising minimal supergravity predictions.

\begin{figure}[tb]
\postscript{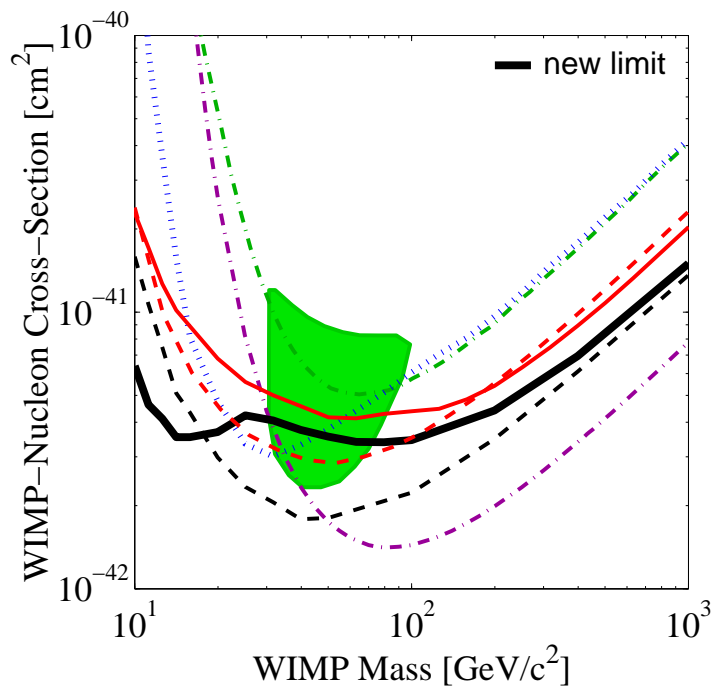}{0.65}
\caption{Regions of dark matter mass and spin-independent proton
scattering cross sections.  The shaded region is the $3\sigma$ favored
region from DAMA.  The dot-dashed line is the exclusion contour from
EDELWEISS, and the thick solid black line is the exclusion contour
from CDMS. From Ref.~\protect\citenum{Akerib:2003px}.
\label{fig:cdms} }
\end{figure}

The DAMA collaboration has reported evidence for direct detection of
dark matter from annual modulation in scattering
rates~\cite{Bernabei:2000qi}.  The favored dark matter mass and proton
spin-independent cross section are shown in \figref{cdms}.  By
comparing \figsref{direct}{cdms}, one sees that the interaction
strength favored by DAMA is very large relative to typical predictions
in minimal supergravity.  Such cross sections may be realized in less
restrictive supersymmetry scenarios.  However, more problematic from
the point of view of providing a supersymmetric interpretation is that
the experiments EDELWEISS~\cite{Benoit:2002hf} and
CDMS~\cite{Akerib:2003px} have also searched for dark matter with
similar sensitivities and have not found signals.  Their exclusion
bounds are also given in \figref{cdms}.\footnote{Recent data from CDMS
in the Soudan mine has pushed the discrepancy to even greater
levels~\cite{Akerib:2004fq}.}  Given standard halo and neutralino
interaction assumptions, these data are inconsistent at a high level.

Non-standard halo models and velocity
distributions~\cite{Copi:2002hm,Freese:2003tt} and non-standard and
generalized dark matter
interactions~\cite{Ullio:2000bv,Smith:2001hy,Kurylov:2003ra,%
Tucker-Smith:2004jv} have been considered as means to bring
consistency to the experimental picture.  The results are mixed.
Given the current status of direct detection experiments, a
supersymmetric interpretation is at best premature.  It is worth
noting, however, that the current results bode well for the future, as
many well-motivated supersymmetry models predict cross sections not
far from current sensitivities.

\subsection{Indirect Detection}
\label{sec:indirect}

After freeze out, dark matter pair annihilation becomes greatly
suppressed.  However, after the creation of structure in the universe,
dark matter annihilation in overdense regions of the universe may
again become significant.  Dark matter may therefore be detected
indirectly: pairs of dark matter particles annihilate {\em somewhere},
producing {\em something}, which is detected {\em somehow}.  There are
a large number of possibilities. Below we briefly discuss three of the
more promising signals.

\subsubsection{Positrons}

Dark matter in our galactic halo may annihilate to positrons, which
may be detected in space-based or balloon-borne
experiments~\cite{Rudaz:1987ry,Tylka:1989xj,Turner:1990kg,%
Kamionkowski:1991ty,Baltz:1999xv,Moskalenko:1999sb}.
(Anti-protons~\cite{Stecker:1988fx,Chardonnet:1996ca,%
Bergstrom:1999jc,Bieber:1999dn} and
anti-deuterium~\cite{Donato:1999gy} have also been suggested as
promising signals.)

The positron background is most likely to be composed of secondaries
produced in the interactions of cosmic ray nuclei with interstellar
gas, and is expected to fall as $\sim E_{e^+}^{-3.1}$.  At energies
below 10 GeV, there are also large uncertainties in the
background~\cite{Baltz:1999xv,Moskalenko:1999sb}.  The most promising
signal is therefore hard positrons from $\chi \chi$ annihilation.

Unfortunately, the monoenergetic signal $\chi \chi \to e^+ e^-$ is
extremely suppressed.  As noted above, $\chi \chi \to f \bar{f}$ is
either $P$-wave suppressed or chirality suppressed.  At present times,
as opposed to during freeze out, $P$-wave suppression is especially
severe, since $v^2 \sim 10^{-6}$, and so direct annihilation to
positrons is effectively absent.\footnote{Note that this suppression
is rather special, in that it follows from the Majorana nature of
neutralinos; it is absent for other types of dark matter, such as dark
matter with spin 1~\protect\cite{Servant:2002aq,Cheng:2002ej}.} The
positron signal therefore results from processes such as $\chi \chi
\to W^+ W^-$ followed by $W^+ \to e^+ \nu$, and is a continuum, not a
line, at the source.

To obtain the positron energy distribution we would observe, the
source energy distribution must be propagated through the halo to us.
The resulting differential positron flux is~\cite{Moskalenko:1999sb}
\begin{equation}
\frac{d\Phi_{e^+}}{d\Omega dE} = \frac{\rho_{\chi}^2}{\mchi^2}
\sum_i \sigma_i v B_{e^+}^i 
\int dE_0\, f_i(E_0)\, G(E_0, E) \ ,
\end{equation}
where $\rho_{\chi}$ is the local neutralino mass density, the sum is over
all annihilation channels, and $B_{e^+}^i$ is the branching fraction
to positrons in channel $i$. The source function $f(E_0)$ gives the
initial positron energy distribution from neutralino annihilation.
$G(E_0, E)$ is the Green's function describing positron propagation in
the galaxy~\cite{Protheroe:1982pp} and contains all the halo model
dependence.

\begin{figure}[tb]
\postscript{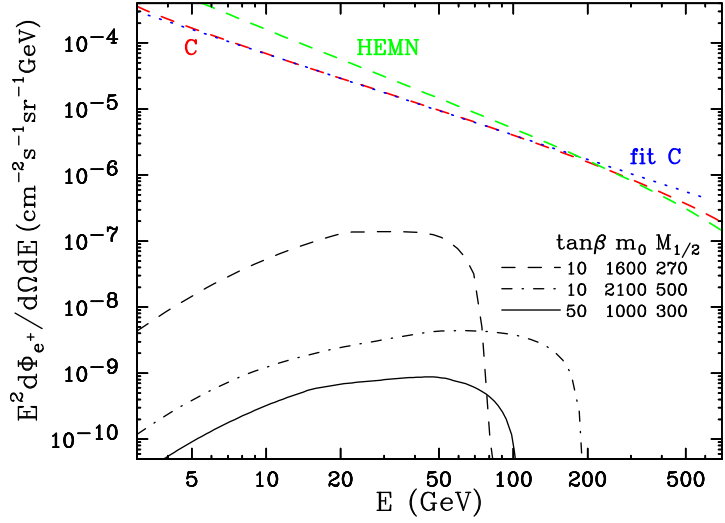}{0.65}
\caption{The differential positron flux for three minimal supergravity
  models.  The curves labeled C and HEMN are background models from
  Ref.~\protect\citenum{Moskalenko:1999sb}. {}From
  Ref.~\protect\citenum{Feng:2000zu}.  }
\label{fig:dFdEPos}
\end{figure}

Three sample positron spectra are given in \figref{dFdEPos}.  For all
of them, $E^2 d\Phi/dE$ peaks at energies $E \sim m_{\chi}/2$.  These
signals are all well below background.  However, a smooth halo
distribution has been assumed.  For clumpy halos, which are well
within the realm of possibility, the signal may be enhanced
significantly.  In the next few years, both PAMELA, a satellite
detector, and AMS-02, an experiment to be placed on the International
Space Station, will provide precision probes of the positron spectrum.
These experiments and other recently completed experiments are listed
in Table~\ref{table:positrons}.

\begin{table}[tb]
\caption{Recent and planned $e^+$ detector experiments. We list each
experiment's start date, duration, geometrical acceptance in
$\cm^2~\sr$, maximal $E_{e^+}$ sensitivity in GeV, and (expected)
total number of $e^+$ detected per GeV at $E_{e^+} = 100~\gev$.
{}From Ref.~\protect\citenum{Feng:2000zu}.  \label{table:positrons} }
\begin{center}
\begin{tabular}{lllrrrrr}
 Experiment \rule[-2.8mm]{0mm}{7mm}
  & Type
   & Date
    & Duration
    & Accept.
     & $E_{e^+}^{\text{max}}$
       & $\frac{dN}{dE}(100)$ \\  \hline
 HEAT94/95
  & Balloon
   & 1994/95
    & 29/26 hr
    & 495
     & 50
       & ---          \\
 CAPRICE94/98
  & Balloon
   & 1994/98
    & 18/21 hr
    & 163
     & 10/30
       & ---          \\
 PAMELA
  & Satellite
   & 
    & 3 yr
    & 20
     & 200
       & 0.7         \\
 AMS-02
  & ISS
   & 
    & 3 yr
    & 6500
     & 1000
       & 250
\end{tabular}
\end{center}
\end{table}

Finally, the High Energy Antimatter Telescope (HEAT) experiment, a
balloon-borne magnetic spectrometer, has found evidence for an excess
of positrons at energy $\sim 8~\gev$ in data from
1994/95~\cite{Barwick:1995gv,Barwick:1997ig} and
2000~\cite{heatnew}. The observed bump in the positron fraction
$N_{e^+}/(N_{e^+} + N_{e^-})$ is not naturally obtained by neutralino
dark matter for two reasons.  First, as noted above, for a smooth
halo, the annihilation cross sections that produce the desired relic
density predict positron fluxes that are far too low to explain the
observed excess.  In principle, this objection may be overcome by a
sufficiently clumpy halo. Second, neutralino annihilation produces
positrons only through cascades, resulting in a smooth positron energy
distribution.  This is an inevitable consequence of the Majorana
nature of neutralinos.  Nevertheless, even the addition of a smooth
component from neutralino annihilation may improve the fit to data,
and the possibility of a supersymmetric explanation for the HEAT
anomaly has been explored in a number of
studies~\cite{Kane:2001fz,Baltz:2001ir,Kane:2002nm}.

\begin{figure}[tb]
\begin{minipage}[t]{0.48\textwidth}
\begin{center}
\postscript{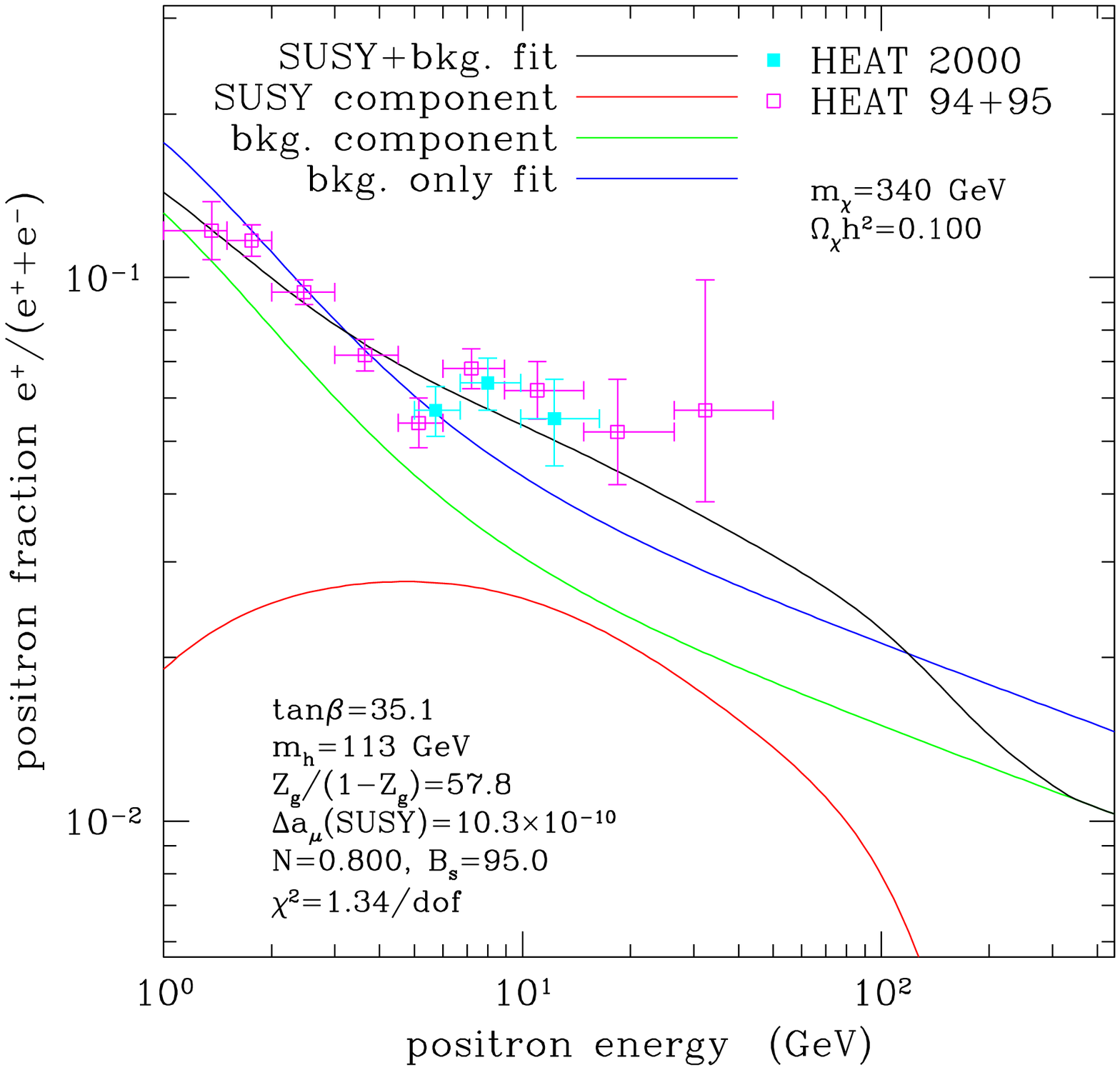}{0.95}
\end{center}
\end{minipage}
\hfil
\begin{minipage}[t]{0.48\textwidth}
\begin{center}
\postscript{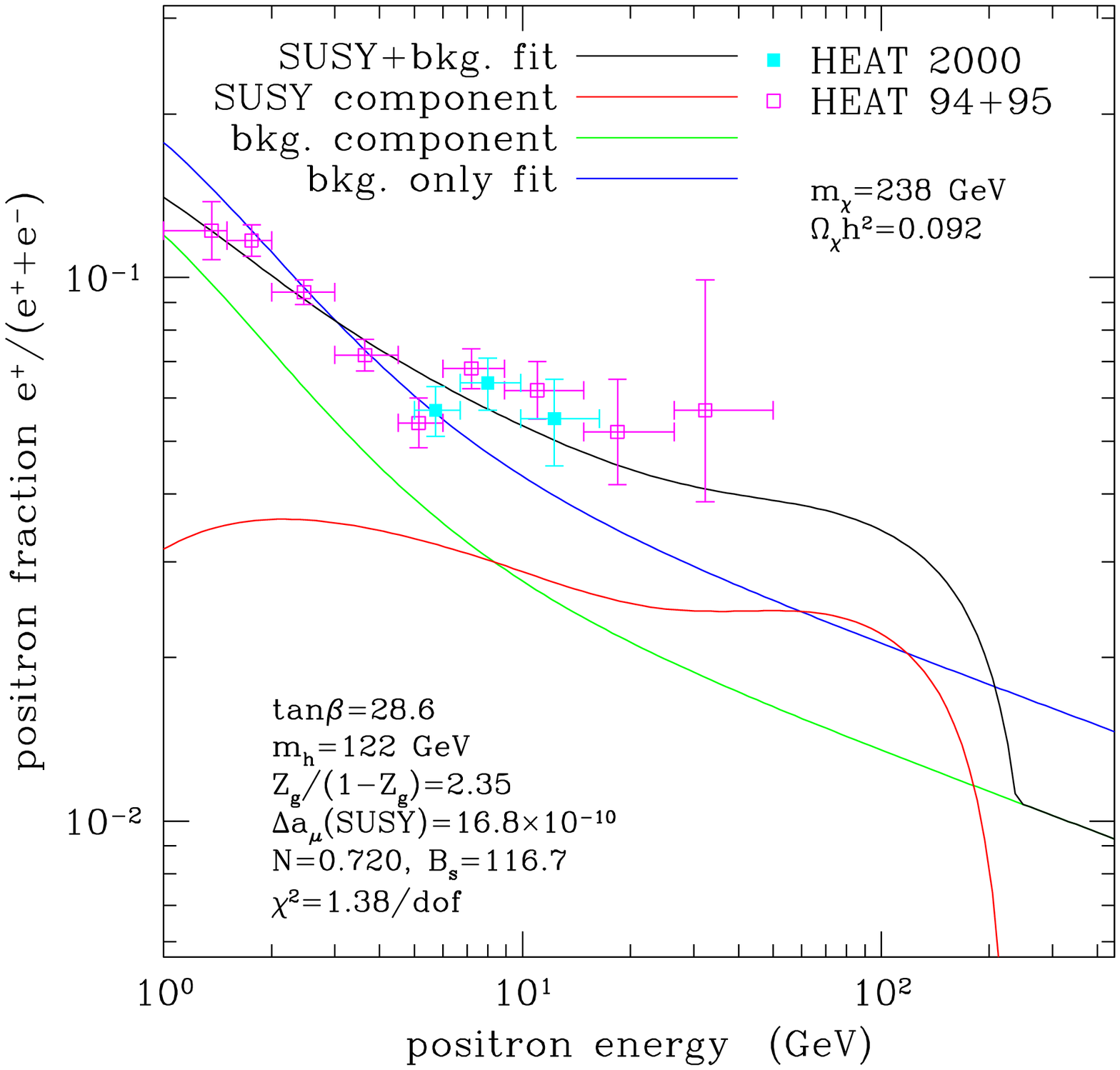}{0.95}
\end{center}
\end{minipage}
\caption{Two positron spectra for which contributions from neutralino
annihilation improve the fit to HEAT data.  In each case, the
contribution from neutralino annihilation has been enhanced by a
factor $B_s \sim 100$ relative to the prediction for a smooth halo.
{}From Ref.~\protect\citenum{Baltz:2001ir}.
\label{fig:positronab} }
\end{figure}

Two ``best fit'' results from Ref.~\citenum{Baltz:2001ir} are shown in
\figref{positronab}.  In this study, the ignorance of subhalo
structure is parameterized by a constant $B_s$, an overall
normalization factor that enhances the positron flux relative to what
would be expected for a smooth halo.  As can be seen in
\figref{positronab}, both spectra give improved fits to the data.
They require substantial boost factors, however, with $B_s \sim 100$.
Such large boost factors may be disfavored by models of halo
formation~\cite{Hooper:2003ad}.

\subsubsection{Photons}

Dark matter in the galactic center may annihilate to photons, which
can be detected in atmospheric Cherenkov telescopes on the ground, or
in space-based detectors~\cite{Rudaz:1987ry,Stecker:1987dz,%
Stecker:1988ar,Bergstrom:1998fj,Berezinsky:1992mx,Berezinsky:1994wv,%
Urban:1992ej,Gondolo:1999ef}.  (Photons from the galactic
halo~\cite{Bergstrom:1998zs,Bergstrom:1999jj}, or even from
extra-galactic sources~\cite{Baltz:2000ra} have also been considered.)

The main source of photons is from cascade decays of other primary
annihilation products.  A line source from loop-mediated processes
such as $\chi \chi \to \gamma
\gamma$~\cite{Rudaz:1990rt,Bergstrom:1997fh,Bern:1997ng} and $\chi
\chi \to \gamma Z$~\cite{Ullio:1998ke} is
possible~\cite{Bergstrom:1998fj}, but is typically highly
suppressed~\cite{Berezinsky:1992sn}.
 
The differential photon flux along a direction that forms an angle
$\psi$ with respect to the direction of the galactic center is
\begin{equation}
\frac{d \Phi_{\gamma}}{d\Omega dE} = 
\sum_i \frac{dN_{\gamma}^i}{dE} 
\sigma_i v \frac{1}{4\pi \mchi^2} \int_{\psi} \rho^2 dl \ ,
\label{photon}
\end{equation}
where the sum is over all annihilation channels, $\rho$ is the
neutralino mass density, and the integral is along the line of sight.
All of the halo model dependence is isolated in the integral.
Depending on the clumpiness or cuspiness of the halo density profile,
this integral may vary by as much as 5 orders of
magnitude~\cite{Bergstrom:1998fj}.

\begin{figure}[tb]
\postscript{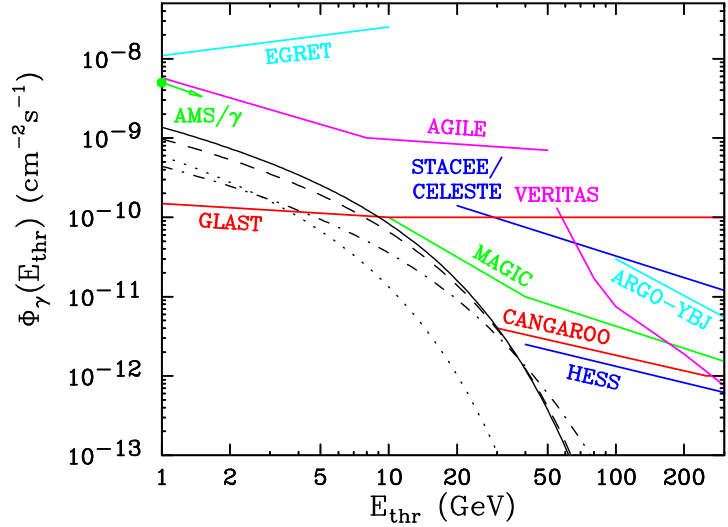}{0.65}
\caption{Integral photon fluxes $\Phi_{\gamma}(\ethr)$ as a function
of threshold energy $\ethr$ for $A_0=0$, $\mu>0$, $m_t = 174~\gev$,
and halo parameter $\bar{J} = 500$.  The four models have relic
density $\Omegachi h^2 \approx 0.15$, and are specified by $(\tb, m_0,
\mgaugino, \mchi, R_{\chi}) = (10, 100, 170, 61, 0.93)$ (dotted),
$(10, 1600, 270, 97, 0.77)$ (dashed), $(10, 2100, 500, 202, 0.88)$
(dot-dashed), and $(50, 1000, 300, 120, 0.96)$ (solid), where all
masses are in GeV. Point source flux sensitivity estimates for several
gamma ray detectors are also shown. {}From
Ref.~\protect\citenum{Feng:2000zu}.  }
\label{fig:photon_explims}
\end{figure}

The integrated photon signal for 4 representative minimal supergravity
models is given in \figref{photon_explims}.  A moderate halo profile
is assumed.  Experiments sensitive to such photon fluxes are listed in
Table~\ref{table:photons}, and their sensitivities are given in
\figref{photon_explims}.

\begin{table}[tb]
\caption{Some of the current and planned $\gamma$ ray detector
experiments with sensitivity to photon energies $10~\gev \protect\alt
E_{\gamma} \protect\alt 300~\gev$.  We list each experiment's start
date and expected $E_{\gamma}$ coverage in GeV.  The energy ranges are
approximate.  For experiments constructed in stages, the listed
threshold energies will not be realized initially.  {}From
Ref.~\protect\citenum{Feng:2000zu}.  \label{table:photons} }
\begin{center}
\begin{tabular}{lllr}
 Experiment \rule[-2.8mm]{0mm}{7mm}
  & \hspace*{2mm} Type
   & Date
    & $E_{\gamma}$ Range
\\  
\hline
 EGRET
  & Satellite
   & 1991-2000
    & 0.02--30
\\
 STACEE
  & ACT array
   & 1998
    & 20--300
\\
 CELESTE
  & ACT array
   & 1998
    & 20--300
\\
 ARGO-YBJ
  & Air shower
   & 2001
    & 100--2,000
\\
 MAGIC
  & ACT
   & 
    & 10--1000
\\
 AGILE
  & Satellite
   & 
    & 0.03--50
\\
 HESS
  & ACT array
   & 
    & 10--1000
\\
 AMS/$\gamma$
  & Space station
   & 
    & 0.3--100
\\
 CANGAROO III
  & ACT array
   & 
    & 30--50,000
\\
 VERITAS
  & ACT array
   & 
    & 50--50,000
\\
 GLAST
  & Satellite
   & 
    & 0.1--300
\end{tabular}
\end{center}
\end{table}

\subsubsection{Neutrinos}

When neutralinos pass through astrophysical objects, they may be
slowed below escape velocity by elastic scattering.  Once captured,
they then settle to the center, where their densities and annihilation
rates are greatly enhanced.  While most of their annihilation products
are immediately absorbed, neutrinos are not.  Neutralinos may
therefore annihilate to high energy neutrinos in the cores of the
Earth~\cite{Freese:1986qw,Krauss:1986aa,Gaisser:1986ha,Gould:1989eq,%
Bottino:1994xp,Berezinsky:1996ga} and Sun~\cite{Gaisser:1986ha,%
Bottino:1994xp,Berezinsky:1996ga,Press:1985ug,Silk:1985ax,%
Srednicki:1987vj,Hagelin:1986gv,Ng:1987qt,Ellis:1988sh} and be
detected on Earth in neutrino telescopes.

The formalism for calculating neutrino fluxes from dark matter
annihilation is complicated but well developed. (See
Ref.~\citenum{Jungman:1995df} for a review.)  In contrast to the
previous two indirect detection examples, neutrino rates depend not
only on annihilation cross sections, but also on $\chi N$ scattering,
which determines the neutralino capture rate in the Sun and Earth.

As with positrons, $\chi \chi \to \nu \bar{\nu}$ is
helicity-suppressed, and so neutrinos are produced only in the decays of
primary annihilation products.  Typical neutrino energies are then
$E_{\nu} \sim \frac{1}{2}\mchi$ to $\frac{1}{3}\mchi$, with the most
energetic spectra resulting from $WW$, $ZZ$, and, to a lesser extent,
$\tau\bar{\tau}$.  After propagating to the Earth's surface, neutrinos
are detected through their charged-current interactions.  The most
promising signal is from upward-going muon neutrinos that convert to
muons in the surrounding rock, water, or ice, producing through-going
muons in detectors.  The detection rate for such neutrinos is greatly
enhanced for high energy neutrinos, as both the charged-current cross
section and the muon range are proportional to $E_{\nu}$.

\begin{figure}[tb]
\postscript{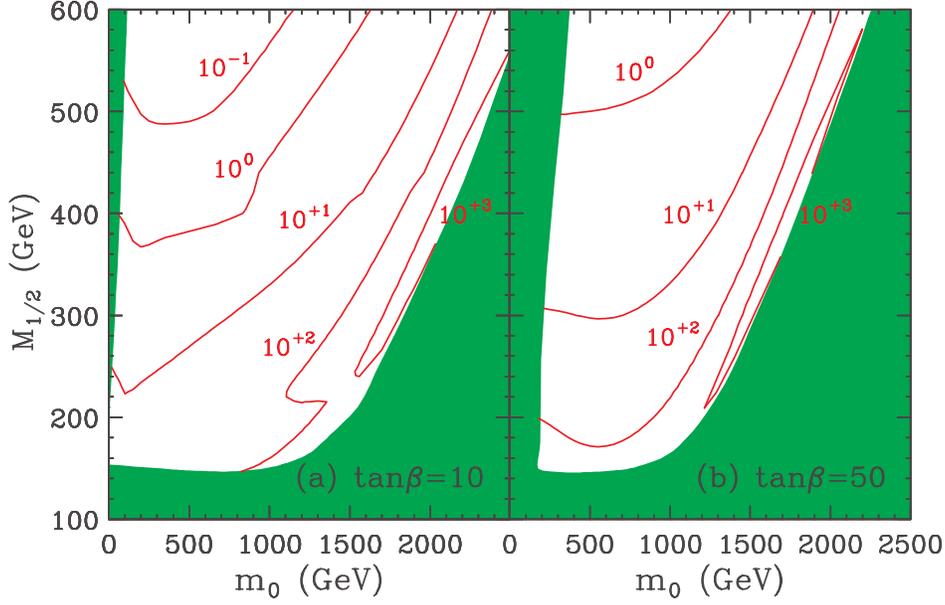}{0.85}
\caption{Muon flux from the Sun in $\km^{-2}~\yr^{-1}$ for $v =
270~\km/\s$ and $\rho_{\chi} = 0.3~\gev/\cm^3$. {}From
Ref.~\protect\citenum{Feng:2000zu}.  }
\label{fig:fluxS}
\end{figure}

The most promising source of neutrinos is the core of the Sun.  Muon
flux rates from the Sun are presented in \figref{fluxS}.  Fluxes as
large as $1000~\km^{-2}~\s^{-1}$ are possible.  Past, present, and
future neutrino telescopes and their properties are listed in
Table~\ref{table:NT}.  Comparing \figref{fluxS} with
Table~\ref{table:NT}, we find that present limits do not significantly
constrain the minimal supergravity parameter space.  However, given
that the effective area of neutrino telescope experiments is expected
to increase by 10 to 100 in the next few years, muon fluxes of order
10--100 $\km^{-2}~\yr^{-1}$ may be within reach.

\begin{table}[tb]
\caption{Current and planned neutrino experiments.  We list also each
experiment's start date, physical dimensions (or approximate effective
area), muon threshold energy $E_{\mu}^{\rm thr}$ in GeV, and 90\% CL
flux limits for the
Sun $\Phi_{\mu}^{\odot}$ in $\km^{-2}~\yr^{-1}$ for half-cone angle
$\theta \approx 15^{\circ}$ when available.  {}From
Ref.~\protect\citenum{Feng:2000zu}.
\label{table:NT} }
\begin{center}
\begin{tabular}{lllrrrr}
 Experiment \rule[-2.8mm]{0mm}{7mm}
 & \hspace*{3mm} Type
  & Date
   & Dimensions\hspace*{2mm}
    & $E_{\mu}^{\rm thr}$
      & $\Phi_{\mu}^{\odot}$\hspace*{2mm}  \\  \hline
 Baksan
 & Ground
  & 1978
   & $17 \times 17 \times 11 ~\text{m}^3$
    & 1  
      & $7.6 \times 10^3$ \\  
 Kamiokande
 & Ground
  & 1983 
   & $\sim 150 ~\text{m}^2$
    & 3
      & $17 \times 10^3$ \\
 MACRO
 & Ground
  & 1989
   & $12 \times 77 \times 9 ~\text{m}^3$
    & 2
      & $6.5 \times 10^3$ \\  
 Super-Kamiokande
 & Ground
  & 1996
   & $\sim 1200~\text{m}^2$
    & 1.6  
       & $5.0 \times 10^3$  \\  
 Baikal NT-96
 & Water
  & 1996
   & $\sim 1000~\text{m}^2$
    & 10  
       &              \\  
 AMANDA B-10
 & Under-ice
  & 1997 
   & $\sim 1000~\text{m}^{2\, \dag}$\hspace*{-1.73mm}
    & $\sim 25$
      &     \\
 Baikal NT-200
 & Water
  & 1998
   & $\sim 2000~\text{m}^2$
     &  \\  
 AMANDA II
 & Ice
  & 2000
   & $\sim 3 \times 10^4~\text{m}^2$
    & $\sim 50$
      &     \\
 NESTOR$^{\S}$
 & Water
  & 
   & $\sim 10^4~\text{m}^{2\, \ddag}$\hspace*{-1.73mm}
    & few
      &     \\
 ANTARES
 & Water
  & 
   &  $\sim 2 \times 10^4~\text{m}^{2\, \ddag}$\hspace*{-1.73mm}
    & $\sim 5$--10
      &     \\    
 IceCube
 & Ice
  & 
   & $\sim 10^6~\text{m}^2$
    & 
      &    
\end{tabular}
\end{center}

$^\dag$ Hard spectrum, $\mchi = 100~\gev$. \
$^{\S}$ One tower. \
$^\ddag$ $E_{\mu} \sim 100~\gev$.
\end{table}

\subsection{Summary}

Neutralinos are excellent dark matter candidates. The lightest
neutralino emerges naturally as the lightest supersymmetric particle
and is stable in simple supersymmetric models.  In addition, the
neutralino is non-baryonic, cold, and weakly-interacting, and so has
all the right properties to be dark matter, and its thermal relic
density is naturally in the desired range.

Current bounds on $\Omega_{\text{DM}}$ are already highly
constraining.  Although these constraints do not provide useful upper
bounds on supersymmetric particle masses, they do restrict
supersymmetric parameter space.  In minimal supergravity, the
cosmologically preferred regions of parameter space include the bulk,
focus point, $A$ funnel, and stau coannihilation regions.

\begin{figure}[tb]
\postscript{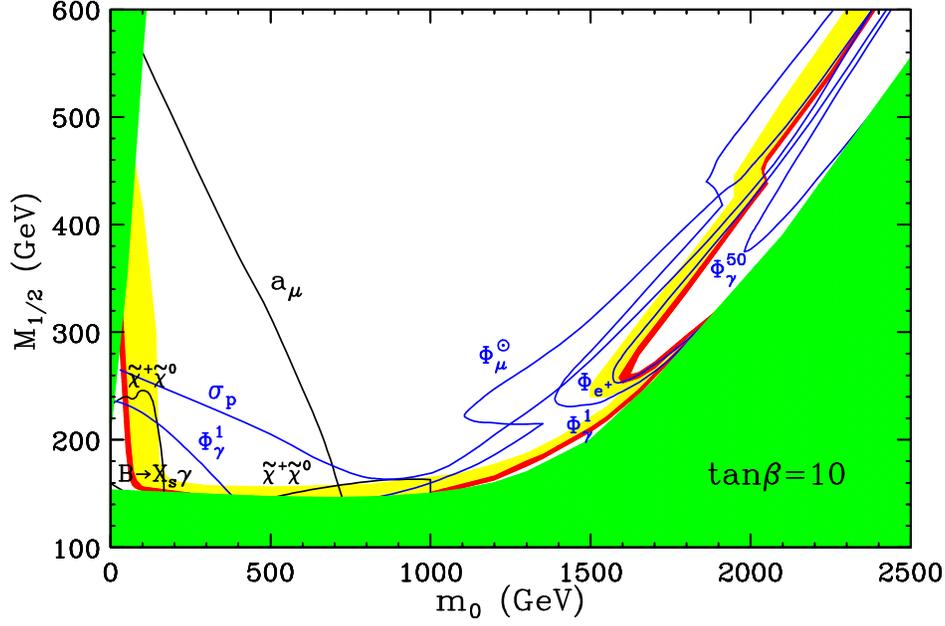}{0.85}
\caption{Estimated reaches of various high-energy collider and
low-energy precision searches (black), direct dark matter searches
(red), and indirect dark matter searches (blue) in the next few years
for minimal supergravity with $A_0 = 0$, $\tb=10$, and $\mu >0$. The
excluded green regions are as in \figref{mSUGRA_LSP}.  The blue
(yellow) shaded region has $0.1 < \OmegaDM h^2 < 0.3$ ($0.025 <
\OmegaDM h^2 < 1$). The regions probed extend the curves toward the
excluded green regions.  {}From Ref.~\protect\citenum{Feng:2000zu}.}
\label{fig:reach10}
\end{figure}

Neutralinos may be detected either directly through their interactions
with ordinary matter or indirectly through their annihilation decay
products.  Null results from direct and indirect dark matter searches
are not yet very constraining.  Future sensitivities of various particle
physics and dark matter detection experiments are shown in
\figref{reach10}.  The sensitivities assumed, and experiments likely
to achieve these sensitivities in the near future, are listed in
Table~\ref{table:comp}.

\begin{table}[tb]
\caption{Constraints on supersymmetric models used in
\figref{reach10}.  We also list experiments likely to reach these
sensitivities in the near future. {}From
Ref.~\protect\citenum{Feng:2000zu}.
\label{table:comp}
}
\begin{tabular}{llll}
 Observable \rule[-2.8mm]{0mm}{7mm}
  & Type
   & Bound
    & Experiment(s)   \\  \hline
 $\tilde{\chi}^{\pm} \tilde{\chi}^0$
  & Collider
   & See Refs.~\protect\citenum{Lykken:2000kp,Matchev:1999nb,Matchev:1999yn}
    & Tevatron Run II \\ 
 $B \to X_s \gamma$
  & Low energy
   & $|\Delta B(B\rightarrow X_s\gamma)| < 1.2\times 10^{-4}$
    & BaBar, BELLE     \\
 Muon MDM
  & Low energy
   & $|a_{\mu}^{\text{SUSY}}| < 8 \times 10^{-10}$
    & Brookhaven E821  \\
 $\sigma_{\text{proton}}$
  & Direct DM
   & \figref{direct}
    & CDMS2, CRESST2 \\
 $\nu$ from Earth
  & Indirect DM
   & $\Phi_{\mu}^{\oplus} < 100~\km^{-2}~\yr^{-1}$
    & AMANDA \\
 $\nu$ from Sun
  & Indirect DM
   & $\Phi_{\mu}^{\odot} < 100~\km^{-2}~\yr^{-1}$
    & AMANDA \\
 $\gamma$ (gal. center)
  & Indirect DM
   & $\Phi_{\gamma}(1) < 1.5\times 10^{-10}~\cm^{-2}~\s^{-1}$
    & GLAST \\
 $\gamma$ (gal. center)
  & Indirect DM
   & $\Phi_{\gamma}(50) < 7\times 10^{-12}~\cm^{-2}~\s^{-1}$
    & HESS, MAGIC \\
 $e^+$ cosmic rays
  & Indirect DM
   & $(S/B)_{\text{max}} < 0.01$
    & AMS-02           
\end{tabular}
\end{table}

Several interesting features are apparent.  First, traditional
particle physics and dark matter searches, particularly indirect
detection experiments, are highly complementary.  Second, at least one
dark matter experiment is predicted to see a signal in almost all of
the cosmologically preferred region.  This illustration is in the
context of minimal supergravity, but can be expected to hold more
generally.  The prospects for neutralino dark matter discovery are
therefore promising.

\section{Gravitino Cosmology}
\label{sec:gravitino}

In \secref{neutralino}, we largely ignored the gravitino.  In this
Section, we will rectify this omission.  Although gravitino
interactions are highly suppressed, gravitinos may have implications
for many aspects of cosmology, including Big Bang nucleosynthesis
(BBN), the cosmic microwave background, inflation, and reheating.
Gravitino cosmology is in many ways complementary to neutralino
cosmology, providing another rich arena for connections between
microscopic physics and cosmology.

\subsection{Gravitino Properties}

The properties of gravitinos may be systematically derived by
supersymmetrizing the standard model coupled to gravity.  Here we will
be content with highlighting the main results.

In an exactly supersymmetric theory, the gravitino is a massless spin
3/2 particle with two degrees of freedom.  Once supersymmetry is
broken, the gravitino eats a spin 1/2 fermion, the Goldstino of
supersymmetry breaking, and becomes a massive spin 3/2 particle with
four degrees of freedom.  As noted in \secref{susybreaking}, the
resulting gravitino mass is
\begin{equation}
m_{\gravitino} = \frac{F}{\sqrt{3} \mstar} \ ,
\end{equation}
where $\mstar \equiv (8\pi G_N)^{-1/2} \simeq 2.4 \times 10^{18}~\gev$
is the reduced Planck mass.  Gravitinos couple standard model
particles to their superpartners through gravitino-gaugino-gauge boson
interactions
\begin{eqnarray}
L = - \frac{i}{8 \mstar}
\bar{\tilde{G}}_{\mu} \left[ \gamma^{\nu}, \gamma^{\rho} \right]
\gamma^{\mu}\tilde{V} F_{\nu\rho} \ ,
\end{eqnarray}
and gravitino-sfermion-fermion interactions
\begin{eqnarray}
L = - \frac{1}{\sqrt{2} \mstar} \partial_{\nu} \tilde{f} \, \bar{f}
\, \gamma^{\mu} \gamma^{\nu} \tilde{G}_{\mu} \ .
\end{eqnarray}

In models with high-scale supersymmetry breaking, such as conventional
supergravity theories, $F \sim \mweak \mstar$, as explained in
\secref{susybreaking}.  The gravitino mass is therefore of the order
of the other superpartner masses, and we expect them all to be in the
range $\sim 100~\gev - 1~\tev$.  The gravitino's effective couplings
are $\sim E/\mstar$, where $E$ is the energy of the process.  The
gravitino's interactions are therefore typically extremely weak, as
they are suppressed by the Planck scale.

We will focus on theories with high-scale supersymmetry breaking in
the following discussion.  Note, however, that in theories with
low-scale supersymmetry breaking, the gravitino may be much lighter,
for example, as light as $\sim \ev$ in some simple gauge-mediated
supersymmetry breaking models.  The gravitino's interactions through
its Goldstino components may also be much stronger, suppressed by
$F/\mweak$ rather than $\mstar$.  For a summary of gravitino cosmology
in such scenarios, see Ref.~\citenum{Giudice:1998bp}.

\subsection{Thermal Relic Density}
\label{sec:gravrelic}

If gravitinos are to play a cosmological role, we must first identify
their production mechanism.  There are a number of possibilities.
Given our discussion of the neutralino thermal relic density in
\secref{neutralino}, a natural starting place is to consider gravitino
production as a result of freeze out from thermal equilibrium.  At
present, the gravitino coupling $E/\mstar$ is a huge suppression.
However, if we extrapolate back to very early times with temperatures
$T \sim \mstar$, even gravitational couplings were strong, and
gravitinos were in thermal equilibrium, with $n_{\gravitino} =
n_{\text{eq}}$.  Once the temperature drops below the Planck scale,
however, gravitinos quickly decouple with the number density
appropriate for relativistic particles.  Following decoupling, their
number density then satisfies $n_{\gravitino} \propto R^{-3} \propto
T^3$.  This has the same scaling behavior as the background photon
number density, however, and so we expect roughly similar number
densities now.

If such gravitinos are stable, they could be dark matter.  In fact,
the first supersymmetric dark matter candidate proposed was the
gravitino~\cite{Pagels:ke}. However, the overclosure bound implies
\begin{equation}
\Omega_{\gravitino} \alt 1 \Rightarrow m_{\gravitino} \alt 1~\kev \ .
\end{equation}
This is not surprising --- relic neutrinos have a similar density, and
the overclosure bound on their mass is similar.  

On the other hand, gravitinos may be unstable~\cite{Weinberg:zq}.
This may be because $R$-parity is broken, or because the gravitino is
not the LSP.  In this case, there is no bound from overclosure, but
there are still constraints.  In particular, the gravitino's decay
products will destroy the successful predictions of BBN for light
element abundances if the decay takes place after BBN.  In the case
where decay to a lighter supersymmetric particle is possible, we can
estimate the gravitino lifetime to be
\begin{equation}
\tau_{\gravitino} \sim \frac{\mstar^2}{m_{\gravitino}^3} \sim 0.1~\yr
\left[ \frac{100~\gev}{m_{\gravitino}} \right] ^3 \ .
\end{equation}
Requiring gravitino decays to be completed before BBN at $t\sim 1~\s$
implies~\cite{Weinberg:zq}
\begin{equation}
m_{\gravitino} \agt 10~\tev \ .
\end{equation}

In both cases, the required masses are incompatible with the most
natural expectations of conventional supergravity theories.
Gravitinos may, however, be a significant component of dark matter if
they are stable with mass $\sim \kev$.  Such masses are possible in
low-scale supersymmetry breaking scenarios, given an appropriately
chosen supersymmetry-breaking scale $F$.

\subsection{Production during Reheating}

In the context of inflation, the gravitino production scenario of
\secref{gravrelic} is rather unnatural.  Between the time when $T \sim
\mstar$ and now, we expect the universe to inflate, which would dilute
any gravitino relic thermal density.  Inflation does provide another
source for gravitinos, however.  Specifically, following inflation, we
expect an era of reheating, during which the energy of the inflaton
potential is transferred to standard model particles and
superpartners, creating a hot thermal bath in which gravitinos may be
produced~\cite{Krauss:1983ik,Nanopoulos:1983up,Khlopov:pf,%
Ellis:1984eq,Juszkiewicz:gg}.

After reheating, the universe is characterized by three hierarchically
separated rates: the interaction rate of standard model particles and
their superpartners with each other, $\sigma_{\text{SM}} n$; the
expansion rate, $H$; and the rate of interactions involving one
gravitino, $\sigma_{\gravitino} n$.  Here $n$ is the number density of
standard model particles.  After reheating, the universe is expected
to have a temperature well below the Planck scale, but still well
above standard model masses.  These rates may then be estimated by
dimensional analysis, and we find
\begin{equation}
\sigma_{\text{SM}} n \sim T \gg H \sim \frac{T^2}{\mstar} \gg 
\sigma_{\gravitino} n \sim \frac{T^3}{\mstar^2} \ .
\end{equation}

The picture that emerges, then, is that after reheating, there is a
thermal bath of standard model particles and their superpartners.
Occasionally these interact to produce a gravitino through
interactions like $g g \to \tilde{g} \gravitino$.  The produced
gravitinos then propagate through the universe essentially without
interacting. If they are stable, as we will assume throughout this
section, they contribute to the present dark matter density.

To determine the gravitino abundance, we turn once again to the
Boltzmann equation
\begin{equation}
\frac{dn}{dt} = -3 H n - \langle \sigma_A v \rangle 
\left( n^2 - \nequ^2 \right) \ .
\end{equation}
In this case, the source term $\nequ^2$ arises from interactions such
as $g g \to \tilde{g} \gravitino$.  In contrast to our previous
application of the Boltzmann equation in \secref{freezeout}, however,
here the $n^2$ sink term, originating from interactions such as
$\tilde{g} \gravitino \to g g$, is negligible.  Changing variables as
before with $t \to T$ and $n \to Y \equiv n/s$, we find
\begin{equation}
\frac{dY}{dT} = - \frac{ \langle \sigma_{\gravitino} v \rangle }
{H T s} n^2 \ .
\end{equation}
The right-hand side is independent of $T$, since $n \propto T^3$, $H
\propto T^2$ and $s \propto T^3$.  We thus find an extremely simple
relation --- the gravitino relic number density is linearly
proportional to the reheat temperature $T_R$.

\begin{figure}[tb]
\postscript{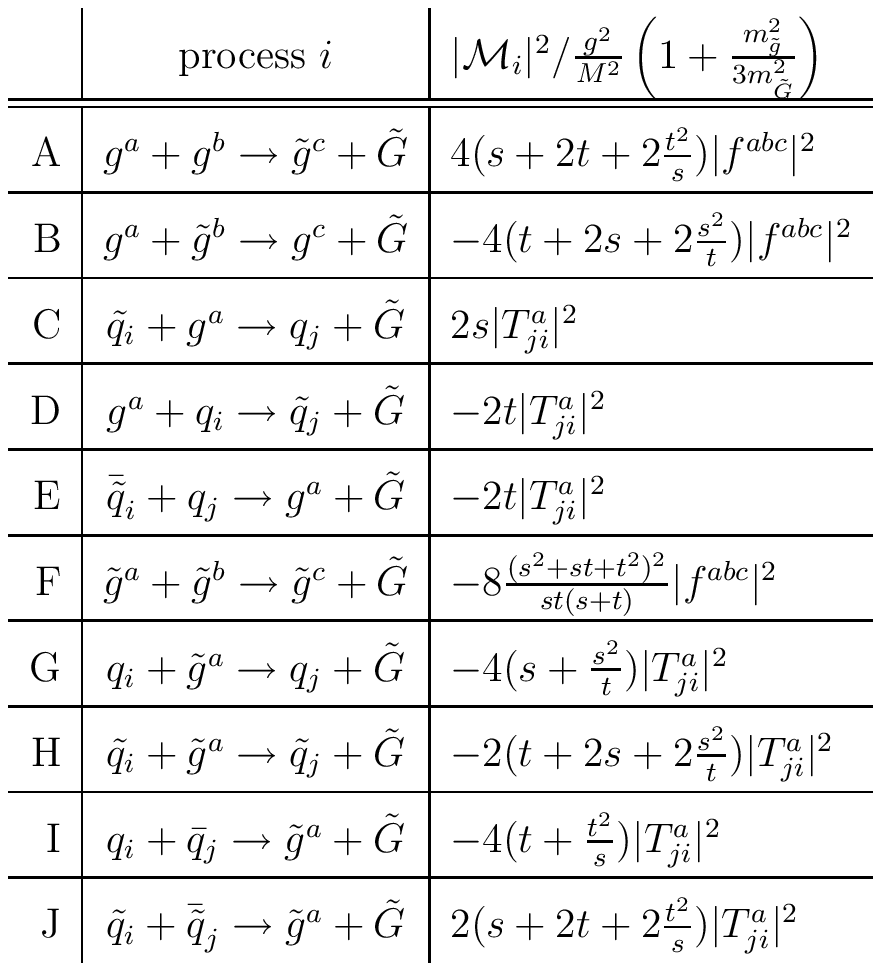}{0.55}
\caption{Processes contributing to gravitino production after
  reheating.  {}From Ref.~\protect\citenum{Bolz:2000fu}.}
\label{fig:reheatprocesses}
\end{figure}

\begin{figure}[tb]
\postscript{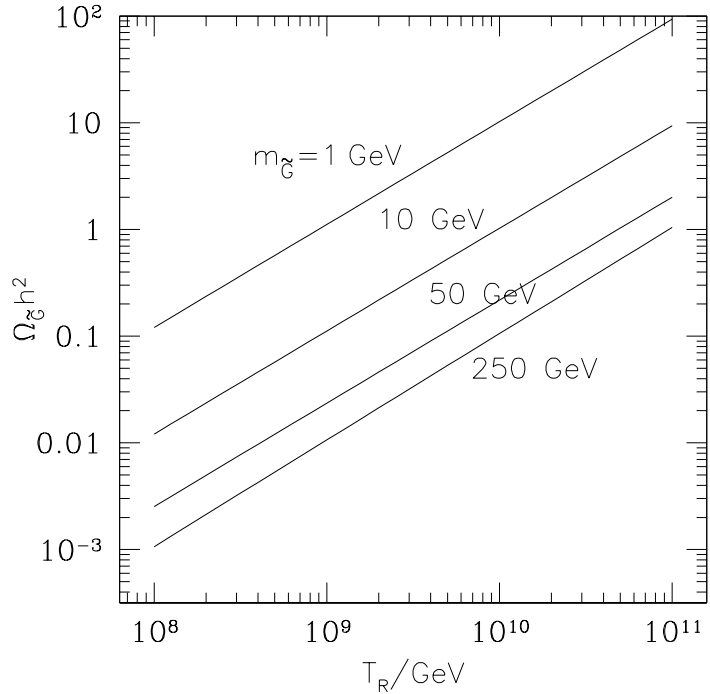}{0.65}
\caption{The gravitino relic abundance $\Omega_{\gravitino} h^2$ as a
function of reheating temperature $T_R$ for various gravitino masses
and gluino mass $m_{\tilde{g}} = 700~\gev$.  {}From
Ref.~\protect\citenum{Bolz:2000fu}.}
\label{fig:reheatbounds}
\end{figure}

The constant of proportionality is the gravitino production cross
section.  The leading $2\to 2$ QCD interactions have been included in
Ref.~\citenum{Bolz:2000fu}.  These are listed in
\figref{reheatprocesses}.  With these results, the gravitino relic
density can be determined as a function of reheating temperature $T_R$
and gravitino mass.  The results are given in \figref{reheatbounds}.
For gravitino mass $m_{\gravitino} \sim 100~\gev$, the constraint on
$\OmegaDM$ requires reheating temperature $T_R \alt 10^{10}~\gev$,
providing a bound on the inflaton potential.  Of course, if this bound
is nearly saturated, gravitinos produced after reheating may be a
significant component of dark matter.

\subsection{Production from Late Decays}
\label{sec:latedecay}

A third mechanism for gravitino production is through the cascade
decays of other supersymmetric particles.  If the gravitino is not the
LSP, cascade decays will bypass the gravitino, given its highly
suppressed couplings.  However, as discussed in \secref{susybreaking},
the gravitino may be the LSP, even in high-scale supersymmetry
breaking models.  If the gravitino is the LSP, all cascades will
ultimately end in a gravitino.

An alternative gravitino dark matter scenario is therefore the
following~\cite{Feng:2003xh,Feng:2003uy}. Assume that the gravitino is
the LSP and stable.  To separate this scenario from the previous two,
assume that inflation dilutes the primordial gravitino density and the
universe reheats to a temperature low enough that gravitino production
is negligible.  Because the gravitino couples only gravitationally
with all interactions suppressed by the Planck scale, it plays no role
in the thermodynamics of the early universe.  The next-to-lightest
supersymmetric particle (NLSP) therefore freezes out as usual; if it
is weakly-interacting, its relic density will be near the desired
value.  However, much later, after
\begin{equation}
\tau \sim \frac{\mstar^2}{\mweak^3} \sim 10^5~\s - 10^{8}~\s \ ,
\label{year}
\end{equation}
the NLSP decays to the gravitino LSP.  The gravitino therefore becomes
dark matter with relic density
\begin{equation}
\Omega_{\gravitino} = \frac{m_{\gravitino}}{m_{\text{NLSP}}}
\Omega_{\text{NLSP}} \ .
\end{equation}
The gravitino and NLSP masses are naturally of the same order in
theories with high-scale supersymmetry breaking.  Gravitino LSPs may
therefore form a significant relic component of our universe,
inheriting the desired relic density from WIMP decay.  In contrast to
the previous two production mechanisms, the desired relic density is
achieved naturally without the introduction of new energy scales.

Given our discussion in \secref{gravrelic}, the decay time of
\eqref{year}, well after BBN, should be of concern.  In the present
case, the decaying particle is a WIMP and so has a density far below
that of a relativistic particle. (Recall \figref{relicabund}.)
However, one must check to see if the light element abundances are
greatly perturbed.  In fact, for some weak-scale NLSP and gravitino
masses they are, and for some they
aren't~\cite{Feng:2003xh,Feng:2003uy}.  We discuss this below, along
with other constraints on this scenario.

Models with weak-scale extra dimensions also provide a similar dark
matter particle in the form of Kaluza-Klein
gravitons~\cite{Feng:2003xh,Feng:2003nr}, with Kaluza-Klein gauge
bosons or leptons playing the role of the decaying
WIMP~\cite{Servant:2002aq,Cheng:2002ej}.  Because such dark matter
candidates naturally preserve the WIMP relic abundance, but have
interactions that are weaker than weak, they have been named
superweakly-interacting massive particles, or
``superWIMPs.''\footnote{A different dark matter candidate that also
predicts late decays is
axinos~\cite{Covi:1999ty,Covi:2001nw,Covi:2004rb,Hooper:2004qf}.}

We see now that our discussion in \secref{neutralino} of WIMP dark
matter was only valid for the ``half'' of parameter space where
$m_{3/2} > m_{\text{LSP}}$.  When the gravitino is the LSP, there are
number of new implications of supersymmetry for cosmology.  For
example, the ``$\tilde{\tau}$ LSP'' region is no longer excluded by
searches for charged dark
matter~\cite{Feng:2003xh,Feng:2003uy,Ellis:2003dn,Ellis:2004qe}, as
the $\tilde{\tau}$ is no longer stable, but only metastable.  There is
therefore the possibility of stable heavy charged particles appearing
in collider detectors~\cite{Feng:1997zr,LHCstable}.  Further, regions
with too much dark matter are no longer excluded, because the
gravitino dark matter density is reduced by $m_{\gravitino} /
m_{\text{NLSP}}$ relative to the NLSP density.  As we will discuss
below, the late decays producing gravitinos may have detectable
consequences for BBN and the cosmic microwave background.
Astrophysical signatures in the diffuse photon
spectrum~\cite{Feng:2003xh}, the ionization of the
universe~\cite{Chen:2003gz}, and the suppression of small scale
structure~\cite{Sigurdson:2003vy} are also interesting possibilities.

\subsection{Detection}

If gravitinos are the dark matter, all direct and indirect searches
for dark matter are hopeless, because all interaction cross sections
and annihilation rates are suppressed by the Planck scale.  Instead,
one must turn to finding evidence for gravitino production in the
early universe.  In the case of gravitinos produced at $T\sim \mstar$
or during reheating, the relevant physics is at such high energy
scales that signals are absent, or at least require strong theoretical
assumptions.  In the case of production by late decays, however, there
are several possible early universe signals.  We consider a few of
these in this Section.

\subsubsection{Energy Release}

If gravitinos are produced by late decays, the relevant reaction is
$\NLSP \to \gravitino + S$, where $S$ denotes one or more standard
model particles.  Because the gravitino is essentially invisible, the
observable consequences rely on finding signals of $S$ production in
the early universe.  Signals from late decays have been considered in
Refs.~\citenum{Ellis:1984er,Ellis:1990nb,%
Kawasaki:1994sc,Holtmann:1998gd,Kawasaki:2000qr,Asaka:1998ju,%
Cyburt:2002uv,Reno:1987qw,Dimopoulos:1988ue,Khlopov:rs,Kohri:2001jx}.
In principle, the strength of these signals depends on what $S$ is and
its initial energy distribution.  It turns out, however, that most
signals depend only on the time of energy release, that is, the NLSP's
lifetime $\tau$, and the average total electromagnetic or hadronic
energy released in NLSP decay.

Here we will consider two possible NLSPs: the photino and the stau.
In the photino case,
\begin{equation}
\Gamma(\photino \to \gamma \gravitino) 
= \frac{1}{48\pi M_*^2}
\frac{m_{\photino}^5}{m_{\gravitino}^2} 
\left[1 - \frac{m_{\gravitino}^2}{m_{\photino}^2} \right]^3 
\left[1 + 3 \frac{m_{\gravitino}^2}{m_{\photino}^2} \right] \ .
\label{photinolifetime}
\end{equation}
In the limit $\Delta m \equiv m_{\photino} - m_{\gravitino} \ll
m_{\gravitino}$, the decay lifetime is
\begin{equation}
\tau(\photino \to \gamma \gravitino) 
\approx 1.8 \times 10^7~\s 
\left[ \frac{100~\gev}{\Delta m} \right]^3  \ ,
\end{equation}
independent of the overall superpartner mass scale.  For the stau
case,
\begin{equation}
 \Gamma(\stau \to \tau \gravitino)
 =\frac{1}{48\pi M_*^2} 
\frac{m_{\stau}^5}{m_{\gravitino}^2} 
\left[1 - \frac{m_{\gravitino}^2}{m_{\stau}^2} \right]^4 .
\label{sleptonlifetime}
\end{equation}
In the limit $\Delta m \equiv m_{\stau} - m_{\gravitino} \ll
m_{\gravitino}$, the decay lifetime is
\begin{equation}
\tau(\stau \to \tau \gravitino) 
 \approx 3.6 \times 10^8~\s 
\left[ \frac{100~\gev}{\Delta m} \right]^4 
\frac{m_{\gravitino}}{1~\tev} \ . 
\end{equation}

The electromagnetic energy release is conveniently written in terms of
\begin{equation}
\zetaEM \equiv \epsEM Y_{\NLSP} \ ,
\end{equation}
where $\epsEM$ is the initial electromagnetic energy released in each
NLSP decay, and $Y_{\NLSP} \equiv n_{\NLSP}/n_{\gamma}^{\text{BG}}$ is
the NLSP number density before they decay, normalized to the number
density of background photons $n_{\gamma}^{\text{BG}} = 2 \zeta(3)
T^3/\pi^2$.  We define hadronic energy release similarly as $\zetahad
\equiv \epshad Y_{\NLSP}$. 

NLSP velocities are negligible when they decay, and so the potentially
visible energy is
\begin{equation}
E_S \equiv \frac{m_{\NLSP}^2 - m_{\gravitino}^2}{2m_{\NLSP}} \ .
\label{ES}
\end{equation}
For the photino case, $S = \gamma$.  At leading order, all of the
initial photon energy is deposited in an electromagnetic shower, and
so
\begin{equation}
\epsEM = E_{\gamma} \ , \quad \epshad \simeq 0 \ .
\label{Egamma}
\end{equation}
For the stau case,
\begin{equation}
\epsEM \approx 
\frac{1}{3} E_{\tau} - E_{\tau} \ , \quad \epshad = 0 \ ,
\label{Etau}
\end{equation}
where the range in $\epsEM$ results from the possible variation in
electromagnetic energy from $\pi^{\pm}$ and $\nu$ decay products.  The
precise value of $\epsEM$ is in principle calculable once the stau's
chirality and mass, and the superWIMP mass, are specified.  However,
as the possible variation in $\epsEM$ is not great relative to other
effects, we will simply present results below for the representative
value of $\epsEM = \frac{1}{2} E_{\tau}$.

\begin{figure}[tbp]
\postscript{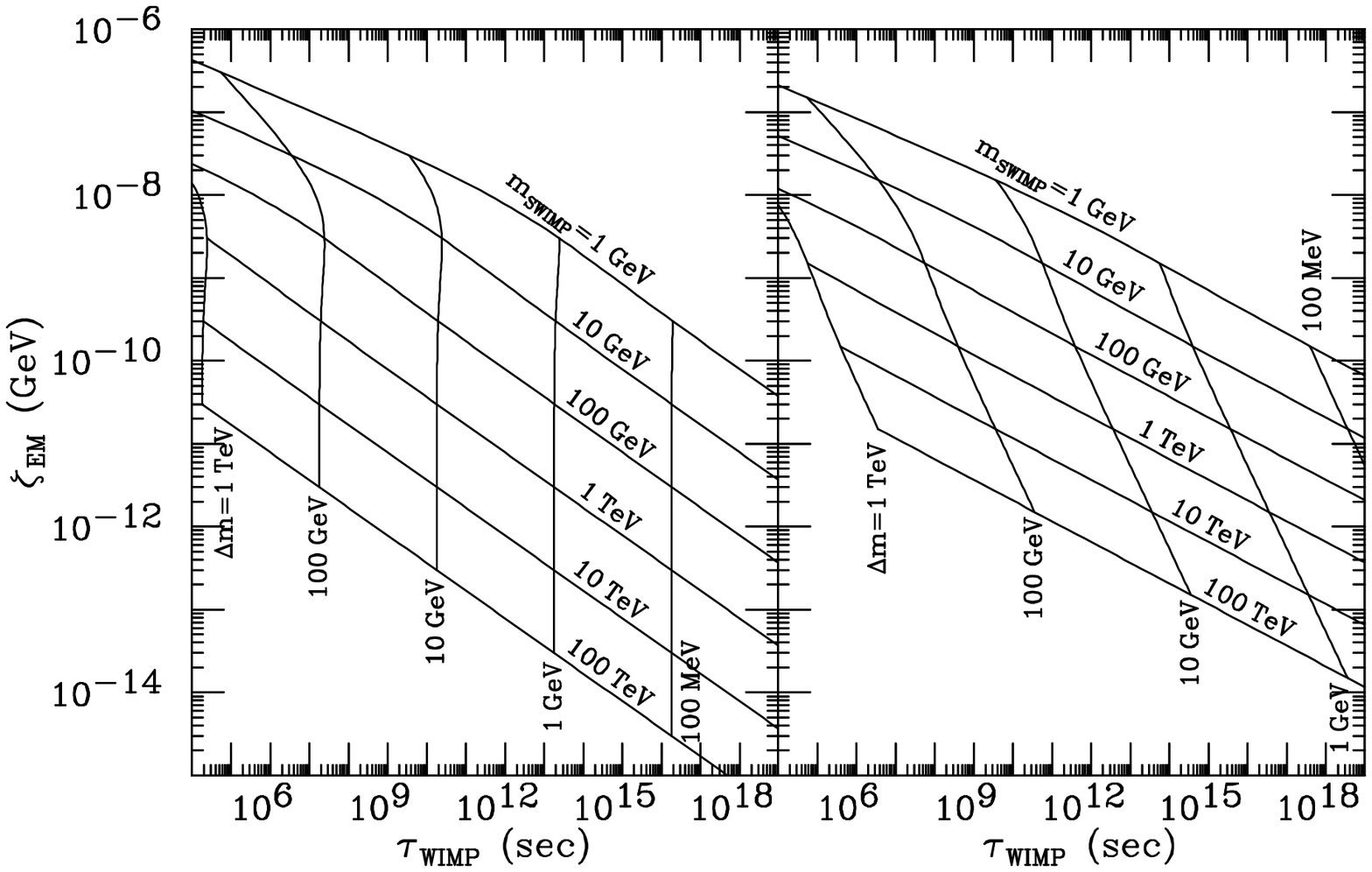}{0.85}
\caption{Predicted values of NLSP lifetime $\tau$ and electromagnetic
  energy release $\zetaEM \equiv \epsEM Y_{\NLSP}$ in the $\photino$
  (left) and $\stau$ (right) NLSP scenarios for $m_{\gravitino} =
  1~\gev$, $10~\gev$, \ldots, $100~\tev$ (top to bottom) and $\Delta m
  \equiv m_{\NLSP} - m_{\gravitino} = 1~\tev$, $100~\gev$, \ldots,
  $100~\mev$ (left to right).  For the $\stau$ NLSP scenario, we
  assume $\epsEM = \frac{1}{2} E_{\tau}$.  {}From
  Ref.~\protect\citenum{Feng:2003uy}.
\label{fig:prediction} }
\end{figure}

The lifetimes and energy releases in the photino and stau NLSP scenarios
are given in Fig.~\ref{fig:prediction} for a range of $(m_{\NLSP},
\Delta m)$.  For natural weak-scale values of these parameters, the
lifetimes and energy releases in the neutralino and stau scenarios are
similar, with lifetimes of about a year, in accord with the rough
estimate of \eqref{year}, and energy releases of
\begin{equation}
\zeta_{\text{EM}} \sim 10^{-9}~\gev \ .
\label{zeta}
\end{equation}
Such values have testable implications, as we now discuss.

\subsubsection{Big Bang Nucleosynthesis}
\label{sec:bbn}

\begin{figure}[tb]
\begin{minipage}[t]{0.48\textwidth}
\begin{center}
\postscript{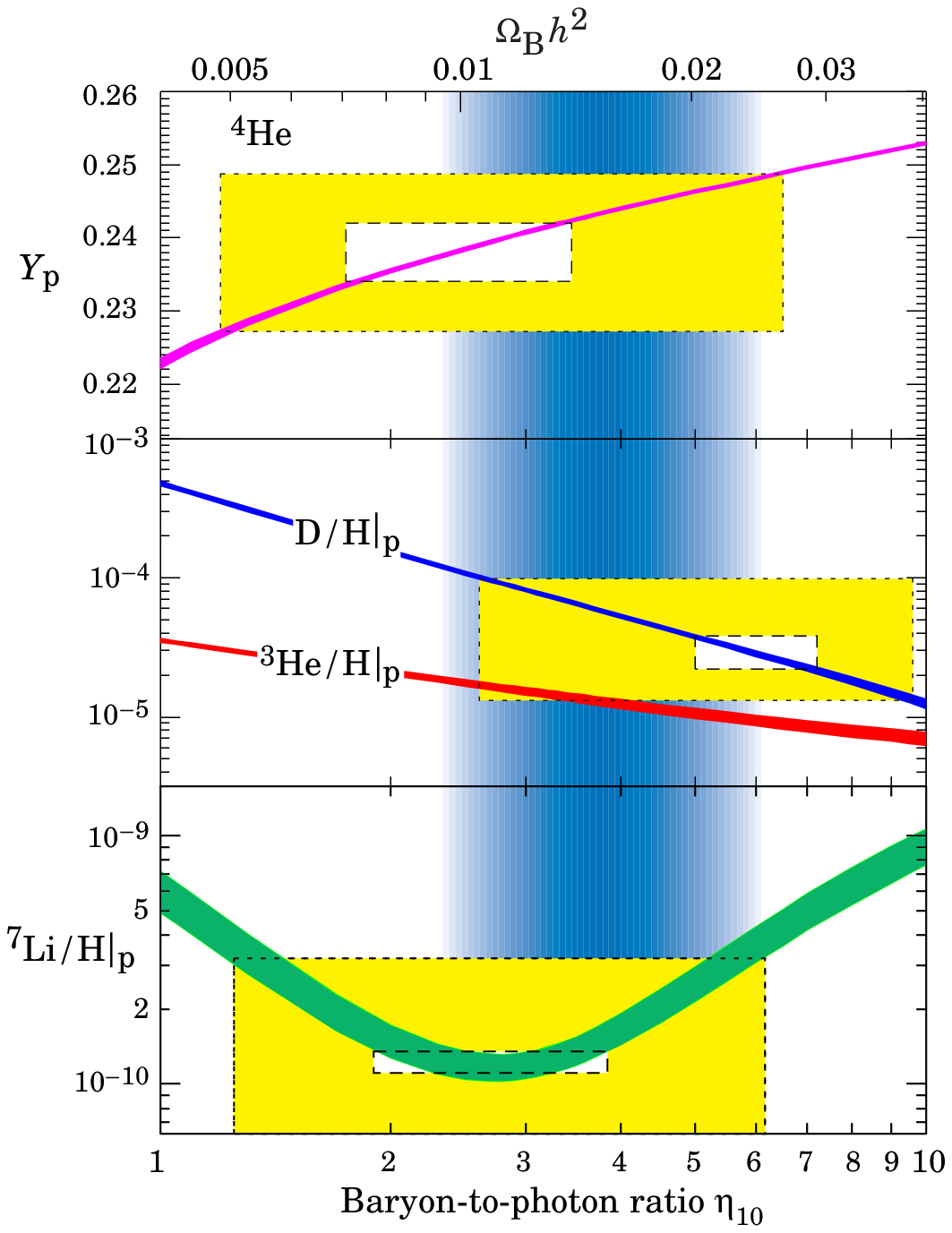}{0.95}
\end{center}
\end{minipage}
\hfil
\begin{minipage}[t]{0.48\textwidth}
\begin{center}
\postscript{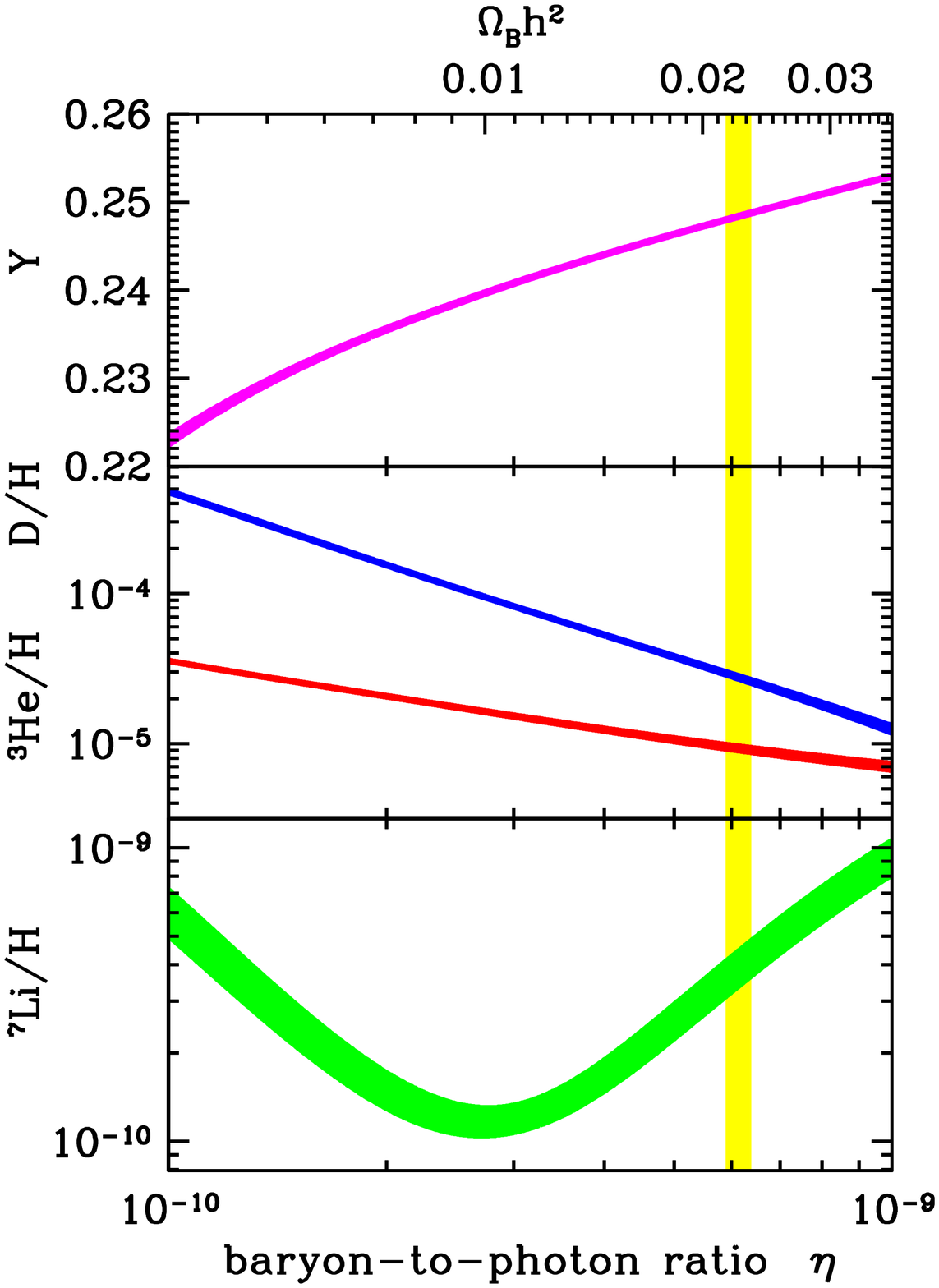}{0.96}
\end{center}
\end{minipage}
\caption{Bounds on the baryon density $\Omega_B h^2$ from BBN (left,
  from Ref.~\protect\citenum{Hagiwara:fs}) and the CMB (right, from
  Ref.~\protect\citenum{Cyburt:2003fe}).  The new and extremely
  precise CMB constraint favors the BBN $\Omega_B h^2$ determination
  from deuterium and implies that the $^7$Li abundance is anomalously
  low.
\label{fig:omegaB} }
\end{figure}

Big Bang nucleosynthesis predicts primordial light element abundances
in terms of one free parameter, the baryon-to-photon ratio $\eta
\equiv n_B / n_{\gamma}$.  At present, the observed D, $^4$He, $^3$He,
and $^7$Li abundances may be accommodated for baryon-to-photon ratios
in the range\cite{Hagiwara:fs}
\begin{equation}
\eta_{10} \equiv  \eta / 10^{-10} = 2.6-6.2 \ .
\label{etarange}
\end{equation}
(See \figref{omegaB}.)  In light of the difficulty of making precise
theoretical predictions and reducing (or even estimating) systematic
uncertainties in the observations, this consistency is a well-known
triumph of standard Big Bang cosmology.

At the same time, given recent and expected advances in precision
cosmology, the standard BBN picture merits close scrutiny. Recently,
BBN baryometry has been supplemented by CMB data, which alone yields
$\eta_{10} = 6.1 \pm 0.4$~\cite{Spergel:2003cb}. Observations of
deuterium absorption features in spectra from high redshift quasars
imply a primordial D fraction of $\text{D/H} = 2.78_{-0.38}^{+0.44}
\times 10^{-5}$~\cite{Kirkman:2003uv}. Combined with standard BBN
calculations~\cite{Burles:2000zk}, this yields $\eta_{10} = 5.9 \pm
0.5$.  The remarkable agreement between CMB and D baryometers has two
new implications for scenarios with late-decaying particles.  First,
assuming there is no fine-tuned cancellation of unrelated effects, it
prohibits significant entropy production between the times of BBN and
decoupling.  Second, the CMB measurement supports determinations of
$\eta$ from D, already considered by many to be the most reliable BBN
baryometer.  It suggests that if D and another BBN baryometer
disagree, the ``problem'' lies with the other light element abundance
--- either its systematic uncertainties have been underestimated, or
its value is modified by new astrophysics or particle physics. At
present BBN predicts a $^7$Li abundance significantly greater
observed.  This disagreement may therefore provide specific evidence
for late-decaying particles in general, and gravitino dark matter in
particular.

\begin{figure}[tb]
\postscript{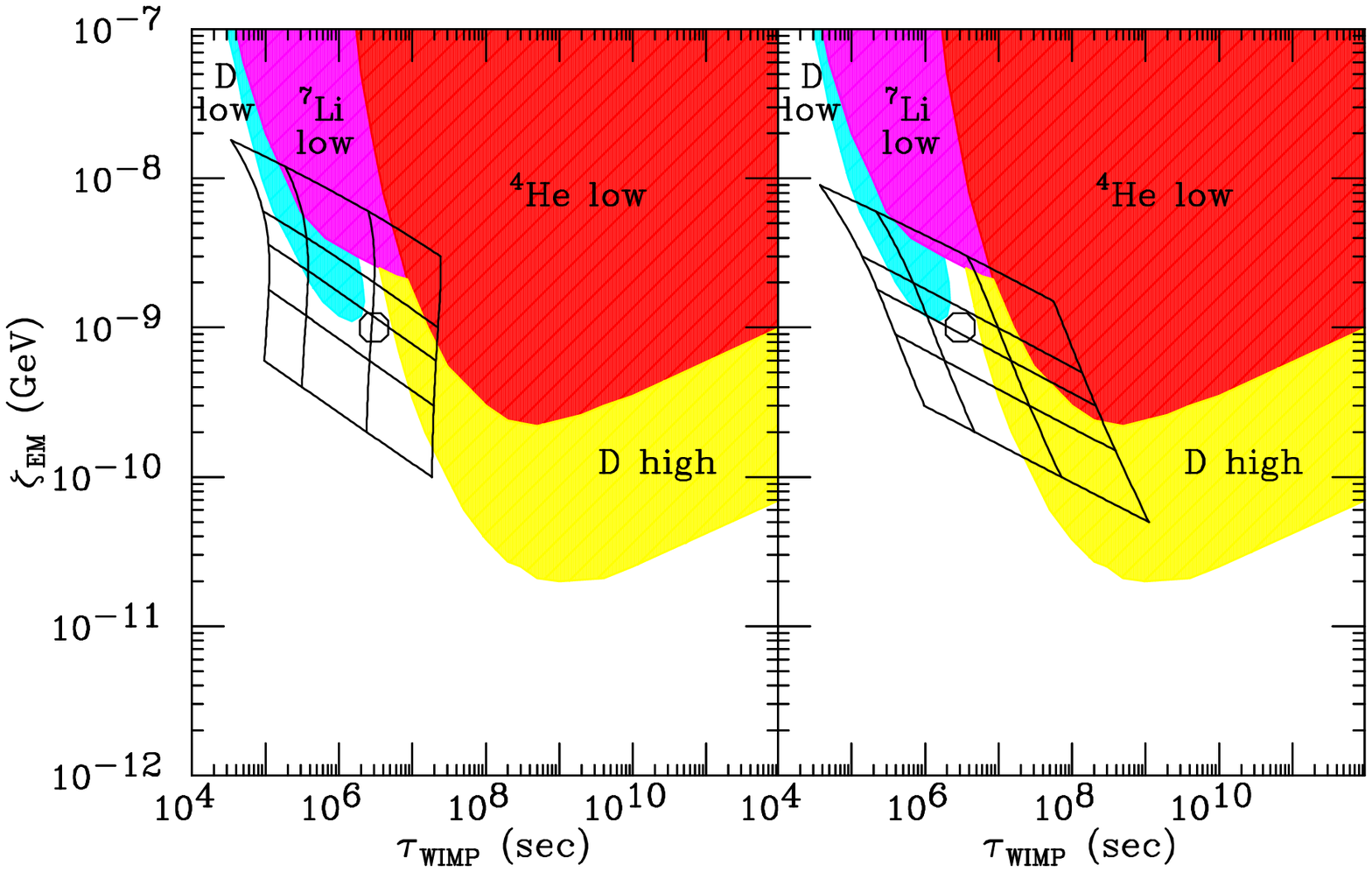}{0.85}
\caption{The grid gives predicted values of NLSP lifetime $\tau$ and
electromagnetic energy release $\zetaEM \equiv \epsEM Y_{\NLSP}$ in
the $\photino$ (left) and $\stau$ (right) NLSP scenarios for
$m_{\gravitino} = 100~\gev$, $300~\gev$, $500~\gev$, $1~\tev$, and
$3~\tev$ (top to bottom) and $\Delta m \equiv m_{\NLSP} -
m_{\gravitino} = 600~\gev$, $400~\gev$, $200~\gev$, and $100~\gev$
(left to right).  For the $\stau$ NLSP scenario, we assume $\epsEM =
\frac{1}{2} E_{\tau}$. BBN constraints exclude the shaded
regions~\protect\cite{Cyburt:2002uv}.  The best fit region with
$(\tau, \zetaEM) \sim (3 \times 10^6~\s, 10^{-9}~\gev)$, where $^7$Li
is reduced to observed levels by late decays of NLSPs to gravitinos,
is given by the circle.  {}From Ref.~\protect\citenum{Feng:2003uy}.
\label{fig:bbn} }
\end{figure}

Given the overall success of BBN, the first implication for new
physics is that it should not drastically alter any of the light
element abundances.  This requirement restricts the amount of energy
released at various times in the history of the universe. A recent
analysis of electromagnetic cascades finds that the shaded regions of
Fig.~\ref{fig:bbn} are excluded by such
considerations~\cite{Cyburt:2002uv}.  The various regions are
disfavored by the following conservative criteria:
\begin{eqnarray}
\mtext{D low}      \ : && \text{D/H} < 1.3 \times 10^{-5} \\
\mtext{D high}     \ : && \text{D/H} > 5.3 \times 10^{-5} \\
\mtext{$^4$He low} \ : && Y_p < 0.227 \\
\mtext{$^7$Li low} \ : && \mtext{$^7$Li/H} < 0.9 \times 10^{-10} \ .
\end{eqnarray}

A subset of superWIMP predictions from Fig.~\ref{fig:prediction} is
superimposed on this plot.  The subset is for weak-scale
$m_{\gravitino}$ and $\Delta m$, the most natural values, given the
independent motivations for new physics at the weak scale.  The BBN
constraint eliminates some of the region predicted by the superWIMP
scenario, but regions with $m_{\NLSP}, m_{\gravitino} \sim \mweak$
remain viable.

The $^7$Li anomaly discussed above may be taken as evidence for new
physics, however.  To improve the agreement of observations and BBN
predictions, it is necessary to destroy $^7$Li without harming the
concordance between CMB and other BBN determinations of $\eta$.  This
may be accomplished for $(\tau, \zetaEM) \sim (3 \times 10^6~\s,
10^{-9}~\gev)$~\cite{Cyburt:2002uv}.  This ``best fit'' point is
marked in Fig.~\ref{fig:bbn}.  The amount of energy release is
determined by the requirement that $^7$Li be reduced to observed
levels without being completely destroyed -- one cannot therefore be
too far from the ``$^7$Li low'' region.  In addition, one cannot
destroy or create too much of the other elements.  $^4$He, with a
binding threshold energy of 19.8 MeV, much higher than Lithium's 2.5
MeV, is not significantly destroyed.  On the other hand, D is loosely
bound, with a binding energy of 2.2 MeV.  The two primary reactions
are D destruction through $\gamma \text{D} \to np$ and D creation
through $ \gamma \, {}^4\text{He} \to \text{DD}$.  These are balanced
in the channel of Fig.~\ref{fig:bbn} between the ``low D'' and ``high
D'' regions, and the requirement that the electromagnetic energy that
destroys $^7$Li not disturb the D abundance specifies the preferred
decay time $\tau \sim 3\times 10^6~\s$.

Without theoretical guidance, this scenario for resolving the $^7$Li
abundance is rather fine-tuned: possible decay times and energy
releases span tens of orders of magnitude, and there is no motivation
for the specific range of parameters required to resolve BBN
discrepancies.  In the superWIMP scenario, however, both $\tau$ and
$\zetaEM$ are specified: the decay time is necessarily that of a
gravitational decay of a weak-scale mass particle, leading to
\eqref{year}, and the energy release is determined by the requirement
that superWIMPs be the dark matter, leading to \eqref{zeta}.
Remarkably, these values coincide with the best fit values for $\tau$
and $\zetaEM$.  More quantitatively, we note that the grids of
predictions for the $\photino$ and $\stau$ scenarios given in
Fig.~\ref{fig:bbn} cover the best fit region.  Current discrepancies
in BBN light element abundances may therefore be naturally explained
by gravitino dark matter.

This tentative evidence may be reinforced or disfavored in a number of
ways. Improvements in the BBN observations discussed above may show if
the $^7$Li abundance is truly below predictions.  In addition,
measurements of $^6$Li/H and $^6$Li/$^7$Li may constrain astrophysical
depletion of $^7$Li and may also provide additional evidence for late
decaying particles in the best fit region~\cite{Holtmann:1998gd,%
Jedamzik:1999di,Kawasaki:2000qr,Cyburt:2002uv,Jedamzik:2004er}. Finally,
if the best fit region is indeed realized by $\NLSP \to \gravitino$
decays, there are a number of other testable implications for
cosmology and particle physics. We discuss one of these in the
following section.  Additional discussion, including diffuse photon
signals, the implications of hadronic energy release, and novel
collider analyses, may be found in Refs.~\citenum{Feng:2003xh,%
Feng:2003uy,Kawasaki:2004yh,Feng:2004zu,Feng:2004mt,%
Buchmuller:2004rq,Buchmuller:2004tm,Wang:2004ib}.

\subsubsection{The Cosmic Microwave Background}
\label{sec:mu}

The injection of electromagnetic energy may also distort the frequency
dependence of the CMB black body radiation.  For the decay times of
interest, with redshifts $z \sim 10^5 - 10^7$, the resulting photons
interact efficiently through $\gamma e^- \to \gamma e^-$, but photon
number is conserved, since double Compton scattering $\gamma e^- \to
\gamma \gamma e^-$ and thermal bremsstrahlung $e X \to e X \gamma$,
where $X$ is an ion, are inefficient.  The spectrum therefore relaxes
to statistical but not thermodynamic equilibrium, resulting in a
Bose-Einstein distribution function
\begin{equation}
f_{\gamma}(E) = \frac{1}{e^{E/(kT) + \mu} - 1} \ ,
\end{equation}
with chemical potential $\mu \ne 0$.

For the low values of baryon density currently favored, the effects of
double Compton scattering are more significant than those of thermal
bremsstrahlung.  The value of the chemical potential $\mu$ may
therefore be approximated for small energy releases by the analytic
expression\cite{Hu:gc}
\begin{equation}
\mu = 8.0 \times 10^{-4} 
\left[ \frac{\tau}{10^6~\s} \right]^{\frac{1}{2}} 
\left[ \frac{\zetaEM}{10^{-9}~\gev} \right] 
e^{-(\tau_{\text{dC}}/\tau)^{5/4}} \ ,
\end{equation}
where
\begin{equation}
\tau_{\text{dC}} = 6.1 \times 10^6~\s
\left[ \frac{T_0}{2.725~\K} \right] ^{-\frac{12}{5}} 
\left[ \frac{\Omega_B h^2}{0.022} \right]^{\frac{4}{5}}
\left[ \frac{1-\frac{1}{2} Y_p}{0.88} \right]^{\frac{4}{5}} \ .
\end{equation}

\begin{figure}[tbp]
\postscript{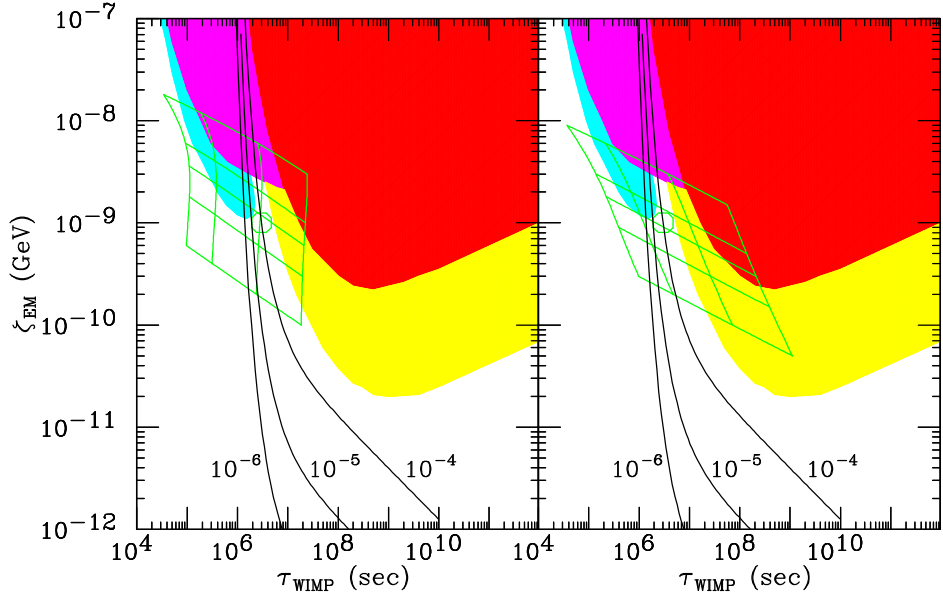}{0.85}
\caption{Contours of $\mu$, parameterizing the distortion of the CMB
  from a Planckian spectrum, in the $(\tau, \zetaEM)$ plane.  Regions
  predicted by the gravitino dark matter scenario, and BBN excluded
  and best fit regions are given as in Fig.~\protect\ref{fig:bbn}.
  {}From Ref.~\protect\citenum{Feng:2003uy}.
\label{fig:mu} }
\end{figure}

In Fig.~\ref{fig:mu} we show contours of chemical potential $\mu$.
The current bound is $\mu < 9\times
10^{-5}$.\cite{Fixsen:1996nj,Hagiwara:fs} We see that, although there
are at present no indications of deviations from black body, current
limits are already sensitive to the superWIMP scenario, and
particularly to regions favored by the BBN considerations described in
Sec.~\ref{sec:bbn}. In the future, the Diffuse Microwave Emission
Survey (DIMES) may improve sensitivities to $\mu \approx 2 \times
10^{-6}$.\cite{DIMES} DIMES will therefore probe further into
superWIMP parameter space, and will effectively probe all of the
favored region where the $^7$Li underabundance is explained by
gravitino dark matter.

\subsection{Summary}

\begin{itemize}

\item The gravitino mass is determined by the scale of supersymmetry
breaking and may be anywhere in the range from eV to TeV.  In
supergravity theories, its mass is at the weak scale and its couplings
are suppressed by the Planck scale, and so extremely weak.

\item If gravitinos are produced as a thermal relic, their mass is
bounded by overclosure to be $m_{\gravitino} \alt \kev$ if they are
stable, and by BBN to be $m_{\gravitino} \agt 10~\tev$ if they are
unstable.

\item Gravitinos may be produced after inflation during reheating. For
stable weak-scale gravitinos, overclosure places an upper bound on the
reheat temperature of the order of $10^{10}~\gev$.

\item Weak-scale gravitinos may also be produced in NLSP decays at
time $t \sim 10^4 - 10^8~\s$.  In this case, gravitinos may be dark
matter, naturally inheriting the desired relic density.  Gravitino
dark matter is undetectable by conventional direct and indirect dark
matter searches, but may be discovered through its imprint on early
universe signals, such as BBN and the CMB.
\end{itemize}

\section{Prospects}
\label{sec:synergy}

We have now discussed a wide variety of cosmological implications of
supersymmetry.  If discoveries are made in astrophysical and
cosmological observations, what are the prospects for determining if
this new physics is supersymmetry?  Put more generally, what are the
prospects for a microscopic understanding of the dark universe?  Such
questions are grand, and their answers speculative.  Nevertheless,
some lessons may be drawn even now.  As we will see, even in the best
of cases, we will need diverse experiments from both particle physics
and cosmology to explore this frontier.

\subsection{The Particle Physics/Cosmology Interface}

As a case study, let us confine our discussion to one topic:
neutralino dark matter.  We assume that non-baryonic dark matter is in
fact neutralinos.  If this is so, what are the prospects for
establishing this, and what tools will we need?

It is first important to recognize the limitations of both cosmology
and particle physics when taken separately:
\begin{itemize}
\item {\em Cosmological observations and astrophysical experiments
cannot discover supersymmetry.}  As noted in \secref{introduction},
cosmological data leaves the properties of dark matter largely
unconstrained.  If dark matter is discovered in direct or indirect
detection experiments, its mass and interaction strengths will be
bounded but only very roughly at first.  (For example, the region
favored by the DAMA signal spans factors of a few in both mass and
interaction strength; see \figref{cdms}.)  These constraints will be
sharpened by follow-up experiments.  However, the microscopic
implications of such experiments are clouded by significant
astrophysical ambiguities, such as the dark matter velocity
distribution, halo profiles, etc.  Even with signals in a variety of
direct and indirect detection experiments, it is unlikely that dark
matter properties will be constrained enough to differentiate
supersymmetry from other reasonable possibilities.

\item {\em Particle experiments cannot discover dark matter.}  If
weak-scale superpartners exist, particle colliders will almost
certainly be able to discover at least some of them.  However, even if
they find all of them, the dark matter candidate will most likely
appear only as missing energy and momentum.  Furthermore, collider
experiments can only test the stability of such particles up to
lifetimes of $\sim 10^{-7}~\s$.  As we have seen in
\secref{gravitino}, lifetimes of a year or more are perfectly natural
in well-motivated models of new physics.  The conclusion that a
particle seen in collider experiments is the dark matter therefore
requires an unjustified extrapolation of 24 orders of magnitude in the
particle's lifetime.
\end{itemize}

\begin{figure}[tb]
\postscript{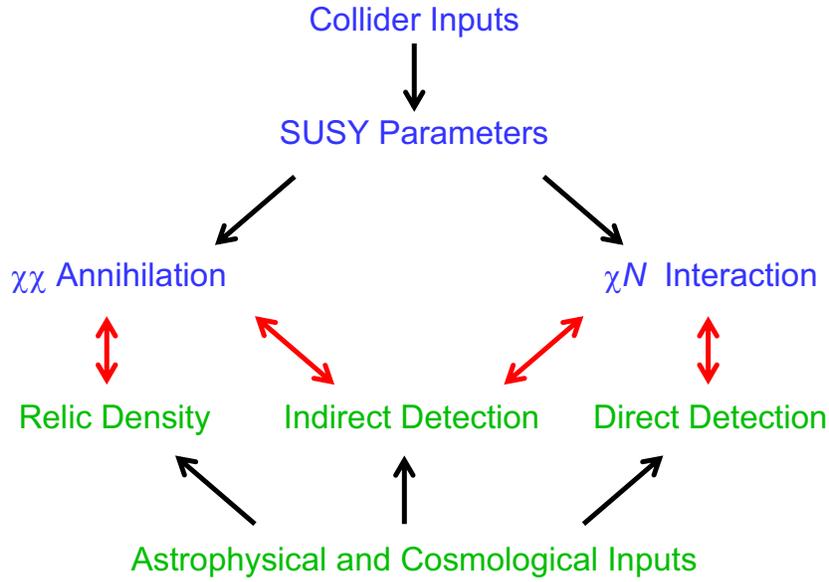}{0.75}
\caption{The road to a microscopic understanding of neutralino dark
matter.}
\label{fig:flowchart}
\end{figure}

Through the combination of both approaches, however, it is possible
that a cohesive and compelling theory of dark matter will emerge.  A
schematic picture of the combined investigation of neutralino dark
matter is given in \figref{flowchart}~\cite{Feng:2002tc}.  Working
from the bottom, cosmological observations have already determined the
relic density with some precision.  Future observations, such as by
the Planck satellite~\cite{Planck}, are likely to reduce uncertainties
in the relic density determination to the 1\% level, given now
standard cosmological assumptions.  Astrophysical experiments may also
detect dark matter either directly through its interactions with
ordinary matter or indirectly through its annihilation decay products.
Such data, combined with astrophysical inputs such as the dark matter
halo profile and local density, will provide information about the
strength of $\chi N$ scattering and $\chi \chi$ annihilation.
 
At the same time, working from the top of \figref{flowchart},
colliders will discover supersymmetry and begin to determine the
parameters of the weak-scale Lagrangian.  These parameters will, in
principle, fix the neutralino's thermal relic density, the $\chi N$
scattering cross section, and the neutralino pair annihilation rates.
Completion of this program at a high level of precision, followed by
detailed comparison with the measured relic density and detection
rates from cosmology and astrophysics will provide a great deal of
information about the suitability of neutralinos as dark matter
candidates.

\subsection{The Role of Colliders}

Clearly data from particle colliders will be required to identify
neutralino dark matter.  The requirements for colliders depend
sensitively on what scenario is realized in nature.  As examples,
consider some of the cosmologically preferred regions of minimal
supergravity discussed in \secref{thermal}.  In the bulk region, one
must verify that the neutralino is Bino-like and must determine the
masses of sfermions that appear in the $t$-channel annihilation
diagrams.  In the focus point region, the neutralino's gaugino-ness
must be precisely measured, whereas in the $A$ funnel region, a high
precision measurement of $m_A - 2 m_{\chi}$ is critical.  Finally, for
the co-annihilation region, there is extreme sensitivity to the
$\tilde{\tau}$--$\chi$ mass splitting.  Measurements below the GeV
level are required to accurately determine the predicted thermal relic
density.

Let us consider the bulk region scenario in more detail.  Not all
sfermion masses need be measured.  For example, if the right-handed
sleptons are light, they typically give the dominant contribution,
since these have the largest hypercharge $Y$ and the annihilation
diagram is proportional to $Y^4$.  In such cases, measurements of
$m_{\tilde{l}_R}$ and lower bounds on left-handed slepton and squark
masses will provide a reasonable starting point.

\begin{figure}[tb]
\postscript{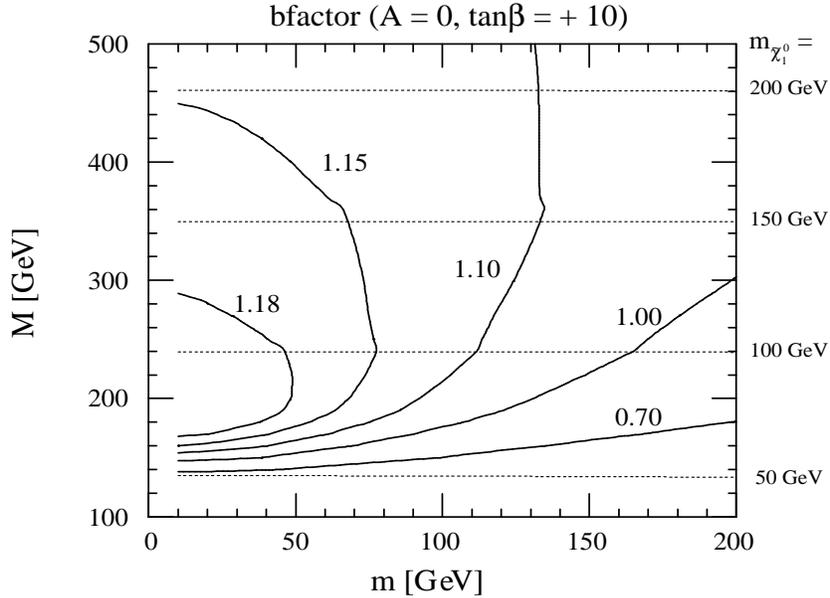}{0.75}
\caption{The ratio of the true $\OmegaDM h^2$ to that calculated with
the $\tilde{l}_R$ $t$-channel diagrams in the $(m,M)$ plane, where
$m$ and $M$ are the universal scalar and gaugino masses of minimal
supergravity, respectively.  {}From
Ref.~\protect\citenum{Drees:2000he}.}
\label{fig:lhcbfactor}
\end{figure}

The possibility of doing this at the LHC has been considered in
Ref.~\citenum{Drees:2000he}.  In much of the bulk region, the cascade
decay $\tilde{q}_L \to \tilde{\chi}^0_2 q \to \tilde{l}_R l q \to l^+
l^- \tilde{\chi}^0_1 q$ is open.  Kinematic endpoints may then be used
to determine the $\tilde{l}_R$ and $\tilde{\chi}^0_1$ masses
precisely.  Assuming that the lightest neutralino is Bino-like, one
may then estimate the relic density, keeping only $\tilde{l}_R$
exchange diagrams.  As shown in \figref{lhcbfactor}, this provides an
estimate accurate to about $\sim 20\%$ in minimal supergravity.
Following this, one would then need to determine the gaugino-ness
of the lightest neutralino and set lower bounds on the other sfermion
masses.

\begin{figure}[tb]
\postscript{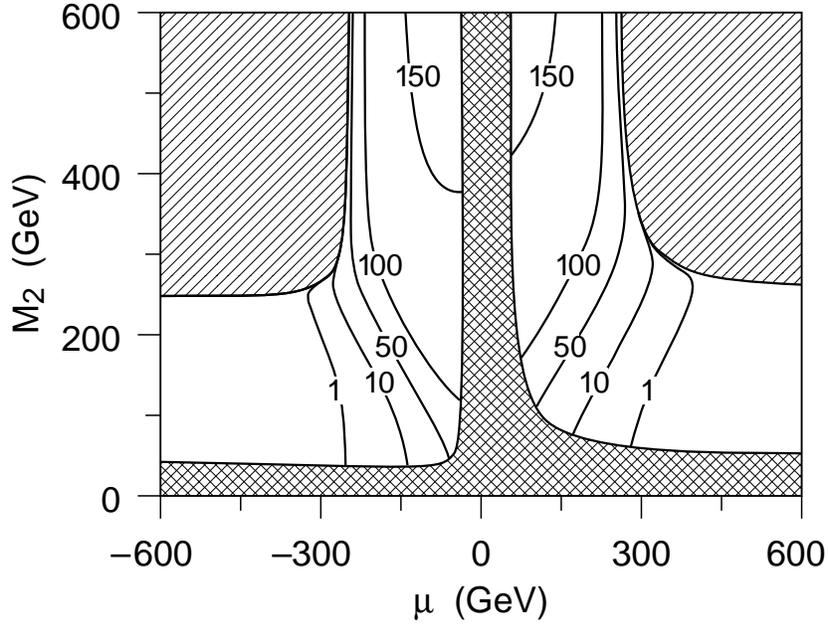}{0.75}
\caption{Contours of constant cross section $\sigma ( e^+ e_R^- \to
\tilde{\chi}^+ \tilde{\chi}^-)$ for a $\sqrt{s} = 500~\gev$ linear
collider.  {}From Ref.~\protect\citenum{Feng:1995zd}.}
\label{fig:lcgauginoness}
\end{figure}

At a linear collider, one may establish that the new particles being
produced are supersymmetric by measuring their dimensionless
couplings.  One may then go on to determine the gaugino-ness of the
LSP in a model-independent manner.  For example, the cross section
$\sigma ( e^+ e_R^- \to \tilde{\chi}^+ \tilde{\chi}^-)$ nearly
vanishes for gaugino-like charginos.  It therefore provides a
sensitive measure of chargino gaugino-ness.  (See
\figref{lcgauginoness}.)  Combined with kinematic measurements of the
chargino mass, the parameters $M_2$ and $\mu$ may be measured
precisely.  Further measurements can use these results to pinpoint
$M_1$ and $\tan\beta$, and thereby the gaugino-ness of the LSP.
Precisions of $\sim 1\%$ or better are possible, translating into
predictions for relic densities and dark matter cross sections that
will match the precision expected from cosmological data.

\subsection{Synthesis}

If the relic density and interaction strengths as determined by
astrophysics and cosmology agree with the predictions of particle
physics with high precision, this agreement will provide strong
evidence that the dark matter is in fact supersymmetric.  It will
imply that we understand the history of the universe back to the
freeze out temperature $\sim 10~\gev$, or times $t \sim 10^{-8}~\s$.
Recall that our current knowledge of the history of the universe is on
sure footing only back to Big Bang nucleosynthesis at temperatures of
$\sim \mev$, or times $t \sim 1~\s$.  Dark matter studies could
therefore provide the necessary evidence to push back our knowledge of
the universe another 8 orders of magnitude in time, a formidable
achievement.

On the other hand, the determinations of relic density and dark matter
interaction strengths by particle physics and cosmology may not agree.
Progress then has many possible paths.  If the disparity is great, one
might look to other dark matter candidates, such as the
axion~\cite{Peccei:1977ur,Weinberg:1977ma,Wilczek:pj} or other
supersymmetric possibilities~\cite{Kusenko:1997si,Kusenko:1997vp,%
Enqvist:2003gh,Covi:1999ty,Covi:2001nw,Covi:2004rb}.  If the relic
density determinations are reasonably close, one might explore the
possibility that the neutralino is not stable, but deposits much of
its relic density in a gravitino LSP, as discussed in
\secref{latedecay}.

Alternatively, one might look to non-standard cosmologies for a
resolution.  The identification of the thermal relic density with the
present day cold dark matter density is subject to cosmological
assumptions.  The calculation of the thermal relic density assumes
that the dominant source of dark matter is from dark matter particles
falling out of thermal equilibrium.  It is possible, however, that the
bulk of the dark matter is created not through thermal equilibrium and
freeze out, but through the out-of-equilibrium decay of a supermassive
particle. The actual relic density would then be greater than the
thermal relic density.  The thermal relic density calculation also
assumes that nothing unusual happens once the dark matter is produced
at temperatures of $T\sim {\cal O}(10)~\gev$.  Large entropy
production by late-decaying particles may dilute calculated relic
densities.  In this case, the actual relic density would be less than
the naive thermal relic density.  The bottom line is that the cold
dark matter density obtained following the path from the bottom of
\figref{flowchart} need not coincide with the thermal relic density
obtained by following the path from the top. Instead, discrepancies
might provide new insights into the history of our universe.

In a similar vein, the neutralino-nucleon cross sections need not
match the dark matter detection rates. As stressed above, this
correspondence requires astrophysical assumptions.  The uncertainties
and problems associated with these issues have been discussed
extensively~\cite{Sikivie:1996nn,Brhlik:1999tt,Gelmini:2000dm,%
Calcaneo-Roldan:2000yt}.  It is possible, however, that the relic
densities, as determined independently by particle physics and
cosmology, agree to 1\%, but the detection rates differ.  One would
then be confident that neutralinos are the dark matter and particle
physics uncertainties would be eliminated, allowing detection
experiments to probe astrophysics.  For example, direct detection
rates would then provide information about the local dark matter
density and velocity distributions, and indirect detection rates would
provide information about halo profiles.  The synergy between
cosmology and particle physics would then truly come full circle.

\subsection{Summary}

A microscopic understanding of the dark universe is a challenging
goal.  As an example, we have focused here on prospects for a
fundamental description of dark matter.  Cosmological measurements,
although able to bound total energy densities, cannot tell us much
about the dark matter's microscopic properties.  On the other hand,
particle physics experiments may produce dark matter and may measure
its properties rather precisely, but cannot never establish its
stability on cosmological time scales.  It is only through the
combination of approaches in particle physics, astrophysics, and
cosmology that the identity of dark matter will be uncovered.  The
task requires many diverse experiments, and will likely take decades
to complete.  Nevertheless, if any of the connections between the weak
scale and cosmology described here are realized in nature, one would
be hard-pressed to envision a more exciting era of discovery than the
coming years.

\section{Acknowledgments}

I am grateful to the organizers of the 2003 SLAC Summer Institute,
where this material was first presented, and to the students and
participants for their enthusiasm and lively discussions.  I thank the
many colleagues, especially Konstantin Matchev, Arvind Rajaraman,
Fumihiro Takayama, and Frank Wilczek, who have helped to shape my
understanding of the topics presented here.  This work was supported
in part by National Science Foundation CAREER Award PHY--0239817 and
in part by the Alfred P.~Sloan Foundation.


\addcontentsline{toc}{section}{References}

\begin{thebibliography}{999}

\bibitem{Spergel:2003cb}
D.~N.~Spergel {\it et al.},
``First Year Wilkinson Microwave Anisotropy Probe (WMAP)
Observations: Determination of Cosmological Parameters,''
astro-ph/0302209.

\bibitem{Tegmark:2003ud}
M.~Tegmark {\it et al.}  [SDSS Collaboration],
``Cosmological parameters from SDSS and WMAP,''
astro-ph/0310723.

\bibitem{susyintros}
For proper introductions to supersymmetry, see, for example,
H.~J.~Muller-Kirsten and A.~Wiedemann,
{\em Supersymmetry: An Introduction With Conceptual And Calculational 
Details},
Print-86-0955 (KAISERSLAUTERN);
J.~Wess and J.~Bagger, {\em Supersymmetry and supergravity}, 2nd
edition (Princeton University Press, Princeton, NJ, 1992);
H.~E.~Haber,
``Introductory low-energy supersymmetry,''
hep-ph/9306207;
X.~Tata,
``Supersymmetry: Where it is and how to find it,''
hep-ph/9510287;
M.~Drees,
``An introduction to supersymmetry,''
hep-ph/9611409;
J.~D.~Lykken,
``Introduction to supersymmetry,''
hep-th/9612114;
S.~P.~Martin,
``A supersymmetry primer,''
hep-ph/9709356;
S.~Weinberg,
{\em The Quantum Theory Of Fields.  Vol. 3: Supersymmetry} 
(Cambridge University Press, Cambridge, UK, 2000);
N.~Polonsky,
{\em Supersymmetry: Structure and phenomena. Extensions of the
standard model},
Lect.\ Notes Phys.\ {\bf M68}, 1 (2001)
[hep-ph/0108236].

\bibitem{Haag:1974qh}
R.~Haag, J.~T.~Lopuszanski and M.~Sohnius,
``All Possible Generators Of Supersymmetries Of The S Matrix,''
Nucl.\ Phys.\ B {\bf 88}, 257 (1975).

\bibitem{Dimopoulos:1981yj}
S.~Dimopoulos, S.~Raby and F.~Wilczek,
``Supersymmetry And The Scale Of Unification,''
Phys.\ Rev.\ D {\bf 24}, 1681 (1981).

\bibitem{Chamseddine:jx}
A.~H.~Chamseddine, R.~Arnowitt and P.~Nath,
``Locally Supersymmetric Grand Unification,''
Phys.\ Rev.\ Lett.\  {\bf 49}, 970 (1982).

\bibitem{Barbieri:1982eh}
R.~Barbieri, S.~Ferrara and C.~A.~Savoy,
``Gauge Models With Spontaneously Broken Local Supersymmetry,''
Phys.\ Lett.\ B {\bf 119}, 343 (1982).

\bibitem{Ohta:1982wn}
N.~Ohta,
``Grand Unified Theories Based On Local Supersymmetry,''
Prog.\ Theor.\ Phys.\  {\bf 70}, 542 (1983).

\bibitem{Hall:iz}
L.~J.~Hall, J.~Lykken and S.~Weinberg,
``Supergravity As The Messenger Of Supersymmetry Breaking,''
Phys.\ Rev.\ D {\bf 27} (1983) 2359.

\bibitem{Alvarez-Gaume:1983gj}
L.~Alvarez-Gaume, J.~Polchinski and M.~B.~Wise,
``Minimal Low-Energy Supergravity,''
Nucl.\ Phys.\ B {\bf 221}, 495 (1983).

\bibitem{Olive:2003iq}
K.~A.~Olive,
``TASI lectures on dark matter,''
astro-ph/0301505.

\bibitem{Feng:2000gh}
J.~L.~Feng, K.~T.~Matchev and F.~Wilczek,
``Neutralino dark matter in focus point supersymmetry,''
Phys.\ Lett.\ B {\bf 482}, 388 (2000)
[hep-ph/0004043].

\bibitem{Goldberg:1983nd}
H.~Goldberg,
``Constraint On The Photino Mass From Cosmology,''
Phys.\ Rev.\ Lett.\  {\bf 50}, 1419 (1983).

\bibitem{Ellis:1983ew}
J.~R.~Ellis, J.~S.~Hagelin, D.~V.~Nanopoulos, K.~A.~Olive and M.~Srednicki,
``Supersymmetric Relics From The Big Bang,''
Nucl.\ Phys.\ B {\bf 238}, 453 (1984).

\bibitem{DMintros}
For proper introductions to dark matter, see, for example,
Refs.~\citenum{Jungman:1995df,Kolb:vq,Olive:2003iq} and
L.~Bergstrom,
``Non-baryonic dark matter: Observational evidence and detection methods,''
Rept.\ Prog.\ Phys.\  {\bf 63}, 793 (2000)
[hep-ph/0002126].

\bibitem{Jungman:1995df}
G.~Jungman, M.~Kamionkowski and K.~Griest,
``Supersymmetric dark matter,''
Phys.\ Rept.\  {\bf 267}, 195 (1996)
[hep-ph/9506380].

\bibitem{Kolb:vq}
E.~W.~Kolb and M.~S.~Turner,
{\em The Early Universe} (Addison-Wesley, Redwood City, CA, 1990). 

\bibitem{Drees:1992am}
M.~Drees and M.~M.~Nojiri,
``The Neutralino relic density in minimal N=1 supergravity,''
Phys.\ Rev.\ D {\bf 47}, 376 (1993)
[hep-ph/9207234].

\bibitem{Gondolo:2000ee}
P.~Gondolo, J.~Edsjo, L.~Bergstrom, P.~Ullio and E.~A.~Baltz,
``DarkSUSY: A numerical package for dark matter calculations in the  MSSM,''
astro-ph/0012234.

\bibitem{Ellis:2003cw}
J.~R.~Ellis, K.~A.~Olive, Y.~Santoso and V.~C.~Spanos,
``Supersymmetric dark matter in light of WMAP,''
Phys.\ Lett.\ B {\bf 565}, 176 (2003)
[hep-ph/0303043].

\bibitem{Feng:1999mn}
J.~L.~Feng, K.~T.~Matchev and T.~Moroi,
``Multi-TeV scalars are natural in minimal supergravity,''
Phys.\ Rev.\ Lett.\  {\bf 84}, 2322 (2000)
[hep-ph/9908309].

\bibitem{Feng:1999zg}
J.~L.~Feng, K.~T.~Matchev and T.~Moroi,
``Focus points and naturalness in supersymmetry,''
Phys.\ Rev.\ D {\bf 61}, 075005 (2000)
[hep-ph/9909334].

\bibitem{Feng:2000bp}
J.~L.~Feng and K.~T.~Matchev,
``Focus point supersymmetry: Proton decay, flavor and CP violation, and the
Higgs boson mass,''
Phys.\ Rev.\ D {\bf 63}, 095003 (2001)
[hep-ph/0011356].

\bibitem{Baer:2002fv}
H.~Baer, C.~Balazs and A.~Belyaev,
``Neutralino relic density in minimal supergravity with
co-annihilations,''
JHEP {\bf 0203}, 042 (2002)
[hep-ph/0202076].

\bibitem{Baer:2002ps}
H.~Baer, C.~Balazs and A.~Belyaev,
``Relic density of neutralinos in minimal supergravity,''
hep-ph/0211213.

\bibitem{Binetruy:1983jf}
P.~Binetruy, G.~Girardi and P.~Salati,
``Constraints On A System Of Two Neutral Fermions From Cosmology,''
Nucl.\ Phys.\ B {\bf 237}, 285 (1984).

\bibitem{Griest:1990kh}
K.~Griest and D.~Seckel,
``Three Exceptions In The Calculation Of Relic Abundances,''
Phys.\ Rev.\ D {\bf 43}, 3191 (1991).

\bibitem{Sikivie:1996nn}
P.~Sikivie, I.~I.~Tkachev and Y.~Wang,
``The secondary infall model of galactic halo formation and the 
spectrum  of cold dark matter particles on earth,''
Phys.\ Rev.\ D {\bf 56}, 1863 (1997)
[astro-ph/9609022].

\bibitem{Brhlik:1999tt}
M.~Brhlik and L.~Roszkowski,
``WIMP velocity impact on direct dark matter searches,''
Phys.\ Lett.\ B {\bf 464}, 303 (1999)
[hep-ph/9903468].

\bibitem{Belli:1999nz} 
P.~Belli, R.~Bernabei, A.~Bottino, F.~Donato, N.~Fornengo, D.~Prosperi
and S.~Scopel,
``Extending the DAMA annual-modulation region by inclusion of the
uncertainties in astrophysical velocities,''
Phys.\ Rev.\ D {\bf 61}, 023512 (2000)
[hep-ph/9903501].

\bibitem{Gelmini:2000dm}
G.~Gelmini and P.~Gondolo,
``WIMP annual modulation with opposite phase in late-infall halo 
models,''
Phys.\ Rev.\ D {\bf 64}, 023504 (2001)
[hep-ph/0012315].

\bibitem{Calcaneo-Roldan:2000yt}
C.~Calcaneo-Roldan and B.~Moore,
``The surface brightness of dark matter: Unique signatures of 
neutralino annihilation in the galactic halo,''
Phys.\ Rev.\ D {\bf 62}, 123005 (2000)
[astro-ph/0010056].

\bibitem{Goodman:1984dc}
M.~W.~Goodman and E.~Witten,
``Detectability Of Certain Dark-Matter Candidates,''
Phys.\ Rev.\ D {\bf 31}, 3059 (1985).

\bibitem{Ellis:2000ds}
J.~R.~Ellis, A.~Ferstl and K.~A.~Olive,
``Re-evaluation of the elastic scattering of supersymmetric dark matter,''
Phys.\ Lett.\ B {\bf 481}, 304 (2000)
[hep-ph/0001005].

\bibitem{Shifman:zn}
M.~A.~Shifman, A.~I.~Vainshtein and V.~I.~Zakharov,
``Remarks On Higgs - Boson Interactions With Nucleons,''
Phys.\ Lett.\ B {\bf 78}, 443 (1978).

\bibitem{Baer:2003jb}
H.~Baer, C.~Balazs, A.~Belyaev and J.~O'Farrill,
``Direct detection of dark matter in supersymmetric models,''
JCAP {\bf 0309}, 007 (2003)
[hep-ph/0305191].

\bibitem{Bernabei:2000qi}
R.~Bernabei {\it et al.}  [DAMA Collaboration],
``Search for WIMP annual modulation signature: Results from DAMA / NaI-3 and
DAMA / NaI-4 and the global combined analysis,''
Phys.\ Lett.\ B {\bf 480}, 23 (2000).

\bibitem{Benoit:2002hf}
A.~Benoit {\it et al.},
``Improved exclusion limits from the EDELWEISS WIMP search,''
Phys.\ Lett.\ B {\bf 545}, 43 (2002)
[astro-ph/0206271].

\bibitem{Akerib:2003px}
D.~S.~Akerib {\it et al.}  [CDMS Collaboration],
``New results from the Cryogenic Dark Matter Search experiment,''
Phys.\ Rev.\ D {\bf 68}, 082002 (2003)
[hep-ex/0306001].

\bibitem{Akerib:2004fq}
D.~S.~Akerib {\it et al.}  [CDMS Collaboration],
``First results from the cryogenic dark matter search in the Soudan
Underground Lab,''
astro-ph/0405033.

\bibitem{Copi:2002hm}
C.~J.~Copi and L.~M.~Krauss,
``Comparing WIMP interaction rate detectors with annual modulation
detectors,''
Phys.\ Rev.\ D {\bf 67}, 103507 (2003)
[astro-ph/0208010].

\bibitem{Freese:2003tt}
K.~Freese, P.~Gondolo and H.~J.~Newberg,
``Detectability of weakly interacting massive particles in the
Sagittarius dwarf tidal stream,''
astro-ph/0309279.

\bibitem{Ullio:2000bv}
P.~Ullio, M.~Kamionkowski and P.~Vogel,
``Spin dependent WIMPs in DAMA?,''
JHEP {\bf 0107}, 044 (2001)
[hep-ph/0010036].

\bibitem{Smith:2001hy}
D.~R.~Smith and N.~Weiner,
``Inelastic dark matter,''
Phys.\ Rev.\ D {\bf 64}, 043502 (2001)
[hep-ph/0101138].

\bibitem{Kurylov:2003ra}
A.~Kurylov and M.~Kamionkowski,
``Generalized analysis of weakly-interacting massive particle
searches,''
Phys.\ Rev.\ D {\bf 69}, 063503 (2004)
[hep-ph/0307185].

\bibitem{Tucker-Smith:2004jv}
D.~Tucker-Smith and N.~Weiner,
``The status of inelastic dark matter,''
hep-ph/0402065.

\bibitem{Rudaz:1987ry}
S.~Rudaz and F.~W.~Stecker,
``Cosmic Ray Anti-Protons, Positrons And Gamma-Rays From Halo Dark Matter
Annihilation,''
Astrophys.\ J.\  {\bf 325}, 16 (1988).

\bibitem{Tylka:1989xj}
A.~J.~Tylka,
``Cosmic Ray Positrons From Annihilation Of Weakly Interacting
Massive Particles In The Galaxy,''
Phys.\ Rev.\ Lett.\  {\bf 63}, 840 (1989).

\bibitem{Turner:1990kg}
M.~S.~Turner and F.~Wilczek,
``Positron Line Radiation From Halo Wimp Annihilations As A Dark
Matter Signature,''
Phys.\ Rev.\  {\bf D42}, 1001 (1990).

\bibitem{Kamionkowski:1991ty}
M.~Kamionkowski and M.~S.~Turner,
``A Distinctive positron feature from heavy WIMP annihilations in the
galactic halo,''
Phys.\ Rev.\  {\bf D43}, 1774 (1991).

\bibitem{Baltz:1999xv}
E.~A.~Baltz and J.~Edsjo,
``Positron propagation and fluxes from neutralino annihilation in the
halo,''
Phys.\ Rev.\  {\bf D59}, 023511 (1999)
[astro-ph/9808243].

\bibitem{Moskalenko:1999sb}
I.~V.~Moskalenko and A.~W.~Strong,
``Positrons from particle dark-matter annihilation in the galactic
halo: Propagation Green's functions,''
Phys.\ Rev.\  {\bf D60}, 063003 (1999)
[astro-ph/9905283].

\bibitem{Stecker:1988fx}
F.~W.~Stecker and A.~J.~Tylka,
``The Cosmic Ray Anti-Proton Spectrum From Dark Matter Annihilation
  And Its Astrophysical Implications: A New Look,''
\APJ{336}{L51}{1989}.

\bibitem{Chardonnet:1996ca}
P.~Chardonnet, G.~Mignola, P.~Salati and R.~Taillet,
``Galactic diffusion and the antiproton signal of supersymmetric dark
matter,''
Phys.\ Lett.\  {\bf B384}, 161 (1996)
[astro-ph/9606174].

\bibitem{Bergstrom:1999jc}
L.~Bergstr\"om, J.~Edsj\"o and P.~Ullio,
``Cosmic antiprotons as a probe for supersymmetric dark matter?,''
astro-ph/9902012.

\bibitem{Bieber:1999dn}
J.~W.~Bieber, R.~A.~Burger, R.~Engel, T.~K.~Gaisser, S.~Roesler and
T.~Stanev,
``Antiprotons at solar maximum,''
Phys.\ Rev.\ Lett.\  {\bf 83}, 674 (1999)
[astro-ph/9903163].

\bibitem{Donato:1999gy}
F.~Donato, N.~Fornengo and P.~Salati,
``Antideuterons as a signature of supersymmetric dark matter,''
Phys.\ Rev.\ D {\bf 62}, 043003 (2000)
[hep-ph/9904481].

\bibitem{Servant:2002aq}
G.~Servant and T.~M.~P.~Tait,
``Is the lightest Kaluza-Klein particle a viable dark matter candidate?,''
Nucl.\ Phys.\ B {\bf 650}, 391 (2003)
[hep-ph/0206071].

\bibitem{Cheng:2002ej}
H.~C.~Cheng, J.~L.~Feng and K.~T.~Matchev,
``Kaluza-Klein dark matter,''
Phys.\ Rev.\ Lett.\  {\bf 89}, 211301 (2002)
[hep-ph/0207125].

\bibitem{Protheroe:1982pp}
R.~J.~Protheroe,
``On The Nature Of The Cosmic Ray Positron Spectrum,''
Astrophys.\ J.\  {\bf 254}, 391 (1982).

\bibitem{Barwick:1995gv}
S.~W.~Barwick {\it et al.}  [HEAT Collaboration],
``Cosmic ray positrons at high-energies: A New measurement,''
Phys.\ Rev.\ Lett.\  {\bf 75}, 390 (1995)
[astro-ph/9505141].

\bibitem{Barwick:1997ig}
S.~W.~Barwick {\it et al.}  [HEAT Collaboration],
``Measurements of the cosmic-ray positron fraction from 1-GeV to 50-GeV,''
Astrophys.\ J.\  {\bf 482}, L191 (1997)
[astro-ph/9703192].

\bibitem{heatnew}
S.~Coutu {\em et al.} [HEAT-pbar Collaboration], 
``Positron Measurements with the HEAT-pbar Instrument,''
in {\em Proceedings of the 27th International Cosmic Ray Conference}
(2001).

\bibitem{Kane:2001fz}
G.~L.~Kane, L.~T.~Wang and J.~D.~Wells,
``Supersymmetry and the positron excess in cosmic rays,''
Phys.\ Rev.\ D {\bf 65}, 057701 (2002)
[hep-ph/0108138].

\bibitem{Baltz:2001ir}
E.~A.~Baltz, J.~Edsjo, K.~Freese and P.~Gondolo,
``The cosmic ray positron excess and neutralino dark matter,''
Phys.\ Rev.\ D {\bf 65}, 063511 (2002)
[astro-ph/0109318].

\bibitem{Kane:2002nm}
G.~L.~Kane, L.~T.~Wang and T.~T.~Wang,
``Supersymmetry and the cosmic ray positron excess,''
Phys.\ Lett.\ B {\bf 536}, 263 (2002)
[hep-ph/0202156].

\bibitem{Hooper:2003ad}
D.~Hooper, J.~E.~Taylor and J.~Silk,
``Can supersymmetry naturally explain the positron excess?,''
hep-ph/0312076.

\bibitem{Stecker:1987dz}
F.~W.~Stecker,
``Gamma-Ray Constraints On Dark Matter Reconsidered,''
Phys.\ Lett.\ B {\bf 201}, 529 (1988).

\bibitem{Stecker:1988ar}
F.~W.~Stecker and A.~J.~Tylka,
``Spectra, Fluxes And Observability Of Gamma-Rays From Dark Matter Annihilation
In The Galaxy,''
Astrophys.\ J.\ {\bf 343}, 169 (1989).

\bibitem{Bergstrom:1998fj}
L.~Bergstr\"om, P.~Ullio and J.~H.~Buckley,
``Observability of gamma rays from dark matter neutralino
  annihilations in the Milky Way halo,''
Astropart.\ Phys.\  {\bf 9}, 137 (1998)
[astro-ph/9712318].

\bibitem{Berezinsky:1992mx}
V.~S.~Berezinsky, A.~V.~Gurevich and K.~P.~Zybin,
``Distribution of dark matter in the galaxy and the lower
  limits for the masses of supersymmetric particles,''
Phys.\ Lett.\  {\bf B294}, 221 (1992).

\bibitem{Berezinsky:1994wv}
V.~Berezinsky, A.~Bottino and G.~Mignola,
``High-energy gamma radiation from the galactic center
  due to neutralino annihilation,''
Phys.\ Lett.\  {\bf B325}, 136 (1994)
[hep-ph/9402215].

\bibitem{Urban:1992ej}
M.~Urban, A.~Bouquet, B.~Degrange, P.~Fleury,
J.~Kaplan, A.~L.~Melchior and E.~Pare,
``Searching for TeV dark matter by atmospheric Cerenkov techniques,''
Phys.\ Lett.\  {\bf B293}, 149 (1992)
[hep-ph/9208255].

\bibitem{Gondolo:1999ef}
P.~Gondolo and J.~Silk,
``Dark matter annihilation at the galactic center,''
Phys.\ Rev.\ Lett.\  {\bf 83}, 1719 (1999)
[astro-ph/9906391].

\bibitem{Bergstrom:1998zs}
L.~Bergstr\"om, J.~Edsj\"o and P.~Ullio,
``Possible Indications of a Clumpy Dark Matter Halo,''
Phys.\ Rev.\  {\bf D58}, 083507 (1998)
[astro-ph/9804050].

\bibitem{Bergstrom:1999jj}
L.~Bergstr\"om, J.~Edsj\"o, P.~Gondolo and P.~Ullio,
``Clumpy neutralino dark matter,''
Phys.\ Rev.\  {\bf D59}, 043506 (1999)
[astro-ph/9806072].

\bibitem{Baltz:2000ra}
E.~A.~Baltz, C.~Briot, P.~Salati, R.~Taillet and J.~Silk,
``Detection of neutralino annihilation photons from external
galaxies,''
Phys.\ Rev.\  {\bf D61}, 023514 (2000)
[astro-ph/9909112].

\bibitem{Rudaz:1990rt}
S.~Rudaz and F.~W.~Stecker,
``On The Observability Of The Gamma-Ray Line Flux From Dark Matter
Annihilation,''
Astrophys.\ J.\  {\bf 368}, 406 (1991).

\bibitem{Bergstrom:1997fh}
L.~Bergstr\"om and P.~Ullio,
``Full one-loop calculation of neutralino annihilation into two
photons,''
Nucl.\ Phys.\  {\bf B504}, 27 (1997)
[hep-ph/9706232].

\bibitem{Bern:1997ng}
Z.~Bern, P.~Gondolo and M.~Perelstein,
``Neutralino annihilation into two photons,''
Phys.\ Lett.\  {\bf B411}, 86 (1997)
[hep-ph/9706538].

\bibitem{Ullio:1998ke}
P.~Ullio and L.~Bergstr\"om,
``Neutralino annihilation into a photon and a Z boson,''
Phys.\ Rev.\  {\bf D57}, 1962 (1998)
[hep-ph/9707333].

\bibitem{Berezinsky:1992sn}
V.~S.~Berezinsky, A.~Bottino and V.~de Alfaro,
``Is it possible to detect the gamma ray line from
  neutralino-neutralino annihilation?,''
Phys.\ Lett.\  {\bf B274}, 122 (1992).

\bibitem{Freese:1986qw}
K.~Freese,
``Can Scalar Neutrinos Or Massive Dirac Neutrinos Be The Missing
Mass?,''
Phys.\ Lett.\  {\bf B167}, 295 (1986).

\bibitem{Krauss:1986aa}
L.~M.~Krauss, M.~Srednicki and F.~Wilczek,
``Solar System Constraints And Signatures For Dark Matter
Candidates,''
Phys.\ Rev.\  {\bf D33}, 2079 (1986).

\bibitem{Gaisser:1986ha}
T.~K.~Gaisser, G.~Steigman and S.~Tilav,
``Limits On Cold Dark Matter Candidates From Deep Underground
Detectors,''
Phys.\ Rev.\  {\bf D34}, 2206 (1986).

\bibitem{Gould:1989eq}
A.~Gould, J.~A.~Frieman and K.~Freese,
``Probing The Earth With Wimps,''
Phys.\ Rev.\  {\bf D39}, 1029 (1989).

\bibitem{Bottino:1994xp}
A.~Bottino, N.~Fornengo, G.~Mignola and L.~Moscoso,
``Signals of neutralino dark matter from earth and sun,''
Astropart.\ Phys.\  {\bf 3}, 65 (1995)
[hep-ph/9408391].

\bibitem{Berezinsky:1996ga}
V.~Berezinsky, A.~Bottino, J.~R.~Ellis, N.~Fornengo, G.~Mignola and S.~Scopel,
``Searching for relic neutralinos using neutrino telescopes,''
Astropart.\ Phys.\  {\bf 5}, 333 (1996)
[hep-ph/9603342].

\bibitem{Press:1985ug}
W.~H.~Press and D.~N.~Spergel,
``Capture by the sun of a galactic population of weakly interacting,
  massive particles,''
Astrophys.\ J.\  {\bf 296}, 679 (1985).

\bibitem{Silk:1985ax}
J.~Silk, K.~Olive and M.~Srednicki,
``The photino, the sun, and high-energy neutrinos,''
Phys.\ Rev.\ Lett.\  {\bf 55}, 257 (1985).

\bibitem{Srednicki:1987vj}
M.~Srednicki, K.~A.~Olive and J.~Silk,
``High-Energy Neutrinos From The Sun And Cold Dark Matter,''
Nucl.\ Phys.\  {\bf B279}, 804 (1987).

\bibitem{Hagelin:1986gv}
J.~S.~Hagelin, K.~W.~Ng and K.~A.~Olive,
``A High-Energy Neutrino Signature From Supersymmetric Relics,''
Phys.\ Lett.\  {\bf B180}, 375 (1986).

\bibitem{Ng:1987qt}
K.~Ng, K.~A.~Olive and M.~Srednicki,
``Dark Matter Induced Neutrinos From The Sun: Theory Versus
Experiment,''
Phys.\ Lett.\  {\bf B188}, 138 (1987).

\bibitem{Ellis:1988sh}
J.~Ellis and R.~A.~Flores,
``Realistic Predictions For The Detection Of Supersymmetric Dark
Matter,''
Nucl.\ Phys.\  {\bf B307}, 883 (1988).

\bibitem{Feng:2000zu}
J.~L.~Feng, K.~T.~Matchev and F.~Wilczek,
``Prospects for indirect detection of neutralino dark matter,''
Phys.\ Rev.\ D {\bf 63}, 045024 (2001)
[astro-ph/0008115].

\bibitem{Lykken:2000kp}
J.~D.~Lykken and K.~T.~Matchev,
``Supersymmetry signatures with tau jets at the Tevatron,''
Phys.\ Rev.\  {\bf D61}, 015001 (2000)
[hep-ph/9903238];
``Tau jet signals for supersymmetry at the Tevatron,''
hep-ex/9910033.

\bibitem{Matchev:1999nb}
K.~T.~Matchev and D.~M.~Pierce,
``Supersymmetry reach of the Tevatron via trilepton, like-sign
  dilepton and dilepton plus tau jet signatures,''
Phys.\ Rev.\  {\bf D60}, 075004 (1999)
[hep-ph/9904282].

\bibitem{Matchev:1999yn}
K.~T.~Matchev and D.~M.~Pierce,
``New backgrounds in trilepton, dilepton and dilepton plus tau jet
SUSY signals at the Tevatron,''
Phys.\ Lett.\  {\bf B467}, 225 (1999)
[hep-ph/9907505].

\bibitem{Giudice:1998bp} 
G.~F.~Giudice and R.~Rattazzi,
``Theories with gauge-mediated supersymmetry breaking,''
Phys.\ Rept.\  {\bf 322}, 419 (1999)
[hep-ph/9801271].

\bibitem{Pagels:ke}
H.~Pagels and J.~R.~Primack,
``Supersymmetry, Cosmology And New Tev Physics,''
Phys.\ Rev.\ Lett.\  {\bf 48}, 223 (1982).

\bibitem{Weinberg:zq}
S.~Weinberg,
``Cosmological Constraints On The Scale Of Supersymmetry Breaking,''
Phys.\ Rev.\ Lett.\  {\bf 48}, 1303 (1982).

\bibitem{Krauss:1983ik}
L.~M.~Krauss,
``New Constraints On 'Ino' Masses From Cosmology. 1. Supersymmetric 'Inos',''
Nucl.\ Phys.\ B {\bf 227}, 556 (1983).

\bibitem{Nanopoulos:1983up}
D.~V.~Nanopoulos, K.~A.~Olive and M.~Srednicki,
``After Primordial Inflation,''
Phys.\ Lett.\ B {\bf 127}, 30 (1983).

\bibitem{Khlopov:pf}
M.~Y.~Khlopov and A.~D.~Linde,
``Is It Easy To Save The Gravitino?,''
Phys.\ Lett.\ B {\bf 138} (1984) 265.

\bibitem{Ellis:1984eq}
J.~R.~Ellis, J.~E.~Kim and D.~V.~Nanopoulos,
``Cosmological Gravitino Regeneration And Decay,''
Phys.\ Lett.\ B {\bf 145}, 181 (1984).

\bibitem{Juszkiewicz:gg}
R.~Juszkiewicz, J.~Silk and A.~Stebbins,
``Constraints On Cosmologically Regenerated Gravitinos,''
Phys.\ Lett.\ B {\bf 158}, 463 (1985).

\bibitem{Bolz:2000fu}
M.~Bolz, A.~Brandenburg and W.~Buchmuller,
``Thermal production of gravitinos,''
Nucl.\ Phys.\ B {\bf 606}, 518 (2001)
[hep-ph/0012052].

\bibitem{Feng:2003xh}
J.~L.~Feng, A.~Rajaraman and F.~Takayama,
``Superweakly-interacting massive particles,''
Phys.\ Rev.\ Lett.\  {\bf 91}, 011302 (2003)
[hep-ph/0302215].

\bibitem{Feng:2003uy}
J.~L.~Feng, A.~Rajaraman and F.~Takayama,
``SuperWIMP dark matter signals from the early Universe,''
Phys.\ Rev.\ D {\bf 68}, 063504 (2003)
[hep-ph/0306024].

\bibitem{Covi:1999ty}
L.~Covi, J.~E.~Kim and L.~Roszkowski,
``Axinos as cold dark matter,''
Phys.\ Rev.\ Lett.\  {\bf 82}, 4180 (1999)
[hep-ph/9905212].

\bibitem{Covi:2001nw}
L.~Covi, H.~B.~Kim, J.~E.~Kim and L.~Roszkowski,
``Axinos as dark matter,''
JHEP {\bf 0105}, 033 (2001)
[hep-ph/0101009].

\bibitem{Covi:2004rb}
L.~Covi, L.~Roszkowski, R.~Ruiz de Austri and M.~Small,
``Axino dark matter and the CMSSM,''
hep-ph/0402240.

\bibitem{Hooper:2004qf}
D.~Hooper and L.~T.~Wang,
``Evidence for axino dark matter in the galactic bulge,''
hep-ph/0402220.

\bibitem{Feng:2003nr}
J.~L.~Feng, A.~Rajaraman and F.~Takayama,
``Graviton cosmology in universal extra dimensions,''
Phys.\ Rev.\ D {\bf 68}, 085018 (2003)
[hep-ph/0307375].

\bibitem{Ellis:2003dn}
J.~Ellis, K.~A.~Olive, Y.~Santoso and V.~C.~Spanos,
``Gravitino dark matter in the CMSSM,''
hep-ph/0312262.

\bibitem{Ellis:2004qe}
J.~Ellis, K.~A.~Olive, Y.~Santoso and V.~C.~Spanos,
``Very constrained minimal supersymmetric standard models,''
hep-ph/0405110.

\bibitem{Feng:1997zr}
J.~L.~Feng and T.~Moroi,
``Tevatron signatures of long-lived charged sleptons in
gauge-mediated supersymmetry breaking models,''
Phys.\ Rev.\ D {\bf 58}, 035001 (1998)
[hep-ph/9712499].

\bibitem{LHCstable} 
D.~Acosta, talk given at the {\em 14th Topical Conference on Hadron
Collider Physics}, September 29 - October 4, 2002, Germany.

\bibitem{Chen:2003gz}
X.~Chen and M.~Kamionkowski,
``Particle decays during the cosmic dark ages,''
astro-ph/0310473.

\bibitem{Sigurdson:2003vy}
K.~Sigurdson and M.~Kamionkowski,
``Charged-particle decay and suppression of small-scale power,''
astro-ph/0311486.

\bibitem{Ellis:1984er}
J.~R.~Ellis, D.~V.~Nanopoulos and S.~Sarkar,
``The Cosmology Of Decaying Gravitinos,''
Nucl.\ Phys.\ B {\bf 259}, 175 (1985).

\bibitem{Ellis:1990nb}
J.~R.~Ellis, G.~B.~Gelmini, J.~L.~Lopez, D.~V.~Nanopoulos and S.~Sarkar,
``Astrophysical Constraints On Massive Unstable Neutral Relic Particles,''
Nucl.\ Phys.\ B {\bf 373}, 399 (1992).

\bibitem{Kawasaki:1994sc}
M.~Kawasaki and T.~Moroi,
``Electromagnetic cascade in the early universe and its application 
to the big bang nucleosynthesis,''
Astrophys.\ J.\  {\bf 452}, 506 (1995)
[astro-ph/9412055].

\bibitem{Holtmann:1998gd}
E.~Holtmann, M.~Kawasaki, K.~Kohri and T.~Moroi,
``Radiative decay of a long-lived particle and big-bang 
nucleosynthesis,''
Phys.\ Rev.\ D {\bf 60}, 023506 (1999)
[hep-ph/9805405].

\bibitem{Kawasaki:2000qr}
M.~Kawasaki, K.~Kohri and T.~Moroi,
``Radiative decay of a massive particle and the non-thermal process 
in primordial nucleosynthesis,''
Phys.\ Rev.\ D {\bf 63}, 103502 (2001)
[hep-ph/0012279].

\bibitem{Asaka:1998ju}
T.~Asaka, J.~Hashiba, M.~Kawasaki and T.~Yanagida,
``Spectrum of background X-rays from moduli dark matter,''
Phys.\ Rev.\ D {\bf 58}, 023507 (1998)
[hep-ph/9802271].

\bibitem{Cyburt:2002uv}
R.~H.~Cyburt, J.~Ellis, B.~D.~Fields and K.~A.~Olive,
``Updated nucleosynthesis constraints on unstable relic particles,''
Phys.\ Rev.\ D {\bf 67}, 103521 (2003)
[astro-ph/0211258].

\bibitem{Reno:1987qw}
M.~H.~Reno and D.~Seckel,
``Primordial Nucleosynthesis: The Effects Of Injecting Hadrons,''
Phys.\ Rev.\ D {\bf 37}, 3441 (1988).

\bibitem{Dimopoulos:1988ue}
S.~Dimopoulos, R.~Esmailzadeh, L.~J.~Hall and G.~D.~Starkman,
``Limits On Late Decaying Particles From Nucleosynthesis,''
Nucl.\ Phys.\ B {\bf 311}, 699 (1989).

\bibitem{Khlopov:rs}
M.~Y.~Khlopov,
{\em Cosmoparticle Physics}, Singapore: World Scientific, 1999. 

\bibitem{Kohri:2001jx}
K.~Kohri,
``Primordial nucleosynthesis and hadronic decay of a massive particle
with a relatively short lifetime,''
Phys.\ Rev.\ D {\bf 64}, 043515 (2001)
[astro-ph/0103411].

\bibitem{Hagiwara:fs}
K.~Hagiwara {\it et al.}  [Particle Data Group Collaboration],
``Review Of Particle Physics,''
Phys.\ Rev.\ D {\bf 66}, 010001 (2002).

\bibitem{Cyburt:2003fe}
R.~H.~Cyburt, B.~D.~Fields and K.~A.~Olive,
``Primordial Nucleosynthesis in Light of WMAP,''
Phys.\ Lett.\ B {\bf 567}, 227 (2003)
[astro-ph/0302431].

\bibitem{Kirkman:2003uv}
D.~Kirkman, D.~Tytler, N.~Suzuki, J.~M.~O'Meara and D.~Lubin,
``The cosmological baryon density from the deuterium to hydrogen ratio towards
QSO absorption systems: D/H towards Q1243+3047,''
Astrophys.\ J.\ Suppl.\  {\bf 149}, 1 (2003)
[astro-ph/0302006].

\bibitem{Burles:2000zk}
S.~Burles, K.~M.~Nollett and M.~S.~Turner,
``Big-Bang Nucleosynthesis Predictions for Precision Cosmology,''
Astrophys.\ J.\  {\bf 552}, L1 (2001)
[astro-ph/0010171].

\bibitem{Jedamzik:1999di}
K.~Jedamzik,
``Lithium-6: A Probe of the Early Universe,''
Phys.\ Rev.\ Lett.\  {\bf 84}, 3248 (2000)
[astro-ph/9909445].

\bibitem{Jedamzik:2004er}
K.~Jedamzik,
``Did something decay, evaporate, or annihilate during big bang
nucleosynthesis?,''
astro-ph/0402344.

\bibitem{Kawasaki:2004yh}
M.~Kawasaki, K.~Kohri and T.~Moroi,
``Hadronic decay of late-decaying particles and big-bang nucleosynthesis,''
astro-ph/0402490.

\bibitem{Feng:2004zu}
J.~L.~Feng, S.~f.~Su and F.~Takayama,
``SuperWIMP gravitino dark matter from slepton and sneutrino decays,''
hep-ph/0404198.

\bibitem{Feng:2004mt}
J.~L.~Feng, S.~Su and F.~Takayama,
``Supergravity with a gravitino LSP,''
hep-ph/0404231.

\bibitem{Buchmuller:2004rq}
W.~Buchmuller, K.~Hamaguchi, M.~Ratz and T.~Yanagida,
``Supergravity at colliders,''
Phys.\ Lett.\ B {\bf 588}, 90 (2004)
[hep-ph/0402179].

\bibitem{Buchmuller:2004tm}
W.~Buchmuller, K.~Hamaguchi, M.~Ratz and T.~Yanagida,
``Gravitino and goldstino at colliders,''
hep-ph/0403203.

\bibitem{Wang:2004ib}
F.~Wang and J.~M.~Yang,
``SuperWIMP dark matter scenario in light of WMAP,''
hep-ph/0405186.

\bibitem{Hu:gc}
W.~Hu and J.~Silk,
``Thermalization Constraints And Spectral Distortions For 
 Massive Unstable Relic Particles,''
Phys.\ Rev.\ Lett.\  {\bf 70}, 2661 (1993).

\bibitem{Fixsen:1996nj}
D.~J.~Fixsen \etal,
``The Cosmic Microwave Background Spectrum from the Full COBE/FIRAS 
 Data Set,''
Astrophys.\ J.\  {\bf 473}, 576 (1996)
[astro-ph/9605054].

\bibitem{DIMES}
http://map.gsfc.nasa.gov/DIMES.

\bibitem{Feng:2002tc}
J.~L.~Feng and M.~M.~Nojiri,
``Supersymmetry and the linear collider,''
hep-ph/0210390.

\bibitem{Planck}
http://www.rssd.esa.int/index.php?project=PLANCK.

\bibitem{Drees:2000he}
M.~Drees, Y.~G.~Kim, M.~M.~Nojiri, D.~Toya, K.~Hasuko and T.~Kobayashi,
``Scrutinizing LSP dark matter at the LHC,''
Phys.\ Rev.\ D {\bf 63}, 035008 (2001)
[hep-ph/0007202].

\bibitem{Feng:1995zd}
J.~L.~Feng, M.~E.~Peskin, H.~Murayama and X.~Tata,
``Testing supersymmetry at the next linear collider,''
Phys.\ Rev.\ D {\bf 52}, 1418 (1995)
[hep-ph/9502260].

\bibitem{Peccei:1977ur}
R.~D.~Peccei and H.~R.~Quinn,
``Constraints Imposed By CP Conservation In The Presence Of Instantons,''
Phys.\ Rev.\ D {\bf 16}, 1791 (1977).

\bibitem{Weinberg:1977ma}
S.~Weinberg,
``A New Light Boson?,''
Phys.\ Rev.\ Lett.\  {\bf 40}, 223 (1978).

\bibitem{Wilczek:pj}
F.~Wilczek,
``Problem Of Strong P And T Invariance In The Presence Of Instantons,''
Phys.\ Rev.\ Lett.\  {\bf 40}, 279 (1978).

\bibitem{Kusenko:1997si}
A.~Kusenko and M.~E.~Shaposhnikov,
``Supersymmetric Q-balls as dark matter,''
Phys.\ Lett.\ B {\bf 418}, 46 (1998)
[hep-ph/9709492].

\bibitem{Kusenko:1997vp}
A.~Kusenko, V.~Kuzmin, M.~E.~Shaposhnikov and P.~G.~Tinyakov,
``Experimental signatures of supersymmetric dark-matter Q-balls,''
Phys.\ Rev.\ Lett.\  {\bf 80}, 3185 (1998)
[hep-ph/9712212].

\bibitem{Enqvist:2003gh}
K.~Enqvist and A.~Mazumdar,
``Cosmological consequences of MSSM flat directions,''
Phys.\ Rept.\  {\bf 380}, 99 (2003)
[hep-ph/0209244].

\end{thebibliography}
\end{document}